\documentclass[times]{nmeauth}
\pdfoutput=1
\newcommand{\sqz}{{\mathrm{sqz}}}
\newcommand{\psl}{{\mathrm{psl}}}
\newcommand{\ctt}{{\mathrm{ctt}}}
\newcommand{\shr}{{\mathrm{shr}}}
\newcommand{\cav}{{\mathrm{cav}}}

\newcommand{\lubconletter}{{\mathrm{lub}}}
\newcommand{\lubconbletter}{{\mathrm{l}}}

\newcommand{\tns}[1]{\boldsymbol{#1}}

\newcommand{\grad}   { \boldsymbol{\nabla}    }
\newcommand{\surfgrad}   { \tilde{\boldsymbol{\nabla}}    }

\newcommand{\unity}{\tns{I}}
\newcommand{\unitydim}{\tns{I}_{ndim}}

\newcommand{\lubletter}{{\mathrm{f}}}
\newcommand{\lubcavletter}{{\mathrm{c}}}

\newcommand{\presf}{\pres}
\newcommand{\roughness}{R_q}
\newcommand{\slaveroughness}{R_{q, \slaveletter}}
\newcommand{\masterroughness}{R_{q, \masterletter}}
\newcommand{\film}{h}
\newcommand{\vis}{\eta}
\newcommand{\young}{E}
\newcommand{\poisson}{\nu}
\newcommand{\density}{\rho}
\newcommand{\pflowfac}{\Phi_p}
\newcommand{\sflowfac}{\Phi_s}
\newcommand{\fflowfac}{\Phi_f}
\newcommand{\velf}{\tns{v}}
\newcommand{\velftilde}{\tilde{\velf}}
\newcommand{\stddev}{\varrho}
\newcommand{\epsilonnn}{\epsilon}
\newcommand{\phatf}{\hat{p}}
\newcommand{\normveclub}{\tns{n}_{\lubconbletter}}

\newcommand{\gap}{g}
\newcommand{\gaphat}{\hat{g}}
\newcommand{\regthick}{\gap_{\mathrm{max}}}
\newcommand{\regstiff}{\kappa}
\newcommand{\prescn}{\pres_n}
\newcommand{\regstrain}{\varepsilon_{\mathrm{Layer}}}
\newcommand{\eprime}{E^\prime}
\newcommand{\modulelasticity}{E}

\newcommand{\disps}{\tns{u}}
\newcommand{\gendisps}{\tns{u}^{(i)}}
\newcommand{\vels}{\tns{v}}
\newcommand{\velshat}{\hat{\vels}}
\newcommand{\velstilde}{\tilde{\vels}}
\newcommand{\dispsddot}{\ddot{\disps}}
\newcommand{\dispsdot}{\dot{\disps}}
\newcommand{\dispshat}{\hat{\disps}}
\newcommand{\refpointslave}{\boldsymbol{X}^{(\slaveletter)}}

\newcommand{\pointslave}{\boldsymbol{x}^{(\slaveletter)}}
\newcommand{\pointslavedot}{\dot{\boldsymbol{x}}^{(\slaveletter)}}

\newcommand{\pointmasterhat}{\hat{\boldsymbol{x}}^{(\masterletter)}}
\newcommand{\pointmasterhatdot}{\hat{\dot{\boldsymbol{x}}}^{(\masterletter)}}
\newcommand{\defgradient}{\tns{F}}
\newcommand{\defgradienttr}{\tns{F}^{\mathrm{T}}}
\newcommand{\firstpiola}{\tns{P}}
\newcommand{\secondpiola}{\tns{S}}
\newcommand{\refnormvec}{\tns{N}}
\newcommand{\straintensor}{\tns{E}}
\newcommand{\normvec}{\tns{n}}
\newcommand{\refbodyforcehat}{\hat{\tns{b}}_0}
\newcommand{\strainenergyi}{\partial \Psi_{\mathrm{NH}}}
\newcommand{\strainenergyii}{\partial^2 \Psi_{\mathrm{NH}}}
\newcommand{\ctensor}{\mathbb{C}}
\newcommand{\stresss}{\boldsymbol{\sigma}}
\newcommand{\genrefpoint}{\tns{X}^{(i)}}

\newcommand{\contactletter}{{\mathrm{c}}}
\newcommand{\traction}{\tns{t}}
\newcommand{\reftractionhat}{\hat{\tns{t}}_0}
\newcommand{\contraction}{\tns{t}_{\contactletter}}
\newcommand{\tractionlambda}{\tns{\lambda}^{\contactletter}}
\newcommand{\tractionlambdan}{\tns{\lambda}_{n}^{\contactletter}}
\newcommand{\tractionlambdat}{\tns{\lambda}_{t}^{\contactletter}}
\newcommand{\contractiontang}{\tns{t}_{\tau}}
\newcommand{\lubtraction}{\tns{t}_{\lubletter}}
\newcommand{\lubtractionparallel}{{\tns{t}_{\lubletter}}_{\parallel}}
\newcommand{\lubtractionnparallel}{{\tns{t}_{\lubletter}}_{\nparallel}}
\newcommand{\lubcontraction}{\tns{t}_{\lubconletter}}
\newcommand{\coulomblaw}{f^{\mathrm{fr}}}
\newcommand{\dryfriction}{\mu}
\newcommand{\coulomblawbeta}{\beta}
\newcommand{\ineqnormalcon}{C_{nj}}
\newcommand{\ineqtanjcon}{C_{\tau j}}
\newcommand{\slaveletter}{{1\text{}}}
\newcommand{\masterletter}{{2\text{}}}

\newcommand{\domain}{\Omega}
\newcommand{\gendomain}{\Omega_{t}^{(i)}}
\newcommand{\genrefdomain}{\Omega_{0}^{(i)}}

\newcommand{\domainlub}{\domain_{\lubletter,\mtime}}

\newcommand{\domainf}{\Omega_{\lubletter}}

\newcommand{\domainfpositive}{\Omega_{\lubletter_{\lubconbletter}}}
\newcommand{\domainfcav}{\Omega_{\lubletter_{\lubcavletter}}}
\newcommand{\dirich}{{\mathrm{D}}}

\newcommand{\reybound}{{\mathrm{R}}}
\newcommand{\refinterface}{\Gamma}
\newcommand{\interface}{\gamma}
\newcommand{\slaveinterface}{\interface_{\lubconbletter}^{(\slaveletter)}}
\newcommand{\masterinterface}{\interface_{\lubconbletter}^{(\masterletter)}}
\newcommand{\refslaveinterface}{\refinterface_{\lubconbletter}^{(\slaveletter)}}
\newcommand{\refmasterinterface}{\refinterface_{\lubconbletter}^{(\masterletter)}}

\newcommand{\reflubDinterface}{\refinterface_{\lubletter_{\dirich}}}
\newcommand{\reflubreyinterface}{\refinterface_{\lubletter_{\reybound}}}

\newcommand{\refgeninterface}{\refinterface_{\lubconbletter}^{(i)}}
\newcommand{\refgenNinterface}{\refinterface_{\stress}^{(i)}}
\newcommand{\refgenDinterface}{\refinterface_{\disp}^{(i)}}
\newcommand{\geninterface}{\interface_{\lubconbletter}^{(i)}}
\newcommand{\genNinterface}{\interface_{\stress}^{(i)}}
\newcommand{\genDinterface}{\interface_{\disp}^{(i)}}

\newcommand{\mtime}{t}
\newcommand{\mendtime}{T}
\newcommand{\disp}{u}

\newcommand{\pres}{p}
\newcommand{\testfkt}[1]{\delta {#1}}
\newcommand{\stress}{\sigma}

\newcommand{\mortard}{\tns{D}}
\newcommand{\mortarm}{\tns{M}}
\newcommand{\massmatrix}{\mathit{M}}

\newcommand{\Real}{\mathbb{R}}
\newcommand{\Realdim}{\Real^{\dimensions}}
\newcommand{\bimapi}{\phi}

\newcommand{\bimapiii}{\chi}
\newcommand{\bimapiiih}{\chi_{\femletter}}

\newcommand{\dimensions}{n}
\newcommand{\solutionspacepres}{\mathcal{P}}
\newcommand{\testspacepres}{\mathcal{Q}}
\newcommand{\discretespacet}{\mathcal{T}}
\newcommand{\discretespacen}{\mathcal{N}}
\newcommand{\slavespace}{\mathcal{S}}
\newcommand{\masterspace}{\mathcal{M}}
\newcommand{\finalsys}{\mathcal{K}}
\newcommand{\solutionspacedisp}{\tns{\mathcal{U}}}
\newcommand{\testspacedisp}{\tns{\mathcal{V}}}

\newcommand{\galerkindisp}{\mathrm{G}_{\disp}}
\newcommand{\galerkinf}{\mathrm{G}_{\lubletter}}

\newcommand{\virtwlub}{\mathrm{W}_{\lubconletter}}
\newcommand{\virtwlubh}{\mathrm{W}_{\lubconletter,\femletter}}
\newcommand{\virtwf}{\mathrm{W}_{\lubletter}}
\newcommand{\virtwc}{\mathrm{W}_{\contactletter}}
\newcommand{\virtwkin}{\mathrm{W}_{\mathrm{kin}}}
\newcommand{\virtwintex}{\mathrm{W}_{\mathrm{int,ext}}}
\newcommand{\internal}{{\mathrm{int}}}
\newcommand{\external}{{\mathrm{ext}}}

\newcommand{\discreteglobalvec}{\boldsymbol{f}}
\newcommand{\tractionlambdah}{\tns{\lambda}_{\femletter}^{\contactletter}}
\newcommand{\tractionlambdaj}{\tns{\lambda}_{j}^{\contactletter}}
\newcommand{\dispd}{\tns{d}}
\newcommand{\prestns}{\tns{p}}
\newcommand{\residual}{\tns{r}}
\newcommand{\femletter}{{\mathrm{h}}}
\newcommand{\slaveinterfaceh}{\interface_{\lubconbletter, \femletter}^{(\slaveletter)}}
\newcommand{\refslaveinterfaceh}{\refinterface_{\lubconbletter, \femletter}^{(\slaveletter)}}
\newcommand{\masterinterfaceh}{\interface_{\lubconbletter,\femletter}^{(\masterletter)}}
\newcommand{\refmasterinterfaceh}{\refinterface_{\lubconbletter,\femletter}^{(\masterletter)}}
\newcommand{\refdomainoneh}{\Omega_{0,\femletter}^{(1)}}
\newcommand{\refdomaintwoh}{\Omega_{0,\femletter}^{(2)}}
\newcommand{\refdomainfh}{\Omega_{\lubletter,0,\femletter}}

\newcommand{\velftildeh}{\velftilde_{\mathrm{h}}}
\newcommand{\velftildek}{\velftilde_k}

\newcommand{\presh}{\pres_{\mathrm{h}}}
\newcommand{\filmh}{\film_{\mathrm{h}}}
\newcommand{\filmk}{\film_k}
\newcommand{\pointoneh}{\boldsymbol{x}_{\femletter}^{(1)}}
\newcommand{\pointonek}{\boldsymbol{x}_{k}^{(1)}}
\newcommand{\pointtwoh}{\boldsymbol{x}_{\femletter}^{(2)}}
\newcommand{\pointtwom}{\boldsymbol{x}_{m}^{(2)}}
\newcommand{\dispsoneh}{\disps_{\femletter}^{(1)}}
\newcommand{\dispstwoh}{\disps_{\femletter}^{(2)}}
\newcommand{\dispsgenh}{\disps_{\femletter}^{(i)}}
\newcommand{\tractionfoneh}{\traction_{\lubletter,\femletter}^{(1)}}

\newcommand{\ndnode}{n_{\mathrm{dnod}}}
\newcommand{\nnodedim}{n_{\mathrm{dim}}}
\newcommand{\nnodedof}{n_{\mathrm{ddof}}}
\newcommand{\nnodeone}{n_{\mathrm{nod}}^{(1)}}
\newcommand{\nnodetwo}{n_{\mathrm{nod}}^{(2)}}
\newcommand{\nbnodeone}{n_{\mathrm{bnod}}^{(1)}}
\newcommand{\nbnodetwo}{n_{\mathrm{bnod}}^{(2)}}
\newcommand{\npnode}{n_{\mathrm{pnod}}}
\newcommand{\kroneckersym}{\delta_{ij}}

\newcommand{\shapef}{N}
\newcommand{\dualshapef}[1]{\Phi_{#1}}

\usepackage[colorlinks,bookmarksopen,bookmarksnumbered,citecolor=red,urlcolor=red]{hyperref}
\usepackage{amsmath, amssymb, amstext, amsfonts, mathtools, mathrsfs, titletoc}
\usepackage{amsthm}
\usepackage{relsize}
\usepackage{nicefrac} 
\usepackage{algorithm}
\usepackage{algorithmic}
\usepackage{anyfontsize}
\newtheorem*{remark}{Remark}
\usepackage[sort&compress,numbers]{natbib}
\usepackage[scaled=0.92]{helvet} 
\usepackage{array}
\usepackage{stmaryrd}
\usepackage{arydshln}
\usepackage{graphicx}
\usepackage{tabularx} 
\usepackage{booktabs} 
\usepackage{multirow} 
\usepackage{upgreek}
\usepackage[font=small, labelfont=md]{caption, subfig}
\graphicspath{{fig/}}
\DeclareGraphicsExtensions{.pdf}

\begin{document}
\runningheads{M.~Faraji et al.}{A preprint - \today}
\title{A Mortar Finite Element Formulation for Large Deformation Lubricated Contact Problems with Smooth Transition Between Mixed, Elasto-Hydrodynamic and Full Hydrodynamic Lubrication}
\author{Mostafa~Faraji\corrauth, Alexander~Seitz, Christoph~Meier, Wolfgang~A.~Wall}
\address{Institute for Computational Mechanics,
Technical University of Munich,\linebreak
Boltzmannstr. 15, 85747 Garching b. M\"u{}nchen, Germany
}

\corraddr{Mostafa~Faraji, Institute for Computational Mechanics, Technical University of Munich, Boltzmannstra{\ss}e 15, D-85747 Garching, Germany. E-mail:~m.faraji@tum.de}

\begin{abstract}
This work proposes a novel model and numerical formulation for lubricated contact problems describing the mutual interaction between two deformable 3D solid bodies and an interposed fluid film. The solid bodies are consistently described based on nonlinear continuum mechanics allowing for finite deformations and arbitrary constitutive laws. The fluid film is modelled as a quasi-2D flow problem on the interface between the solids governed by the (thickness-)averaged Reynolds equation, which relates pressure to velocity and film thickness, with the latter two fields being provided by the averaged surface velocity and gap profile of the interacting solid bodies. The averaged Reynolds equation accounts for surface roughness utilizing spatially homogenized, effective fluid parameters and for cavitation through a positivity constraint imposed on the pressure field. In contrast to existing approaches, the proposed model accounts for the co-existence of frictional contact tractions and hydrodynamic fluid tractions at every local point on the contact surface of the interacting bodies and covers the entire range from boundary lubrication to mixed, elastohydrodynamic, and eventually to full film hydrodynamic lubrication in one unified modelling framework with smooth transition between these different regimes. Critically, the model relies on a recently proposed regularization scheme for the mechanical contact constraint combining the advantages of classical penalty and Lagrange multiplier approaches by expressing the mechanical contact pressure as a function of the effective gap between the solid bodies while at the same time limiting the minimal gap value occurring at the (theoretical) limit of infinitely high contact pressures. From a methodological point of view, this is the key ingredient to regularize the pressure field in the averaged Reynolds equation, i.e., to avoid the pressure field's singularity in the limit of vanishing fluid film thickness, and thus to enable a smooth transition between all relevant lubrication regimes. From a physical point of view, this approach can be considered as a model for the elastic deformation of surface asperities, with a bounded magnitude depending on the interacting solids' surface roughness. The finite element method is applied for spatial discretization of the 3D solid-mechanical problems and the 2D interface effects, consisting of the averaged Reynolds equation governing the fluid film and the non-penetration constraint of the mechanical contact problem, the latter relying on variationally consistent mortar methods for contact traction discretization. A consistent and accurate model behavior is demonstrated and validated by employing several challenging and practically relevant benchmark test cases. The ability of the model to accurately represent the velocity-dependent friction coefficient of the different lubrication regimes (i.e. mixed, elasto-hydrodynamic and full film lubrication) according to the well-known Stribek curve is also demonstrated via test cases. Eventually, a parametric study is performed to analyze the effect of regularization parameter choice.
\end{abstract}

\maketitle

\section{Introduction}
The development of a new model and numerical formulation that allows for the investigation of the transition from boundary lubrication to elastohydrodynamic lubrication (EHL) of lubricated contact problems is the focus of this contribution. The interaction of contacting surfaces separated by a thin fluid film is of great importance in various engineering and biomechanical applications. It is related both to the large field of contact problems as well as to the wide field of FSI (fluid-structure interaction) problems involving the relative motion, and possibly, the deformations of solids upon their interaction with fluids. Many areas that correspond to these type of problems, mainly vary in space scales, time scales, operating conditions, and material properties. This broad range prohibits the use of one type of approach fitting for all these problems, and requires the use of specific methods tailored for the questions and problems of interest. One approach would be to try to tackle lubricated contact problems starting from a full FSI approach, i.e. using a 3D fluid model goverend by the Navier-Stokes equations. Among others, this would allow to handle scenarios where both lubricated contact regions as well as larger flow fields, that are connected to it and are also interacting with deformable solids, need to be handled. However, most FSI approaches are unable to handle topology changes or contact scenarios. Recently, novel approaches and models that allow such scenarios have been introduced for example in \cite{ager_seitz_2019, ager_porous}. Classically, however, lubrication type problems are based on a reduced, thickness-averaged fluid model defined on the 2D interface between the interacting bodies. Therefore, the lubricated contact problem is a particular kind of FSI problem, in which the fluid part is modeled using the thickness-averaged Reynolds equation upon adopting the thin-film approximation. The continuum mechanics problem underlying the solid domain in the existing lubrication approaches is typically modeled using linear elasticity frameworks by assuming small deformations and a linear elastic material behavior, further simplified often in combination with additional linear elastic half-space approximations (\cite{dowson_higginson_1977, hamrock_1994}). The use of linear elasticity and half-space approximations works well for hard EHL problems, while in soft EHL problems, this approach may not be suitable due to the large deformations and associated geometrical and material nonlinearities occuring in this application. Formulations investigating the EHL of two high-stiffness elements, such as spur gears or ball bearings, are known as hard EHL. Relatively high pressures feature those applications, consequently making the effect of pressure-dependent viscosity (piezo-viscosity) important in hard EHL. On the other hand, soft EHL is considered for applications in which one or both of the lubricated bodies are characterized by a soft material behavior, such as rubber seals or wet tires. Another major application area for soft EHL models are biotribological systems, with examples being synovial joints, contact-lens lubrication, eye eyelid contact, human skin contacts, and oral processing of food (e.g., \cite{dowson_1995,vicente,adams_2007,jones_2007}). Thereby, large elastic deformations take place despite low fluid pressure, making the problem more complex from a modeling point of view. The focus of this paper is on the more challenging case of soft EHL problems possibly including the above mentioned different lubrication regimes and, in particular, on the consistent representation of finite deformations and the (possibly nonlinear) material behavior in the solid domain. \\
By definition, application areas in the lubrication field are multi-disciplinary in nature combining aspects, e.g., from solid mechanics to tribology and further to hydrodynamics and different application areas like, for example, biomechanics. This vast range of fields involving lubrication phenomena urges an in-depth theoretical understanding of the fundamental physics. However, only a small portion of the lubrication problems can be studied analytically as usually significant simplifications, e.g. of the geometry, are required. Experimental investigations in this field comprise initial studies on the lubricant-roller bearing interactions \cite{crook_1963}. Although helpful for studying various material characteristics like lubricant viscosity \cite{crouch} and counter checking the theoretical analysis, the high costs and the limited accessibility of certain quantities and physical fields in experimental investigations are major hindrances that do not favor their extensive application. Therefore, numerical modeling can be considered as a highly promising approach to study the general class of lubricated contact problems. The focus in this paper is set on the macroscopic continuum perspective of describing the interaction between the fluid lubricant and the confining solid bodies, rather than computationally expensive microscopic approaches, explicitly resolving the length scale of individual surface asperities and the associated 3D flow problem by direct numerical simulation of the Navier-Stokes equation.\\
In recent times, the Tribology community is paying increased attention to mixed lubrication models. In typical engineering applications, the nominal contact area is different from the real contact area, which means that the contacting surfaces interact at discrete points due to the asperities' presence (attributed to surface roughness). Hence, when mechanical contact is established between the peaks of surface asperities on the contacting interfaces, resulting e.g. from the high applied load, the high surface roughness or the low relative sliding velocities, the problem enters into the mixed lubrication regime. At this point, the load-carrying role is shared between the surface asperities as well as the lubricant. In this case, the effect of surface roughness also needs to be considered. Researchers have proposed various stochastic models to deal with surface roughness by statistical parameters, with the first approaches dating back to the pioneering work of Patir and Cheng (PC) \cite{patir1,patir2}. They solved the rough surface model problems and derived an averaged flow model based on representative flow factors. These flow factors are included as coefficients in a modified Reynolds equation solved on a smooth macroscale domain without resolving asperities. This distinguishes the problem in view of the averaged effects from the deterministic roughness. Furthermore, the mixed lubrication necessitates modeling the asperity contact constraints.
First fully-coupled and monolithic system approaches to resolve lubricated contacts were presented in \cite{houpert1986}. Later on, a semi-system approach was presented, which could handle EHL problems in a wide range of operating conditions by enhancing the contribution of the right-hand side of the Reynolds equation \cite{ai}. A model to manage the EHL region and the asperity contact region simultaneously for the first time was proposed in \cite{jiang1999}. They solved the Reynolds equation in the EHL region by utilizing a multi-grid scheme and treated negative film thickness as penetration to get asperity contact pressure by de-convolution in the asperity contact region. Afterwards a unified model in succession was published, where the Reynolds equation was employed in the whole interaction area \cite{huzhu}. In 2000, EHL formulations were improved towards higher numerical efficiency due to localized couplings and, therefore sparsely populated matrices \cite{evans2000}. Azam et al. presented a model to simulate the tribofilm growth within the unified mixed lubrication framework \cite{ghanbar2019}. The limitation of all these studies was the fact that all of them relied on the linear elastic half-space approximation.\\
The first contributions towards the nonlinear finite deformation regime were made in \cite{nikas_2002,stup2009, ongun} for the soft EHL problem underlying elastomeric seals. In those works, the sliding rod and housing of the seal are assumed to be rigid bodies, which circumvents the interface coupling modeling. In \cite{schmidt_2010} the transient averaged Reynolds equation for the computation of soft EHL was solved based on the strong coupling of a nonlinear finite element model. Their method is an extension of the study in \cite{ongun} which realize the mixed lubrication case by an exponential contact model using the critical film thickness and contact pressure estimated from experiments and is restricted to planar or axisymmetric geometries. In the meanwhile, new robust techniques for interface discretization such as the mortar method were developed to tie non-matching meshes and also in the context of frictional contact mechanics. The first application of mortar finite element discretizations to lubricated contact problems was performed in \cite{yang2009}. Therein, the lubricant film thickness is directly related to the gap between the deforming bodies' surfaces by means of mortar projection. In turn, the fluid forces are prescribed to the solids' surfaces, leading to a flexible formulation, applicable to a wide range of lubricated contact problems in the full respectively elastohydrodynamic lubrication regime. Recently in \cite{andrei_2019} a monolithic finite-element framework was proposed to solve thin-film flow in a contact interface between a deformable solid with resolved asperities and a rigid flat surface. In their model, identifying the local status of each interface element is required to distinguish between contact and fluid flow and then, the respected domains are solved separately as either dry contact or lubrication problem. The fluid flow is solved using a simplified version of the Reynolds equation without considering the tangential relative motion of the solid walls. All the works mentioned above have tried to introduce a suitable approach to solve the lubricated contact problems. However, they lack contact between two deformable solids with arbitrary surface geometries and do not address the mixed lubrication regime along with other lubrication regimes in a unified manner. Therefore, developing a comprehensive tool is necessary to treat all the complexities involved in lubricated contact problems simultaneously.\\
The present paper now closes the gap of existing approaches by developing an averaged model for the lubricated contact between deformable 3D solid bodies based on a novel approach. The solid bodies are consistently characterized using nonlinear continuum mechanics granting consideration of finite deformations and arbitrary constitutive laws. The fluid film is described as a quasi 2D flow problem on the interface between the solids governed by the averaged Reynolds equation. The averaged Reynolds equation accounts implicitly for surface roughness employing spatially homogenized, effective fluid parameters. Contrary to the existing approaches, the proposed model considers the co-existence of frictional contact tractions and hydrodynamic fluid tractions at every local point on the contact surface of the interacting solids, leading to a unified framework capable of modeling the full range of lubrication regimes from boundary to full film lubrication in a continuous way with smooth transition. Furthermore, it combines the advantages of classical penalty and Lagrange multiplier methods by expressing the mechanical contact pressure as a function of the effective gap between the solid bodies while limiting the minimal gap value taking place at the limit of infinitely high contact pressures. The finite element method is applied for spatial discretization of the problem allowing the possible application of the model to very general and complex geometries. The mechanical contact tractions are discretized by variationally consistent mortar methods. A consistent and accurate model behavior is demonstrated and validated by employing several challenging and practically relevant benchmark test cases. The ability of the model to accurately represent the velocity-dependent friction coefficient of the different lubrication regimes (i.e. mixed, elasto-hydrodynamic and full film lubrication) according to the well-known Stribek curve is also demonstrated via test cases. Eventually, a parametric study is performed to analyze the effect of regularization parameter choice.\\
The outline of this paper is as follows. In Section 2, the lubricated contact model for rough, thin-film FSI is depicted. In Section 3, the solid domain's governing equations, lubrication domain, and the conditions on their coupling and contact interface are discussed. A particular focus is set on the contact behavior of rough surfaces, the choice of the regularization function, and regularized contact conditions. This is followed by presenting the proposed finite element formulations in Section 4, including a derivation of the weak form and the discretization of the individual fields. The computational results of several challenging examples are presented in Section 5, demonstrating the capability of the proposed computational model to serve as a valuable tool for complex applications in lubricated contact problems including solids undergoing large deformation and nonlinear material behavior.
\section{Lubricated contact model}\label{model}
\begin{figure}[h!]
  \centering
  {\includegraphics[width=1.0\textwidth]{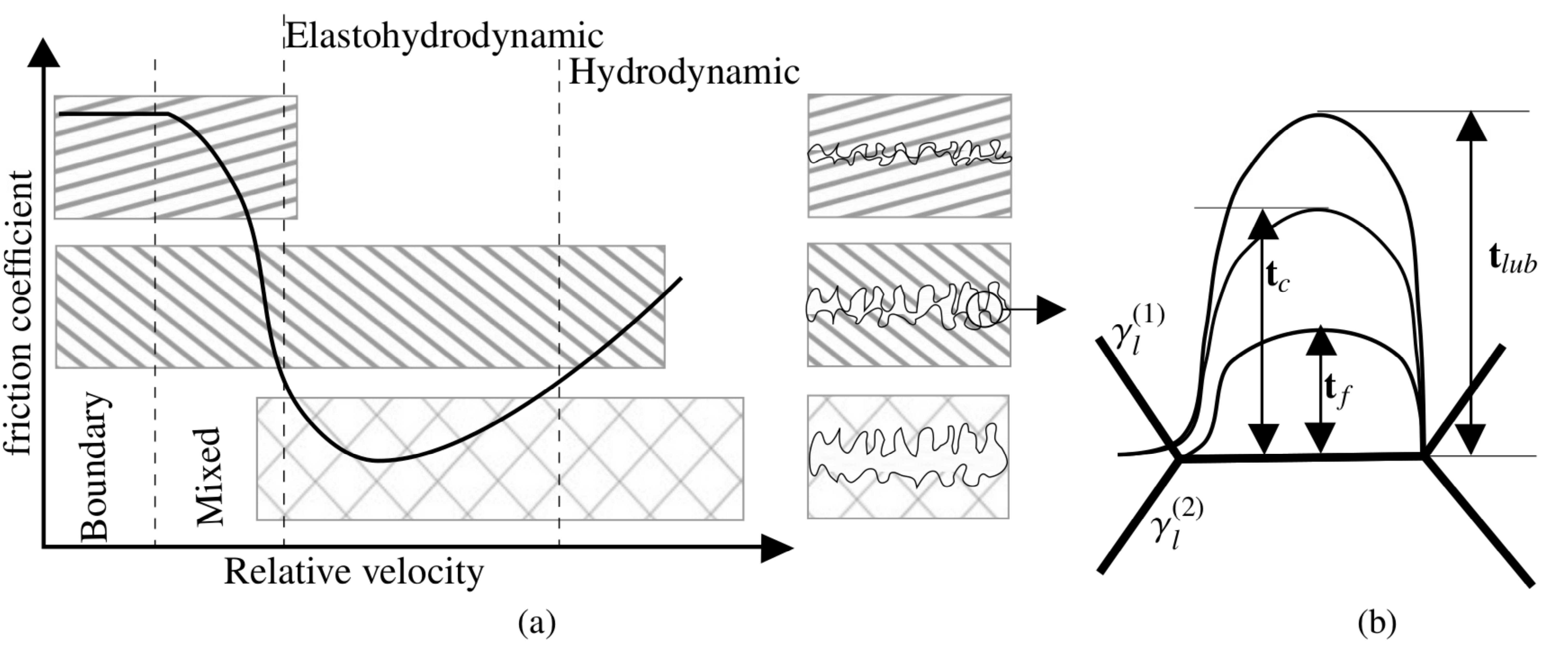}}
  \caption{(a) Lubrication regimes. (b) Schematic of Total pressure in mixed lubrication}
  \label{fig:stribek-mixmodel}
\end{figure}

\begin{figure}[ht]
  \centering
  {\includegraphics[width=1.0\textwidth]{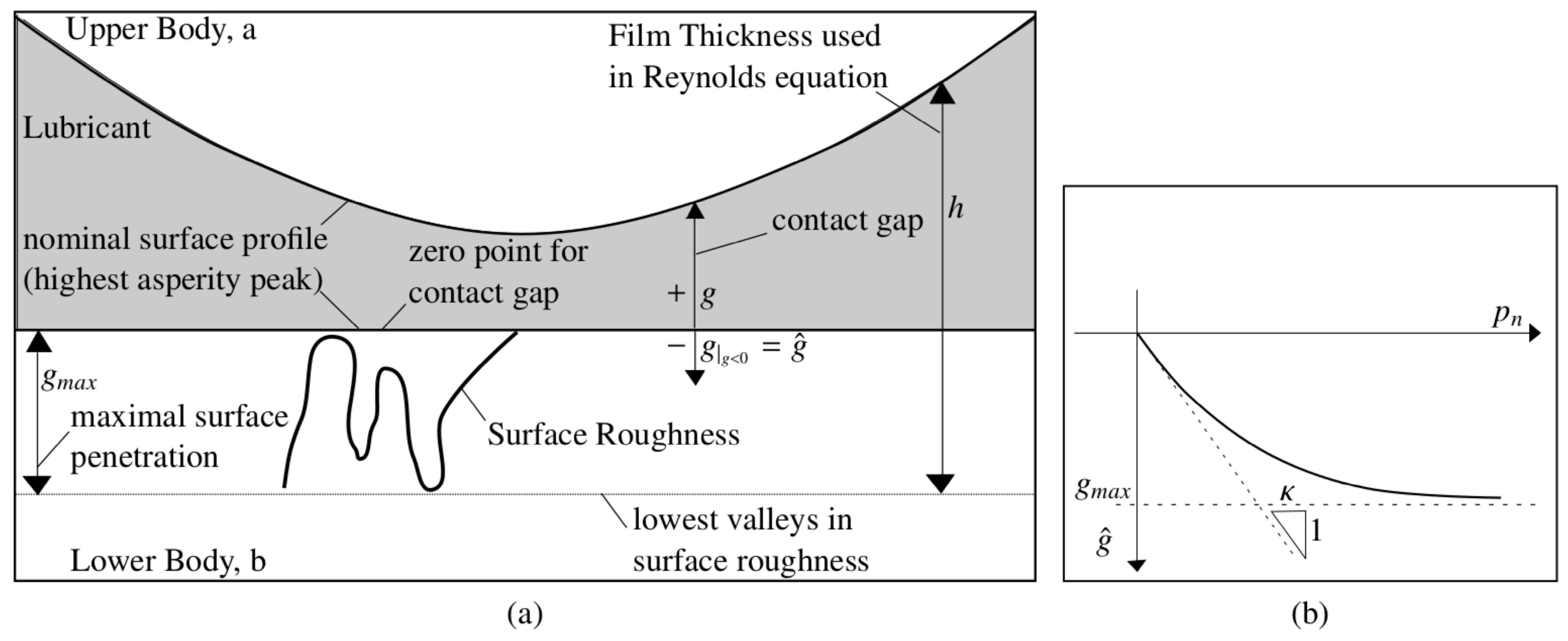}}
  \caption{(a) Schematic of the thin film flow between surfaces in relative motion. (b) Regularization function: relation between asperity contact pressure and the regularized film thickness}
  \label{fig:regcontact}
\end{figure}

In this section, we introduce the general modeling approach for the proposed lubricated contact framework.
The different lubrication regimes shall be explained with the help of the Stribek curve \cite{hamrock_1994}, depicted in Fig.~\ref{fig:stribek-mixmodel}. It schematically shows the transition of the lubrication condition along with a change of the frictional coefficient with respect to increasing relative sliding velocity of the contacting surfaces. For a more detailed presentation of the distinct lubrication regimes, the interested reader is referred to the corresponding literature \cite{hamrock_1994}. To understand the lubricated contact problem better, we focus on mixed lubrication.
The friction condition, where solid as well as hydrodynamic friction components in lubricated contacts are present at the same time, is called mixed lubrication. Generally, the transition from hydrodynamic lubrication to mixed lubrication is defined by the asperity contact. In mixed lubrication, the total traction of lubricated contact $\lubcontraction$ at a local interface point additively splits into the asperity contact traction $\contraction$ and the lubricant traction $\lubtraction$ as shown in Fig.~\ref{fig:stribek-mixmodel}. The asperity contact occurs when the surfaces of solid bodies approach each other and the thickness of the fluid gap in between gets very small or even vanishes, which depends on the roughness parameters of the contacting surfaces. In this context, the film thickness is also termed as contact gap $\gap$ as it defines the relative normal distance between the nominal surface profile of potentially contacting bodies. In the limiting case when the contact gap $\gap$ tends to zero, the Reynolds equation is no longer valid. Therefore, we define a positive regularization thickness $\regthick$ as regularization of our mathematical model equations, which can be interpreted as the maximal possible surface penetration and occurs when surface asperities are completely flattened in the limit of infinite contact pressures. Fortunately, there is also a physical interpretation of this regularization value: It describes the length scale, across which the transition from hydrodynamic to mixed lubrication takes place, thus, it is a measure for the (spatially averaged) surface roughness.
Thus, we propose to relate the regularization thickness $\regthick$ value to the combined root mean square of the roughness of the contacting surfaces
\begin{equation}
\roughness=\sqrt{\slaveroughness^2+ \masterroughness^2}
\end{equation}
Here $\slaveroughness$ and $\masterroughness$ are the root-mean-square roughnesses $\roughness$ of the contacting surfaces $1$ and $2$, respectively. In this work, under the assumption of a Gaussian distribution of the surface profile, the regularization thickness $\regthick\approx 3 \cdot \roughness$ is used. However, we do not resolve individual asperities in contact and only the effective influence of the microscale surface roughness on the macroscale mechanics are taken into account in terms of statistical parameters incorporated in the large-scale model by using the Patir and Cheng average flow model \cite{patir1,patir2}. Following this analogy, the film thickness $\film$ used in the Reynolds equation reads as
\begin{equation}\label{equ:reyfilm}
\film=\gap + \regthick
\end{equation}
By these definitions, now the film thickness holds always positive $\film>0$, although the contact gap $\gap$ can be positive or negative. A negative value $\gap<0$ means penetration, which is the case when contact normal pressure $\prescn$ is acting. A positive value $\gap>0$ means the bodies are not in contact, i.e. contact normal pressure $\prescn$ is zero.
In Fig.~\ref{fig:regcontact}, a schematic view of the final regularized model for the lubricated contact problem is presented, where all involved physical and numerical parameters are shown. It should be mentioned that we consider a combined surface roughness on one side, equivalent to the roughness of two contacting surfaces.
In this work, a regularization function $\gaphat(\prescn)$ for the contact constraint in the regularized layer in the case of negative contact gap $\gap<0$ is defined by an exponential relation between regularized film thickness $\gaphat$ and asperity contact normal pressure $\prescn$ according to
\begin{equation}\label{equ:regfunc}
\gaphat(\prescn)=\regthick (1-e^{-\frac{\regstiff}{\regthick}\cdot \prescn}) 
\end{equation}
While the regularization thickness $\regthick$ is a measure for the height of surface asperities, in a similar fashion, regularization stiffness $\regstiff$ can be interpreted as a measure for the stiffness of surface asperitis. More precisely, the regularization stiffness $\regstiff$ represents the initial gradient of the asperity contact pressure curve, Fig.~\ref{fig:regcontact}.
\begin{equation}
\frac{\partial \prescn}{\partial \gaphat}|_{\gaphat=0}=\regstiff
\end{equation}
Considering the strain measure $\regstrain$ in the regularized layer which can be defined as
\begin{equation}
\regstrain=\frac{\gap}{\regthick}-1
\end{equation}
Then, the stiffness of surface asperities in the regularized layer reads as
\begin{equation}
\regstiff=\frac{\partial \prescn}{\partial \gaphat}=\frac{\partial \prescn}{\partial \regstrain}\cdot \frac{\partial \regstrain}{\partial \gaphat}=\underbrace{\frac{\partial \prescn}{\partial \regstrain}}_{\eprime}\cdot \frac{1}{\regthick}
\end{equation}
where $\eprime$ is the physical stiffness of the layer and is denoted as
\begin{equation}
\eprime=\regstiff \cdot \regthick
\end{equation}
There are two limit cases which need to be considered in order to determine $\eprime$. First case is when $\eprime$ value is equal to or higher than $\modulelasticity$, representing the bodies with very stiff asperities, which will lead to infinite value of $\prescn$ as soon as the bodies come into contact. On the other hand, the second case is when $\eprime$ value is far smaller than $\modulelasticity$. This case means that the asperities are too elastic and very small $\prescn$ can deform them largely. Therefore, as soon as contact appears, the surface asperities are completely flattened which again leads to infinite $\prescn$. By these definitions, the regularization stiffness $\regstiff$ in this work is chosen as
\begin{equation}
\regstiff \approx \underbrace{\frac{\modulelasticity}{10}}_{\eprime} \cdot \frac{1}{\regthick}
\end{equation}
Indeed, physically speaking the regularization is representing the compression of the surface asperities in occurrence of asperity contact. In this sense, $\lubcontraction$ will be an additional contributing to the interface traction. Alternatively the regularization thickness $\regthick$ and the regularization stiffness $\regstiff$ can also be determined by fitting these parameters to experimental data (see e.g. \cite{sitzmann2014,sitzmann_2015}).
\begin{remark}
This model is made for lubrication problems, i.e. boundary lubrication is the limit where still local fluid domains are present, meaning the resulting asperity contact traction $\lubcontraction$ is considered the same as for dry contact in the limit of vanishing velocity and will be determined solely using the regularized contact constraints. Therefore, the theoretical case that the surface asperities are completely flattened by (infinitely) high external forces, which would lead to a vanishing film thickness $\film=0$ and, thus, to infinite pressure values in the Reynolds equation, are beyond the scope of the presented model. This means, for practically relevant loading, the minimal values of the film thickness $\film$ will always be well above zero. However, to make the numerical model more robust it is recommended to add a small tolerance $0 < TOL \leq 1$ to ${\regthick}$ in Eq. (\ref{equ:regfunc}) comparing to $\regthick$ in Eq. (\ref{equ:reyfilm}) such that
\begin{equation}\label{equ:maxhat}
{\regthick}_{|_{\text{Eq. (\ref{equ:regfunc})}}}=(1-TOL){\regthick}_{|_{\text{Eq. (\ref{equ:reyfilm})}}}
\end{equation}
According to Eq. (\ref{equ:maxhat}), the scenario of infinitely high mechanical contact pressures, which leads to $\gaphat=\regthick$ according to Eq. (\ref{equ:regfunc}), results in a non-zero film thickness
\begin{equation}
\film=\gap|_{{\gap}<0}  +\regthick=\gaphat + \regthick= -\regthick(1-TOL)+ \regthick= TOL \ast \regthick
\end{equation}
This can ensure that also in presence of numerical discretization and round-off errors as well as in non-equilibrium configurations arising during the iterations of the non-linear solver (e.g. Newton-Raphson), fluid film thickness values close/equal to zero can be avoided resulting in a more robust solution scheme.
\end{remark}
\section{Governing equations for large deformation lubricated contact problem}\label{ps}
The lubricated contact problem statement involves characterising equations for the solid field, the lubrication field, their respective coupling and the contact interaction. The solid equations are based on the initial boundary value problem (IBVP) of finite deformation elastodynamics, which can be derived from considerations of non-linear kinematics, stress and strain measures and the balances of linear and angular momentum.\\
The IBVP is supplemented by the lubricated-contact-specific traction boundary condition on the lubrication interface which includes the fluid tractions related to the pressure solution of the Reynolds equation and in the case of mixed lubrication and mechanical contact occurrence, the asperity contact contribution to the interface traction vector. \\
The lubricant behaviour is determined by the averaged Reynolds equation in conjunction with the cavitation condition, which will be detailed in Section~\ref{lub}. As, in our case, the displacement solution is time-dependent, the lubrication field is not stationary in the context of such coupled problems.
\begin{figure}[h!]
  \centering
  {\includegraphics[width=1.0\textwidth]{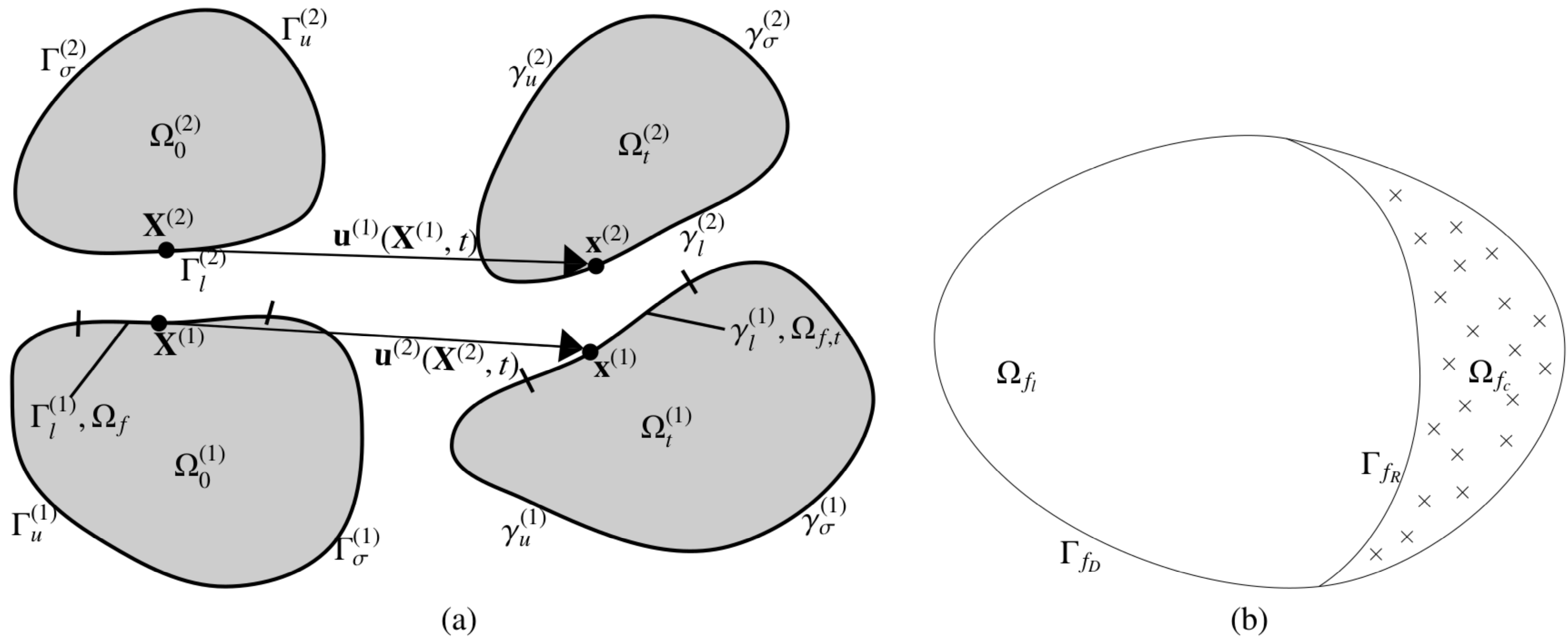}}
  \caption{(a) 3D domain of the two-body lubricated contact problem visualized in 2D. (b) Lubrication domain, 2D interface domain of a 3D problem}
  \label{fig:contactdomain-lubdomain}
\end{figure}
A lubricated contact problem involving two deformable solid bodies and a thin lubricant film in between is considered. Fig.~\ref{fig:contactdomain-lubdomain} gives an overview of the two-body lubricated contact problem setup. The open sets $\genrefdomain \subset \Realdim$ and $\gendomain \subset \Realdim, i=1,2, \dimensions=2,3$ represent the two domains of the solid bodies in the reference and current configuration respectively, traced by a bijective and orientation preserving mapping $\gendomain=\bimapi_t(\genrefdomain)$. As usual, upper case letters refer to quantities in the refrence configuration and lower case letters to the current configuration. The boundaries $\partial \genrefdomain$ are divided into three distinct subsets
\setlength\abovedisplayskip{2pt}
\setlength\belowdisplayskip{2pt}
\begin{equation}
\partial\genrefdomain=\refgenDinterface \cup \refgenNinterface\cup \refgeninterface
\end{equation}
\begin{equation}
\refgenDinterface\cap\refgenNinterface=\refgenDinterface\cap\refgeninterface=\refgeninterface\cap\refgenNinterface=\emptyset
\end{equation}
where $\refgenDinterface$ and $\refgenNinterface$ are the Dirichlet and Neumann boundaries with prescribed displacements and tractions respectively. $\refgeninterface$ represent the lubricated boundaries which are specific to lubricated contact problems and specifies the part of the boundary where the contact (in this framework, presumed to be either lubricated or dry) develops. \\
The counterparts of the boundaries in the current configuration are denoted as $\genDinterface$ ,$\genNinterface$ and $\geninterface$. In accordance to the notation in contact mechanics, $\refslaveinterface$ will be referred to as the slave surface and $\refmasterinterface$ is the master surface.
The lubrication domain, which is a manifold with one dimension less than the solid domains and, in 3D problems, which is the underlying surface for solving the averaged Reynolds equation, equals the slave surface. In the reference configuration, it is denoted as
\begin{equation}
\domainf=\refslaveinterface
\end{equation}
This particular definition will become relevant for the definition of the discrete coupling in Section~\ref{fef}. For the moment, it can be interpreted as a tied coupling of the lubrication domain and the slave surface. On the contrary, the location of the master surface $\refmasterinterface$ is unknown a priori and needs to be determined by normal projection of the slave surface.\\
As shown in Fig.~\ref{fig:contactdomain-lubdomain}, the lubricated contact domain $\domainf$, taken as the part of the slave surface $\refslaveinterface$ is divided into two different subdomains, $\domainfpositive$ with a positive pressure and $\domainfcav$ where the lubricant has cavitations and is ruptured. $\reflubDinterface$ is the surface where the Dirichlet boundary conditions (on pressure) are prescribed for the averaged Reynolds equation.
$\reflubreyinterface$, which is called the Reynolds boundary, is the boundary between $\domainfpositive$ and $\domainfcav$. These definitions result in:
\setlength\abovedisplayskip{2pt}
\setlength\belowdisplayskip{2pt}
\begin{equation}
\domainf=\domainfpositive\cup \domainfcav
\end{equation}
\subsection{Solid part}\label{solid}
The solid phase of a lubricated contact problem is governed by the well-known initial boundary value problem (IBVP) of finite deformation elastodynamics formulated in the reference configuration $\genrefdomain$, which reads as follows:
\setlength\abovedisplayskip{2pt}
\setlength\belowdisplayskip{2pt}
\begin{equation}\label{equ:1}
\density^{(i)}\dispsddot^{(i)} = \grad \cdot (\defgradient^{(i)} \cdot \secondpiola^{(i)}) + \refbodyforcehat^{(i)} \qquad \text{in}\quad\genrefdomain \times (0,\mendtime],
\end{equation}
\begin{equation}\label{equ:dbcstrong}
\gendisps = \dispshat_0^{(i)} \qquad \text{on} \quad\refgenDinterface \times (0,\mendtime],
\end{equation}
\begin{equation}\label{equ:nbcstrong}
\firstpiola^{(i)} \cdot \refnormvec^{(i)} = \reftractionhat^{(i)} \qquad \text{on} \quad \refgenNinterface \times (0,\mendtime],
\end{equation}
Herein, $\defgradient^{(i)}$, $\firstpiola^{(i)}$ and $\secondpiola^{(i)}$ are the material deformation gradient as well as first and second Piola-Kirchhoff stress tensor. $\refnormvec^{(i)}$ is the outward pointing normal vector on $\refgenNinterface$. $\refbodyforcehat^{(i)}$ and $\reftractionhat^{(i)}$ refer to the external body forces and tractions, which are defined with respect to the undeformed unit volume and surface respectively. $\mendtime$ is the end of the considered time interval. Eq. (\ref{equ:1}) involves partial derivatives with respect to time, i.e. the accelerations $\dispsddot^{(i)}$. Hence, additional conditions on the displacements $\gendisps$ and velocities $\vels^{(i)}=\dispsdot^{(i)}$ at the initial time $\mtime=0$ need to be defined:
\begin{equation}
\gendisps(\genrefpoint,0)=\dispshat^{(i)}(\genrefpoint) \qquad \text{in} \quad \genrefdomain \times 0,
\end{equation}
\begin{equation}
\vels^{(i)}(\genrefpoint,0)=\velshat^{(i)}(\genrefpoint) \qquad \text{in} \quad \genrefdomain \times 0,
\end{equation}
The IBVP needs to be supplemented by a suitable constitutive model in order to define a relation between the stresses and strains. For simplicity, a widely used, isotropic, hyperelastic constitutive law, known as the Neo-Hookean model, has been utilized for all the analyses in this work. For such hyperelastic materials the constitutive relations are fully defined by the strain energy function $\strainenergyi$, and the fourth-order constitutive tensor $\ctensor$, the Green-Lagrange strain tensor $\straintensor$ and the second Piola-Kirchhoff stress tensor are defined as:
\begin{equation}\label{equ:NH}
\secondpiola^{(i)}=\frac{\strainenergyi}{\partial\straintensor} ,\quad \ctensor=\frac{\strainenergyii}{\partial\straintensor^{2}}, \quad \straintensor=\frac{1}{2}(\defgradienttr\defgradient-\unity)
\end{equation}
However, it is emphasized that the formulation in this work is independent of the material model and no limitation has been placed on the solid phase constitutive law.
In the "Lubricated Contact" context, the solid bodies are subjected to the lubricated contact traction $\lubcontraction^{{(i)}}$, i.e.
\begin{equation}\label{equ:lubinterfacestrong} 
\stresss \cdot \normvec^{(i)} = {\lubcontraction^{{(i)}}} \qquad \text{on} \quad \geninterface \times (0,\mendtime],
\end{equation}
\begin{equation}\label{equ:couplingtc}
\lubcontraction^{{(i)}}=\contraction^{{(i)}} + \lubtraction^{{(i)}}
\end{equation}
where $\normvec^{(i)}$ denotes the outward pointing normal on $\geninterface$. The lubricated contact traction $\lubcontraction^{{(i)}}$ includes contact pressure $\contraction^{{(i)}}$ and fluid pressure $\lubtraction^{{(i)}}$ contributions and couples the mechanical problems of the two solid bodies interacting via lubricated contact. The contact and fluid pressure contributions will be specified in Sections~\ref{contact} and \ref{couplingstrong}.
\subsection{Lubrication part}\label{lub}
 by many authors [21, 22]. These works confirm an approximately asymptotic relationship between the separation of the surfaces and the contact pressure.\\
Motivated by Section~\ref{model}, the lubricant behaviour is characterized by the averaged Reynolds equation which is assumed to be valid over a surface domain, that in the current formulation is considered to be the slave side of the contacting interface, $\refslaveinterface$, and includes the cavitation contribution and boundary conditions prescribing the pressure. The governing equation is given as
\begin{equation}\label{equ:rey}
\frac{\partial{(\density \film)}}{\partial{\mtime}} + \surfgrad \cdot \left( -\frac{\density \film^{3}}{12\vis} \pflowfac \surfgrad{\presf}+ \frac{\density(\velstilde^{(1)} - \velstilde^{(2)}) \stddev}{2} \sflowfac +
\frac{\density(\velstilde^{(1)} + \velstilde^{(2)})}{2} \film \right) = \epsilonnn\langle-\presf\rangle, \qquad\text{in} \quad \domainf\times (0,\mendtime],
\end{equation}
where $\density$ denotes the lubricant current mass density, which is assumed constant, $\film$ is the fluid film thickness, which directly relates to solid phase deformation Eq. (\ref{equ:reyfilm}), $\vis$ marks the viscosity of the lubricant, $\velstilde^{(1)}$ and $\velstilde^{(2)}$ are (tangential) projections of slave and master surface velocity vectors onto the contact surface fo the interacting solids, the positive scalar $\epsilonnn$ is the penalty parameter and the Macaulay brackets $\langle.\rangle$ denotes a ramp function, and $\presf$ stands for the primary unknown of Eq. (\ref{equ:rey}), namely the fluid pressure. $\pflowfac$ and $\sflowfac$ are the pressure and shear flow factors respectively, that are the statistical parameters from the Patir and Cheng average model \cite{patir1,patir2}. For isotropic topographies with Gaussian roughness height distribution these quantities were obtained through numerical simulations carried out for representative domains at the microscopic scale and are given as
\begin{equation}
\pflowfac = 1+3(\frac{\stddev}{\film})^2
\end{equation}
\begin{equation}
\sflowfac = \frac{-3(\frac{\stddev}{\film})-30(\frac{\stddev}{\film})^3}{1+6(\frac{\stddev}{\film})^2}
\end{equation}
\begin{equation}
\fflowfac = 1+(\frac{\stddev}{\film})^2
\end{equation}
where $\stddev$ is the standard deviation of surface roughness. $\fflowfac$ is the correction factor for roughness and will be used in the average shear stress expression in the coupling Section.
The first term of Eq. (\ref{equ:rey}) includes $\frac{\partial}{\partial \mtime}$ denoting the local time derivative, while $\surfgrad$ in the second term expresses the surface gradient operator, i.e. the projection of the total gradient onto the slave surface.\\
The boundary of the lubrication domain is a Dirichlet boundary with prescribed conditions on the pressure $\presf$:
\begin{equation}
\partial\domainf=\reflubDinterface
\end{equation}
The considerations on the cavitation region $\domainfcav\subseteq \domainf$ and the associated boundary $\reflubreyinterface$ apply accordingly. Their locations are not explicitly determined but indirectly found as part of the solution which adds yet another non-linearity to the problem and makes it more difficult to solve. In this work, a standard penalty regularization method similar to the literature (cf. \cite{cryer1971,rohde_mcallister_1975,wu_1986,yang2009}) is followed which transfers the original problem to an equivalent complementary problem. The term on the right hand side of Eq. (\ref{equ:rey}) adds a contribution to the flow in case of negative pressures and its magnitude is directly related to the extent of constraint violation. This type of constraint enforcement goes along with the advantage that no additional unknowns need to be determined. Moreover, the location of the Reynolds boundary is automatically found as part of the solution without the need of an iterative boundary search and adjustment.
Dirichlet boundary conditions on the pressure read
\begin{equation}\label{equ:dbcrey}
\presf = \phatf, \qquad \text{on} \quad \reflubDinterface \times (0,\mendtime],
\end{equation}
The fluid pressure in the cavitation domain $\domainfcav$ is close to the atmosphere pressure, making it negligible in comparison to the fluid pressure in the lubrication domain $\domainfpositive$. As a result, it reads
\begin{equation}\label{equ:cav}
\presf = 0, \qquad \text{on} \quad \domainfcav \times (0,\mendtime],
\end{equation}
Finally, it has to be noted that the lubricant equation is not stationary in the "Lubricated Contact" context. The pressure $\presf$ is a function of time since the lubrication coupling quantities, namely the film thickness $\film$ and surface velocities $\velstilde^{(i)}$, are determined from the solid dynamic behaviour. In order to make them accessible for the lubrication domain, they need to be defined with respect to the coordinates of the slave surface. For a point $\pointslave$ on the slave surface (in spatial configuration), one can find an associated point $\pointmasterhat$ on the master surface by projecting $\pointslave$ along its current outward normal vector $\normveclub^{(1)}$.
\begin{equation} \label{equ:non21}
\film|_{\slaveinterface} = \underbrace{-\normveclub^{(1)} \cdot (\pointslave-\pointmasterhat(\refpointslave))}_{\gap} + \regthick
\end{equation}
\begin{equation} \label{equ:non22}
\velstilde^{(1)}|_{\pointmasterhat}=(\unitydim - \normveclub^{(1)} \otimes \normveclub^{(1)}) \cdot \pointslavedot
\end{equation}
\begin{equation} \label{equ:non23}
\velstilde^{(2)}|_{\pointmasterhat}=(\unitydim - \normveclub^{(1)} \otimes \normveclub^{(1)}) \cdot \pointmasterhatdot(\refpointslave)
\end{equation}
where $\pointslavedot$ and $\pointmasterhatdot$ are the material velocities of $\pointslave$ and $\pointmasterhat$, respectively. Special considerations are required for $\pointmasterhat$ and $\pointmasterhatdot$ , which are master side quantities associated with coordinates $\refpointslave$ of the slave side via projection. A suitable interface map $\bimapiii : \pointmasterhat \rightarrow \masterinterface$ , needs to be defined, which will be discussed in more detail later on in Section~\ref{fef}. It should be noted that following the model description in Section~\ref{model}, the value of $\regthick$ is considered as constant across the entire slave side.
\subsection{Contact interface}\label{contact}
Having mentioned the solid and lubrication problems, we turn our focus to the contact interface part. We introduce regularized contact conditions which can be interpreted as constitutive contact laws on the contact interfaces representing the elastic deformation of surface asperities instead of enforcing a strict zero-penetration constraint \cite{laursen_2002,wriggers_2006,puso_2008, hueber_2008,hesch_2009,popp2009,popp2010,gitterle2010,fischer_2005, sitzmann2014}.\\
We consider again the two body finite deformation lubricated contact problem as mentioned in Fig.~\ref{fig:contactdomain-lubdomain}. Both bodies are governed by the IBVP described in Section~\ref{solid}, enhanced with the constraints of frictional contact at the potential contact boundary $\refgeninterface$. Since the focus is on finite deformations, the geometrical contact constraints such as the non-penetration condition have to be satisfied in the current configuration, i.e. they have to be enforced between the potential contact surfaces $\refgeninterface=\bimapi_t(\refgeninterface)$. As mentioned in the beginning of this Section, we will refer to $\refslaveinterface$ as the slave surface, and to $\refmasterinterface$ as the master surface.
The slave contact traction $\lubcontraction^{(1)}$ acts on the entire contacting interface $\domainf$ in addition to the fluid pressure and is decomposed as follows to obtain the normal contact pressure $\prescn$ and the tangential contact traction $\contractiontang$,
\begin{equation}\label{equ:normalcontactmain}
\begin{split}
\prescn= \normvec \cdot \lubcontraction^{(1)}, \\
\contractiontang= (\mathbf{1}-\normvec\otimes\normvec) \cdot \lubcontraction^{(1)}
\end{split}
\end{equation}
\paragraph{\textbf{\textit{Regularized contact conditions}}}
Now, we introduce regularized Karush-Kuhn-Tucker (KKT) conditions for normal and tangential contact according to the lubricated contact model proposed in Section~\ref{model}.
The regularized normal contact law is as follows
\begin{equation}\label{equ:normalcontact}
\prescn \geq 0 \qquad \regthick\geq -\gaphat(\prescn) \qquad \prescn \cdot (\regthick+\gaphat(\prescn)) = 0 \qquad \text{on} \quad \slaveinterface
\end{equation}
where $\gaphat(\prescn)$ is the regularization function presented in Section~\ref{model}. 
The contact constraints in tangential direction can be formulated via Coulomb friction law on the slave contact surface,
\begin{equation}\label{equ:friccontact}
\coulomblaw:=\|\contractiontang\| - \dryfriction \cdot \prescn\leq 0 \qquad \velstilde + \coulomblawbeta \contractiontang=0 \qquad \coulomblawbeta \geq 0 \qquad \coulomblawbeta \cdot \coulomblaw=0 \qquad  \text{on}\quad \slaveinterface
\end{equation}
where $\dryfriction$ is the dry solid friction coefficient. Eq. (\ref{equ:friccontact}) requires that the magnitude of the tangential stress vector does not exceed the coefficient of friction times the normal contact pressure. When the tangential stress is less than the Coulomb limit ($\|\contractiontang\| < \dryfriction \cdot \prescn$), the continuity equation ($\coulomblawbeta \cdot [\|\contractiontang\| - \dryfriction \cdot \prescn]=0$) forces $\coulomblawbeta$ to be zero and accordingly the tangential relative velocity must be zero. This is called the stick state. When the tangential stress is at the Coulomb limit ($\|\contractiontang\| = \dryfriction \cdot \prescn$), $\coulomblawbeta$ may be greater than zero in the continuity equation and therefore the tangential stress is forced to oppose the relative tangential velocity in ($\velstilde + \coulomblawbeta  \contractiontang=0$). This is called the slip state.
\begin{remark}
Our framework currently uses Coulomb classic pressure independent coefficient of friction. Nevertheless, there is no general limitation in using a dry solid friction model that e.g. depends on contact pressure, sliding velocity, or temperature. There is ongoing research on how this coefficient of friction depends on contact pressure and also potentially on lubricant in dry friction scenario, since the asperities are still filled by lubricant even in case of very low tangential movement meaning that the lubricant still can influence the dry solid friction coefficient\cite{popov,ozaki,KRAUS2021}.
\end{remark}
\subsection{Coupling}\label{couplingstrong}
The strong coupling of the two subproblems i.e., the fluid problem and the contact problem, is the crux behind the lubricated contact problem. Primarily, the lubricant film thickness $\film$ depends on the solid field deformation. In the present formulation, the local film thickness $\film$ of the fluid problem is characterized by the projection algorithm that defines the specific geometrical dependence of $\film$ on displacements of the solid field.
Secondly, the hydrodynamic pressure $\presf$ of the fluid problem defines the fluid contribution $\lubtraction^{{(i)}}$ of the lubricated contact traction $\lubcontraction^{(i)}$, acting on the interface of the solid bodies. According to Eq. (\ref{equ:couplingtc}), lubricated contact traction $\lubcontraction^{(i)}$ consists of fluid traction $\lubtraction^{{(i)}}$ and the asperity contact traction $\contraction^{{(i)}}$, where the latter is already explained in Section~\ref{contact}. It is worth to mention again that both of the fluid traction as well as potential contact traction acts on the entire lubrication domain $(\domainf)$, however the asperity contact traction $\contraction^{(i)}$ can be still zero depending on if the gap $\gap$ is positive or not. According to Eq. (\ref{equ:rey}), $\lubtraction^{(i)}$ is defined by the fluid pressure, its gradient and the current solid geometry, thus
\begin{equation}
\lubtraction^{(i)}=\lubtraction^{(i)}(\presf,\grad \presf,\disps)
\end{equation}
where $\presf$ and $\disps$ are governed by the averaged Reynolds equation and the IBVP of the solid phase respectively, see Eq. (\ref{equ:rey}) and Eq. (\ref{equ:1}). Particularly, the parabolic velocity profile assumed by the averaged Reynolds equation leads to the following definition for the fluid traction on the slave side:
\begin{equation}\label{equ:couplingstrong}
\lubtraction^{(1)}=\lubtractionparallel^{(1)} + \lubtractionnparallel^{(1)}
\end{equation}
\begin{equation}
\lubtractionparallel^{(1)}=-\frac{\film}{2}\pflowfac\surfgrad{\presf}
\end{equation}
\begin{equation}
\lubtractionnparallel^{(1)}=-\presf\normveclub^{(1)}-\frac{\vis}{\film}\frac{(\velstilde^{(1)} - \velstilde^{(2)})}{2}(\fflowfac+\sflowfac)
\end{equation}
The term $\lubtractionparallel^{(1)}$ consists of the shear stress owing to the Poiseuille flow contribution originated by the pressure gradient. The term $\lubtractionnparallel^{(1)}$ includes the normal traction due to the hydrodynamic pressure contribution $\presf$, where $\normveclub^{(1)}$ is the outward normal of the lubricated boundary $\refslaveinterface$ on the slave interface in the deformed configuration, and the shear (friction) stress resulting from a Couette flow contribution associated with the tangential relative velocity $\frac{(\velstilde^{(1)} - \velstilde^{(2)})}{2}$.\\
On the master side, the fluid traction is denoted as
\begin{equation}
\lubtraction^{(2)}=[\lubtractionnparallel^{(1)} - \lubtractionparallel^{(1)}]\circ \bimapiii
\end{equation}
where the expression $\bimapiii :\slaveinterface \rightarrow \masterinterface$ represents a suitable mapping from the slave to the master surface.\\
The averaged Reynolds equation is formulated in an arbitrary Lagrangian Eulerian frame on the contact boundary of the solid, and this includes an extra coupling \cite{stupseal2009,stup_2016} due to the finite configuration changes. The surface on which the averaged Reynolds equation is defined and solved, is not known a priori and comprises a part of the solution of the problem. Practically, upon finite element discretization, the positions of the nodes of the finite element mesh used to solve the averaged Reynolds equation Eq. (\ref{equ:rey}) depend on the deformation of the solid field. 
\section{Finite element formulation}\label{fef}
To formulate the finite element discretization of the lubricated contact problem, we follow the classical finite element approach and derive the weak form of the problem with standard procedures.
Therefore, we consider $\solutionspacepres$ and $\testspacepres$ as the solution and weighting function spaces for the fluid pressure field $\presf$ and its variations $\testfkt{\presf}$, respectively.
$\solutionspacedisp^{(i)}$ and $\testspacedisp^{(i)}$ are the corresponding function spaces for the displacement field $\gendisps$, $(i=1,2)$ and its variation $\testfkt{\gendisps}$ respectively.\\ 
The following sections start with the derivation of the weak formulation of the individual subproblems and then, the basics of finite element discretization are introduced and applied. Finally, the solution strategy is presented for the resulting non-linear system of equations.\\
The temporal discretization, based on the generalized-$\alpha$ time-integration scheme for the IBVP governing the solid domain is not explicitly presented. The reader is referred to the following publications for details \cite{chung_1993}.
\subsection{Solid part}\label{fefsolids}
\subsubsection{Weak form}
The focus here is on the derivation of the weak form of the IBVP of the solid bodies; the asperity contact contribution and the fluid part of the lubricated contact traction will be handled separately in Section~\ref{fefcontact}.
If the particular interpretation of the weighting functions $\testfkt{\gendisps}$ as virtual displacements is made, the weak form of the IBVP can be identified as the principle of virtual work. After weighting the residuals of the balance equation Eq. (\ref{equ:1}) and the boundary conditions on the Neumann Eq. (\ref{equ:nbcstrong}) and lubrication boundaries Eq. (\ref{equ:lubinterfacestrong}) respectively for both bodies $\genrefdomain,i=1,2$, applying Gauss divergence theorem and invoking that the virtual displacements $\testfkt{\gendisps}$ on the Dirichlet boundaries $\refgenDinterface$, the weak form of the IBVP of the solid bodies in the reference configuration is obtained.
\begin{multline}\label{equ:solidweak}
\galerkindisp(\disps,\presf,\testfkt{\disps}):= \sum_{i=1}^{2} \left(\galerkindisp^{(i)}(\gendisps,\presf,\testfkt{\gendisps})\right):=
- \sum_{i=1}^{2} \underbrace{ \int_{\geninterface} \testfkt{\gendisps} \cdot \lubcontraction^{(i)}d\interface}_{\text{lubrication traction}} +\\ \sum_{i=1}^{2} \left\lbrace \underbrace{\int_{\genrefdomain}\density^{(i)}\testfkt{\gendisps} \cdot \dispsddot^{(i)} d\domain}_{\mathrm{kinetic}} +  \underbrace{  \int_{\genrefdomain} \left[-\testfkt{\gendisps}\cdot\refbodyforcehat^{(i)} + \testfkt{\straintensor^{(i)}} : \secondpiola^{(i)}\right] d\domain - \int_{\refgenNinterface} \testfkt{\gendisps} \cdot \reftractionhat^{(i)} d\refinterface}_{\mathrm{int,ext}} \right\rbrace=0
\end{multline}
The first term corresponds to the virtual work $\testfkt{\virtwlub}$ of the lubrication traction including $\testfkt{\virtwf}$ and $\testfkt{\virtwc}$ due to the fluid film traction $\lubtraction^{(i)}$ and contact traction $\contraction^{(i)}$ respectively, which will be explained in Section~\ref{fefcontact}. The second and third terms are well-known from the classic virtual work principle of elastodynamics and denote the kinetic virtual work contribution $\testfkt{\virtwkin}$ and the sum of the internal and external virtual works $\testfkt{\virtwintex}$. 
Summing up, the weak form of the solid part of the lubricated contact problem is defined as follows: Find $\gendisps \in \solutionspacedisp^{(i)}$ such that
\begin{equation}\label{equ:non7}
-\testfkt{\virtwkin}-\testfkt{\virtwintex}-\testfkt{\virtwlub}=0
\end{equation}
is satisfied for all weighting functions $\testfkt{\gendisps} \in \testspacedisp^{(i)}$.
It is noteworthy, that the standard terms of solid dynamics, namely the kinetic, internal and external virtual work contributions are expressed in the reference configuration whereas the virtual work due to the lubrication loads and the lubrication equation are formulated with respect to the current configuration.
\subsubsection{Discrete form}
The weak form of the solid part of lubricated contact problem Eq. (\ref{equ:non7}) is continuous with respect to space and time and requires discretization. Spatial discretization is performed by employing standard isoparametric finite elements. The finite element meshes of the two solid subdomains are non-conforming, resulting in non-matching meshes at the lubricated interface. The mesh of the lubrication domain coincides with the mesh on slave surface.
The discretization of the virtual work contributions $\testfkt{\virtwkin}$ and $\testfkt{\virtwintex}$ is standard and will not be treated here. Instead, it is referred to the available literature, such as \cite{hughes_2000,zienkiewicz_fox_2013,zienkiewicz_zhu_2013}. The focus of this work is set on the discretization of the interface phenomena related to lubricated contact problems which is explained in Section~\ref{fefcontact}.
\subsection{Lubrication part}\label{lubweak}
\subsubsection{Weak form}
The finite element formulation of equations Eqs. (\ref{equ:rey}-\ref{equ:cav}) necessitates a transformation of the problem such that weaker differentiability requirements are imposed on the solution functions for $\presf$. In particular, second spatial derivatives of $\presf$ appear in Eq. (\ref{equ:rey}), which will be eliminated in the following derivation of the weak form. 
The weak form is obtained by weighting the residuals of the balance equation Eq. (\ref{equ:rey}) and integrating them over the respective domain. The Dirichlet condition Eq. (\ref{equ:dbcrey}) does not appear in the weak formulation, but will be respected by restricting the solution and weighting function spaces later on. Employing the compact notation from Eq. (\ref{equ:rey}), the weak form reads
\begin{multline}\label{equ:reyweak}
\galerkinf(\disps,\presf,\testfkt{\presf}) := \int_{\domainf}\frac{\partial \film}{\partial \mtime} \cdot \testfkt{\presf} d\domain + \int_{\domainf}\frac{\film^{3}}{12\vis}\pflowfac \surfgrad{p} \cdot\surfgrad\testfkt{\presf} d\domain - \int_{\domainf}\varepsilon_{\presf}\langle {-\presf}\rangle \testfkt{\presf} d\domain \\
 - \int_{\domainf}(\frac{\velstilde^{(1)} + \velstilde^{(2)}}{2} \film) \cdot\surfgrad\testfkt{\presf} d\domain - \int_{\domainf}(\frac{\velstilde^{(1)} - \velstilde^{(2)}}{2} \stddev \sflowfac) \cdot\surfgrad\testfkt{\presf} d\domain = 0
\end{multline}
where $\frac{\partial \film}{\partial \mtime}$ is the time derivative of film thickness which gives the squeeze term. The hydrodynamic lubrication problem has been transformed into the equivalent weak form, which can be stated as: Find $\presf\in\solutionspacepres$ such that Eq. (\ref{equ:reyweak}) is satisfied for all weighting functions $\testfkt{\presf} \in \testspacepres$.
\subsubsection{Discrete form}\label{spatiallub}
As explained in Section~\ref{ps}, the lubrication domain coincides with the slave surface in the continuous setting. This choice is also retained after discretization, i.e. the discretized lubrication domain in the reference configuration is given by $\refdomainfh=\refslaveinterfaceh$ and geometrically coinciding lubrication nodes and elements are defined. The same shape functions as introduced in Eq. (\ref{equ:non8}) are employed to interpolate the geometry, pressure and weighting function values. The discrete version of the weak lubricant equation can be written as
\begin{multline}\label{equ:non17} 
\sum_{j=1}^{\npnode} [ \int_{\domainlub}\frac{\filmh^3}{12\vis}\pflowfac\surfgrad \shapef_j \cdot\surfgrad \presh d\domain + \int_{\domainlub}\frac{\partial \filmh}{\partial \mtime} \cdot \shapef_j d\domain - \int_{\domainlub}\varepsilon_{\presf}\langle {-\presh}\rangle \shapef_j d\domain \\
 - \int_{\domainlub}(\frac{\velftildeh^{(1)} + \velftildeh^{(2)}}{2} \filmh) \cdot\surfgrad \shapef_j d\domain - \int_{\domainlub}(\frac{\velftildeh^{(1)} - \velftildeh^{(2)}}{2} \stddev \sflowfac) \cdot\surfgrad \shapef_j d\domain ] \testfkt{\presf_j} = 0
\end{multline}
with $\npnode=\nnodeone$ denoting the number of lubrication nodes. These terms can be identified (from left to right) as Poiseuille, squeeze, Couette, shear and cavitation term. It should be noted that the integrals are evaluated with respect to the deformed geometry $\domainlub$ and the film thickness $\filmh$ and surface velocities $\velftildeh^{(i)}$ are spatially discretized weighted quantities as follow
\begin{equation}\label{equ:non18}
\filmh=\sum_{k=1}^{\nnodeone} \shapef_k \filmk
\end{equation}
\begin{equation}\label{equ:non19}
\velftildeh^{(i)}=\sum_{k=1}^{\nnodeone} \shapef_k \velftildek^{(i)}
\end{equation}
where the shape function $\shapef_k$ is associated to the slave node $k$ and also to the matching lubrication node respectively. The weighted gap at slave node $k$ is determined as
\begin{equation} \label{equ:non24}
\filmk = \frac{\int_{\slaveinterfaceh}\dualshapef{k} \film d\interface}{\int_{\slaveinterfaceh}\dualshapef{k} d\interface}
\end{equation}
where $\dualshapef{k}$ is the dual base shape function associated to the slave node $k$ and $\film$ is given in Eq. (\ref{equ:non21}). The weighted relative tangential velocity $\velftildek^{(i)}$ at slave node $k$ is determined as
\begin{equation} \label{equ:non24}
\velftildek^{(i)} = \frac{\int_{\slaveinterfaceh}\dualshapef{k} \frac{(\velstilde^{(1)} - \velstilde^{(2)})}{2} d\interface}{\int_{\slaveinterfaceh}\dualshapef{k} d\interface}
\end{equation}
such that $\frac{(\velstilde^{(1)} - \velstilde^{(2)})}{2}$ satisfies the requirement of frame indifference, see e.g. \cite{Curnier1995} for further explanations. 
With those definitions at hand, the final discretized lubrication equation Eq. (\ref{equ:reyweak}) can be obtained in term of global vectors.
All terms are integrated element-wise and assembled into the global residual vector, which is now dependent on displacements and pressure
\begin{equation}\label{equ:non27}
\residual_{\pres}(\dispd,\prestns)=\residual_{\sqz}(\dispd)+\residual_{\psl}(\dispd,\prestns)+\residual_{\ctt}(\dispd)+\residual_{\shr}(\dispd)+\residual_{\cav}(\dispd,\prestns)=0.
\end{equation}
\subsection{Fluid and contact interface coupling}\label{fefcontact}
\subsubsection{Weak form}
To preapre the subsequent mortar finite element discretization of the lubricated contact problem, the virtual work $\testfkt{\virtwlub}$ of the lubrication traction in Eqs. (\ref{equ:solidweak}), including $\testfkt{\virtwf}$ and $\testfkt{\virtwc}$ due to the fluid film traction $\lubtraction^{(i)}$ and contact traction $\contraction^{(i)}$ respectively, is now explained. The contact traction $\contraction^{(i)}$, Eq. (\ref{equ:normalcontactmain}) contributes to $\lubcontraction^{(i)}$ in case of asperity contact occurrence. Therefore, a vector-valued Lagrange multiplier field is introduced at the slave side of the contact interface, to enforce the mechanical contact constraints Eqs. (\ref{equ:normalcontact}) and (\ref{equ:friccontact}), which can be identified as the negative slave side contact traction $\tractionlambda=-\lubcontraction^{(1)}$. The contact Lagrange multiplier is decomposed into a normal part $\tractionlambdan$ and a tangential part $\tractionlambdat$ analogously to the contact traction in Eq. (\ref{equ:normalcontact}).
As mentioned in Section~\ref{couplingstrong}, the fluid film traction $\lubtraction^{(i)}$ consist of $\lubtractionparallel^{(1)}$ contribution due to the Poiseuille flow and $\lubtractionnparallel^{(1)}$ contribution due to the hydrodynamic pressure and a viscous shear stress. Since the lubricant equation is solved solely on the slave surface, the pressure distribution $\presf$ is only known there. Neglecting inertia of the lubricant, inserting $\tractionlambda$, $\lubtractionparallel^{(1)}$ and $\lubtractionnparallel^{(1)}$ into the virtual work expression of the lubrication traction in Eq. (\ref{equ:solidweak}) and reformulation with respect to the current configuration yields
\begin{equation}\label{equ:non2}
\testfkt{\virtwlub}=\int_{\slaveinterface} (\lubtractionnparallel^{(1)} + \tractionlambda) \cdot( \testfkt{\disps}^{(1)}-\testfkt{\disps}^{(2)} \circ \bimapiii) d\interface + \int_{\slaveinterface} \lubtractionparallel^{(1)} \cdot( \testfkt{\disps}^{(1)}+\testfkt{\disps}^{(2)} \circ \bimapiii) d\interface
\end{equation}
The expression $\bimapiii :\slaveinterface \rightarrow \masterinterface$ represents a suitable mapping from the slave to the master surface. As the two surfaces are subjected to relative motion in lubricated contact problems, the mapping is deformation dependent. Thus, the integral in Eq. (\ref{equ:non2}) and the fluid traction $\lubtraction^{(1)}$ need to be evaluated in the current, i.e. deformed configuration. Consequently, the weak form of the averaged Reynolds equation, derived in Section~\ref{lubweak}, needs to be solved on the deformed lubrication boundary to allow for a consistent evaluation of the fluid film (contact) traction.
\subsubsection{Discrete form}
As already mentioned in Section~\ref{couplingstrong}, the lubricated contact problem is coupled in the sense that the displacement solution $\gendisps$ of the solid field is dependent on the lubricated contact traction $\lubcontraction^{(1)}$ and the pressure solution $\presf$ in the lubricant is associated with the film thickness $\film$ and the surface velocities $\velstilde^{(i)}$ defined by the solid problem. Those quantities need to be interchanged among the lubrication domain and both solid surfaces. The interchange of quantities between the slave surface and the lubrication field is straight forward, due to the coinciding nodes and shape functions. The coupling between the slave and master surface, which are characterized by non-matching meshes, is established via mortar discretization, which will be introduced in this section. Lastly, the lubrication field and the master surface are indirectly coupled. Kinematic master surface quantities are mortar projected onto the slave surface.\\
In the following, the geometry interpolations are introduced for the discretization of the lubricated surfaces:
\begin{align}\label{equ:non8}
\pointoneh|_{\refslaveinterfaceh}&=\sum_{k=1}^{\nnodeone}\shapef_k^{(1)} \pointonek & \pointtwoh|_{\refmasterinterfaceh}&=\sum_{m=1}^{\nnodetwo}\shapef_m^{(2)} \pointtwom
\end{align}
Following the isoparametric concept, the same shape functions will be used for discretization of the displacement $\dispsgenh$ and virtual displacement $\testfkt{\dispsgenh}$. Here $\nnodeone$ and $\nnodetwo$ correspond to the number of nodes on the slave and master surface respectively. Discrete nodal positions $\pointonek$ and $\pointtwom$, displacements $\dispd_k^{(1)}$ and $\dispd_m^{(2)}$, and virtual displacements $\testfkt{\dispd_k^{(1)}}$ and $\testfkt{\dispd_m^{(2)}}$ are introduced on both surfaces. They are vectors of size $\nnodedim$ and locally interpolated by $\shapef_k^{(1)}$ and $\shapef_m^{(2)}$. Those shape functions are associated with the discretization of the bulk domains
\begin{align}\label{equ:non9}
\pointoneh|_{\refdomainoneh}&=\sum_{k=1}^{\nbnodeone}\shapef_k^{(1)} \pointonek & \pointtwoh|_{\refdomaintwoh}&=\sum_{m=1}^{\nbnodetwo}\shapef_m^{(2)} \pointtwom\\
\dispsoneh|_{\refdomainoneh}&=\sum_{k=1}^{\nbnodeone}\shapef_k^{(1)} \dispd_k^{(1)} & \dispstwoh|_{\refdomaintwoh}&=\sum_{m=1}^{\nbnodetwo}\shapef_m^{(2)} \dispd_m^{(2)}\\
\testfkt{\dispsoneh}|_{\refdomainoneh}&=\sum_{k=1}^{\nbnodeone}\shapef_k^{(1)} \testfkt{\dispd_k^{(1)}} & \testfkt{\dispstwoh}|_{\refdomaintwoh}&=\sum_{m=1}^{\nbnodetwo}\shapef_m^{(2)} \testfkt{\dispd_m^{(2)}}
\end{align}
where $\nbnodeone$ and $\nbnodetwo$ correspond to the number of nodes of the bulk solids.
The fluid film traction $\lubtraction^{(1)}$ is solely evaluated on the slave side, and therefore an adequate coupling needs to be established in order to obtain its virtual work contribution on the master side, too.  In this paper, the mortar discretization as presented in \cite{yang2009} is the method of choice. To discretize the fluid film traction
\begin{equation}\label{equ:non11}
\tractionfoneh=\sum_{j=1}^{\nnodeone} \dualshapef{j} \traction_j^{(1)}
\end{equation}
where $\traction_j^{(1)}$ is the traction vector at node $j$. It contains the surface gradient of pressure $\surfgrad \presf$ at the slave node $j$ which introduces additional complexity for the lubricated contact problem considering that the pressure gradient is not continuous at each individual node commonly. Thus some weighting treatments has to be carried out, which is done here following the smoothing procedure applied by Yang and Laursen \cite{yang2009}.
It should be noted, that the choice of the discrete function space for the fluid traction and the associated shape functions $\dualshapef{j}$ is of particular importance regarding the mathematical properties and numerical efficiency of the mortar coupling.
\begin{remark}
In fact, the definition of dual shape functions based on a biorthogonality relation with the slave displacement shape functions $\shapef_k^{(1)}$ will heavily facilitate projections between the master and slave side. The dual basis functions are constructed such that they fulfill a biorthogonality condition \cite{wohlmuth2000}
\begin{equation}\label{equ:non35}
\int_{\slaveinterfaceh}\dualshapef{i} \shapef_j d\interface=\kroneckersym\int_{\slaveinterfaceh}\shapef_j d\interface
\end{equation}
where $\kroneckersym$ denotes the Kronecker symbol, i.e. $\kroneckersym=1$ if $i = j$ and $\kroneckersym=0$ otherwise. In practice, the easiest way to define those dual basis functions is via an element-wise linear combination of the standard shape functions $\shapef_i$, see e.g. \cite{wohlmuth2000}. For details on their construction, linearization and application to contact mechanics with small and large deformations including friction, the reader is referred, for instance, to \cite{hueber_2005,hueber_2008,popp2009,popp2010,gitterle2010,wohlmuth2007}.\\
\end{remark}
The Lagrange multiplier $\tractionlambda$ representing the asperity contact traction is interpolated by dual basis functions $\dualshapef{j}$, similar to Eq. (\ref{equ:non11}) and discrete nodal values $\mathlarger{\tractionlambdaj}$:
\begin{equation}\label{equ:non31}
\tractionlambdah=\sum_{j=1}^{\nnodeone}\dualshapef{j} \mathlarger{\tractionlambdaj}
\end{equation}
Inserting the virtual displacement interpolation according to Eq. (\ref{equ:non8}), and Eq. (\ref{equ:non11}) and (\ref{equ:non31}) into Eq. (\ref{equ:non2}) gives
\begin{multline}\label{equ:non13}
\testfkt{\virtwlubh}=\sum_{j=1}^{\nnodeone}\sum_{k=1}^{\nnodeone}({\traction_j}_{\nparallel}^{(1)T} + \mathlarger{\tractionlambdaj})\left(\underbrace{ \int_{\slaveinterfaceh} \dualshapef{j} \shapef_k^{(1)}d\interface \unitydim }_{\mortard[j,k]}
\right)\testfkt{\dispd_k^{(1)}}-\\\sum_{j=1}^{\nnodeone}\sum_{m=1}^{\nnodetwo}({\traction_j}_{\nparallel}^{(1)T} + \mathlarger{\tractionlambdaj})\left(\underbrace{ \int_{\slaveinterfaceh} \dualshapef{j} (\shapef_m^{(2)}\circ\bimapiiih) d\interface \unitydim}_{\mortarm[j,m]} \right)\testfkt{\dispd_m^{(2)}} +\\
\sum_{j=1}^{\nnodeone}\sum_{k=1}^{\nnodeone}{\traction_j}_{\parallel}^{(1)T}\left(\underbrace{ \int_{\slaveinterfaceh} \dualshapef{j} \shapef_k^{(1)}d\interface \unitydim }_{\mortard[j,k]}
\right)\testfkt{\dispd_k^{(1)}}  +   \sum_{j=1}^{\nnodeone}\sum_{m=1}^{\nnodetwo}{\traction_j}_{\parallel}^{(1)T}\left(\underbrace{ \int_{\slaveinterfaceh} \dualshapef{j} (\shapef_m^{(2)}\circ\bimapiiih) d\interface \unitydim}_{\mortarm[j,m]} \right)\testfkt{\dispd_m^{(2)}}
\end{multline}
where $\bimapiiih$ denotes a discrete version of the mapping operator $\bimapiii$. The integrals appearing in Eq. (\ref{equ:non13}) are deformation-dependent since they are evaluated on the deformed slave surface and also due to the discrete mapping operator $\bimapiiih:\slaveinterfaceh\rightarrow \masterinterfaceh$. They make up the nodal blocks of the so-called mortar matrices $\mortard[j,k]$ and $\mortarm[j,m]$, with $j,k=1,...,\nnodeone$ and $m=1,...,\nnodetwo$ which obviously become diagonal when using the dual basis from Eq. (\ref{equ:non35}). $\unitydim$ denotes an identity matrix of size $\nnodedim$. 
The evaluation of the $\mortarm[j,m]$ integral in Eq. (\ref{equ:non13}) requires special considerations, since $\shapef_m^{(2)}$ , which is a shape function with respect to the parameter space of master side elements, is integrated over the slave surface, for which different integration techniques exist \cite{farah2014}.\\
Going back to the discrete virtual work contribution of the lubrication traction, Eq. (\ref{equ:non13}) can be expressed in global matrix notation as given below. Therein, $\dispd_{\slavespace}^{\discretespacet}$ and $\dispd_{\masterspace}^{\discretespacet}$ denote discrete vectors containing slave and master displacement DOFs. A more convenient notation can be achieved, when sorting the global discrete displacement vector $\dispd=(\dispd_{\discretespacen},\dispd_{\slavespace},\dispd_{\masterspace})$ with $\dispd_{\discretespacen}$ containing all the DOFs which are not part of the lubricated boundaries.
\begin{equation}\label{equ:non14}
\testfkt{\virtwlubh}=\testfkt{\dispd}_{\slavespace}^{\discretespacet}\mortard^T(\lubtractionnparallel^{(1)}+ \mathlarger{\tractionlambdaj})-\testfkt{\dispd}_{\masterspace}^{\discretespacet}\mortarm^T(\lubtractionnparallel^{(1)}+ \mathlarger{\tractionlambdaj}) + \testfkt{\dispd}_{\slavespace}^{\discretespacet}\mortard^T\lubtractionparallel^{(1)}+\testfkt{\dispd}_{\masterspace}^{\discretespacet}\mortarm^T\lubtractionparallel^{(1)}=\testfkt{\dispd}^T\discreteglobalvec_{\lubconletter}(\dispd,\prestns,\mathlarger{\tractionlambda)}
\end{equation}
$\discreteglobalvec_{\lubconletter}$ can be interpreted as the discrete global vector of the forces applied by the lubricant and the asperity contact. Next, this lubrication force has to be incorporated in the discrete balance of linear momentum, which is done in a fully implicit way, and complements the discrete equilibrium with a lubrication traction contribution. Applying similar finite element discretizations to the other virtual work terms in Eq. (\ref{equ:non7}) yields
\begin{equation}\label{equ:non16}
\residual_{d}(\dispd,\prestns)=\massmatrix\dispsddot+\discreteglobalvec_{\internal}(\dispd)-\discreteglobalvec_{\external}-\discreteglobalvec_{\lubconletter}(\dispd,\prestns,\mathlarger{\tractionlambda)}=0
\end{equation}
with the global mass matrix $\massmatrix$ and the vector of non-linear internal forces $\discreteglobalvec_{\internal}$. The external forces $\discreteglobalvec_{\external}$ are assumed to be deformation-independent for the sake of simplicity. All the discrete vectors are of size $\nnodedof = \nnodedim \cdot \ndnode$, where $\nnodedof$ refers to the number of displacements DOFs.\\
The discretization of the interface constraint is not being detailed in this paper and interested readers are refered to \cite{hueber_2005,hueber_2008,sitzmann_2015}. The discrete system derived afterwards includes discrete inequality constraints for normal contact $\ineqnormalcon$ and Coulomb friction $\ineqtanjcon$ at all slave nodes $\slavespace$ which can be treated using the nonlinear complementarity (NCP) functions, see e.g. \cite{hueber_2005,hueber_2008,sitzmann_2015}.
\subsubsection{Overall formulation for the coupled lubricated contact problem}
The fully coupled nonlinear system of equations of the lubricated contact problem to be solved for each time step comprises the solid Eq. (\ref{equ:non16}) and lubrication equilibrium Eq. (\ref{equ:non27}), and finally, the contact NCP functions, $\ineqnormalcon$ and $\ineqtanjcon$. All in all, we obtain
\setlength\abovedisplayskip{2pt}
\setlength\belowdisplayskip{2pt}
\begin{equation}\label{equ:new1}
\residual_d(\dispd,\prestns,\tractionlambda)=\massmatrix\dispsddot+\discreteglobalvec_{\internal}(\dispd)-\discreteglobalvec_{\external}-\discreteglobalvec_{\lubconletter}(\dispd,\prestns,\tractionlambda)=0
\end{equation}
\begin{equation}\label{equ:new2}
\residual_p(\dispd,\prestns)=0
\end{equation}
\begin{equation}\label{equ:new3}
\ineqnormalcon(\dispd,\tractionlambda)=0 
\end{equation}
\begin{equation}\label{equ:new4}
\ineqtanjcon(\dispd,\tractionlambda)=0 
\end{equation}
The two complementarity functions are semi-smooth due to the max-function and the Euclidean norm therein. This justifies to solve this coupled system of equations monolithically using a non-smooth version of Newton’s method \cite{qi_sun_1993}.\\
To reduce the computational effort, the discrete contact Lagrange multipliers, are eliminated from the global system of equations Eqs. (\ref{equ:new1}-\ref{equ:new4}) via condensation at a global level. Details on the condensation procedure and the consistent linearization of the contact complementarity functions in the finite deformation case are omitted here, but a step-by-step derivation can be found in \cite{popp2009,popp2010,gitterle2010}. Accordingly, the remaining linear system to be solved consists of displacement and pressure degrees of freedom only:
\begin{equation}
\begin{bmatrix}
\finalsys_{dd} & \finalsys_{dp} \\
\finalsys_{pd} & \finalsys_{pp}
\end{bmatrix}\begin{bmatrix}
\Delta \dispd \\
\Delta \prestns
\end{bmatrix}=-\begin{bmatrix}
\mathfrak{r}_{d} \\
\mathfrak{r}_{p}
\end{bmatrix}
\end{equation}
\section{Numerical results}\label{results}
In the following sections, three numerical examples with focus on different aspects of computationally solving lubricated contact problems are presented. To start with, the relative motion of a cylinder on a rigid, flat surface with zero dry contact traction contribution is analyzed to demonstrate the principle processes present in the Elastohydrodynamic regime including large deformations. Next, an elastic pin on rigid plane example is studied which includes dry contact traction contribution. The examination of the elastic pin on rigid plane problem proves the ability of the framework to correctly model the continuous transition from mixed to full film lubrication.
Finally, a ball-on-disk Tribometer is analyzed, which includes large contact areas, and validates the capability of our lubricated contact algorithm to accurately represent the lubricated contact problem during the full range of the Stribeck curve by comparing the results to experimental studies. For all examples presented in this section, eight-noded linear hexahedron elements (HEX8) are applied for the spatial discretization of all solid domains and four-noded linear quadrilaterals elements (QUAD4), which consistently result from the HEX8 elements evaluated at element surfaces, for the lubrication domain. All presented algorithms have been implemented in our parallel in-house multiphysics research code BACI\cite{baciweb}. A semi-smooth Newton scheme is applied to solve the nonlinear equations and the convergence criterion for solution of the resulting linearized system of equations and incremental update of the unknowns was chosen as 1e-8.
\subsection{Cylinder on a rigid flat surface}
\begin{figure}[h!]
  \centering
  \subfloat[]{\includegraphics[width=0.4\textwidth]{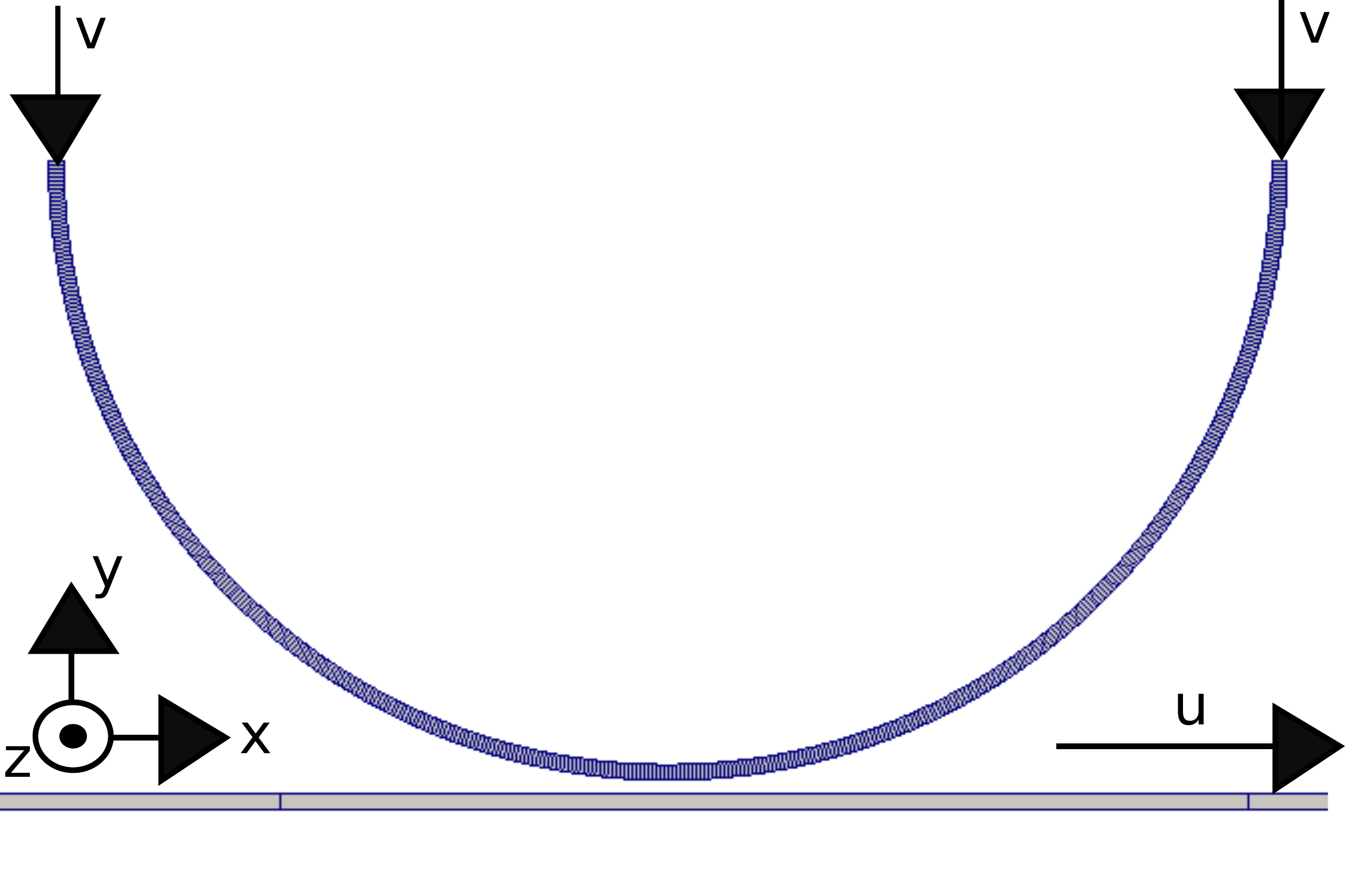}\label{fig:cylonflatschematic}}
  \hfill
  \subfloat[]{\includegraphics[width=0.4\textwidth]{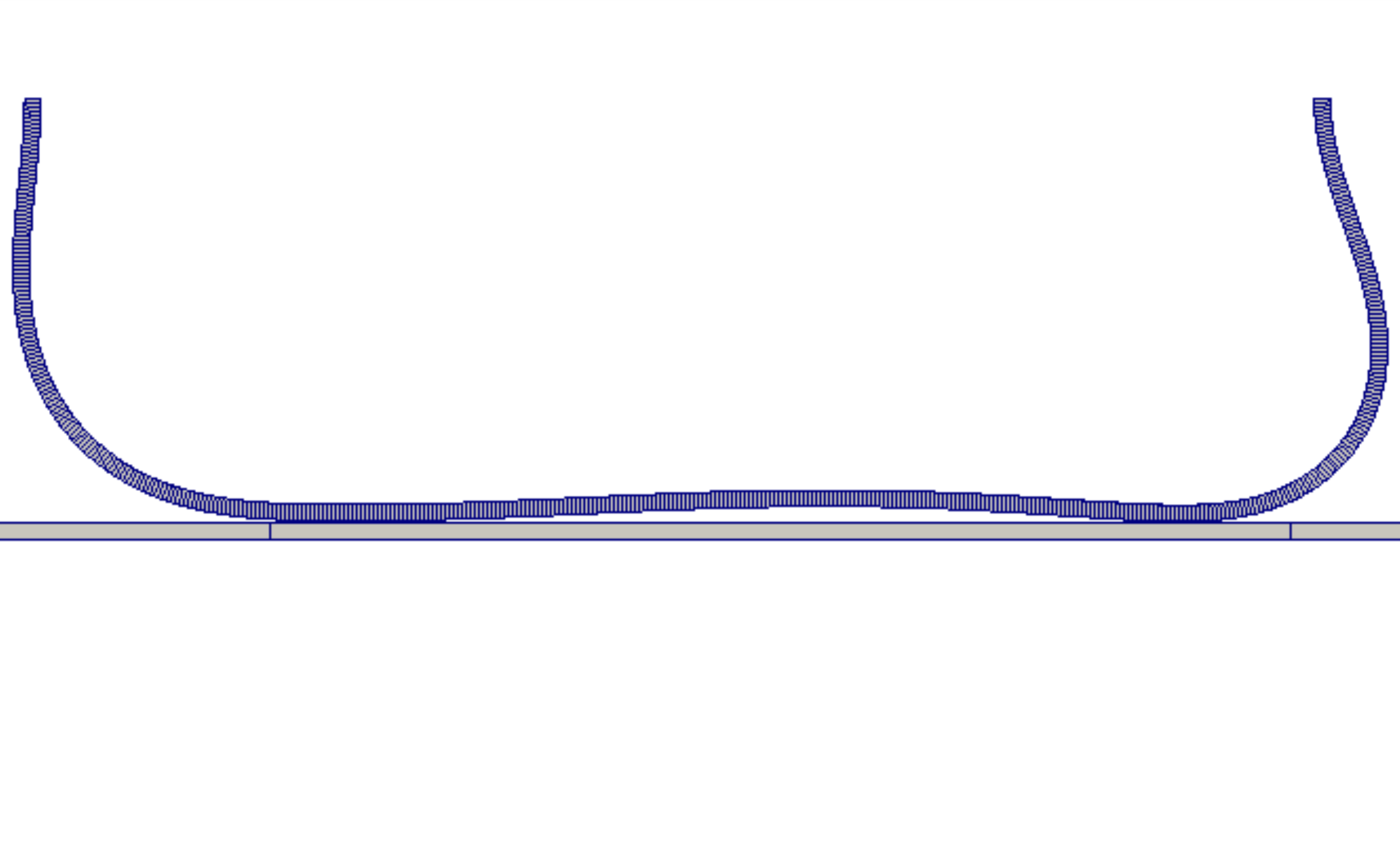}\label{fig:cylmesh1}}
  \caption{The quasi two dimensional elastohydrodynamic lubricated contact problem: (a) Initial state along with problem schematic; (b) deformed configuration at time t=7.5}
  \label{fig:cylmesh}
\end{figure}

\begin{figure}[h!]
  \centering
  \subfloat[timestep 55]{\includegraphics[width=0.5\textwidth]{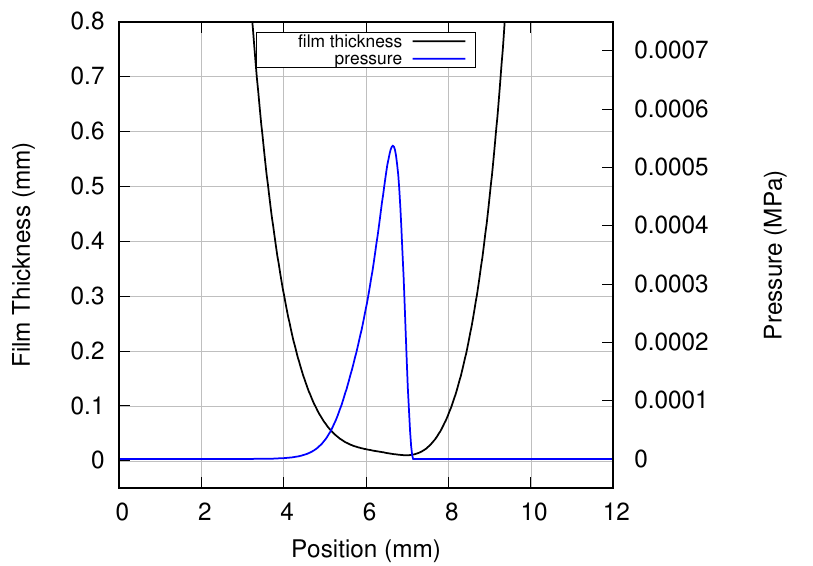}\label{fig:cyl55}}
  \hfill
  \subfloat[timestep 150]{\includegraphics[width=0.5\textwidth]{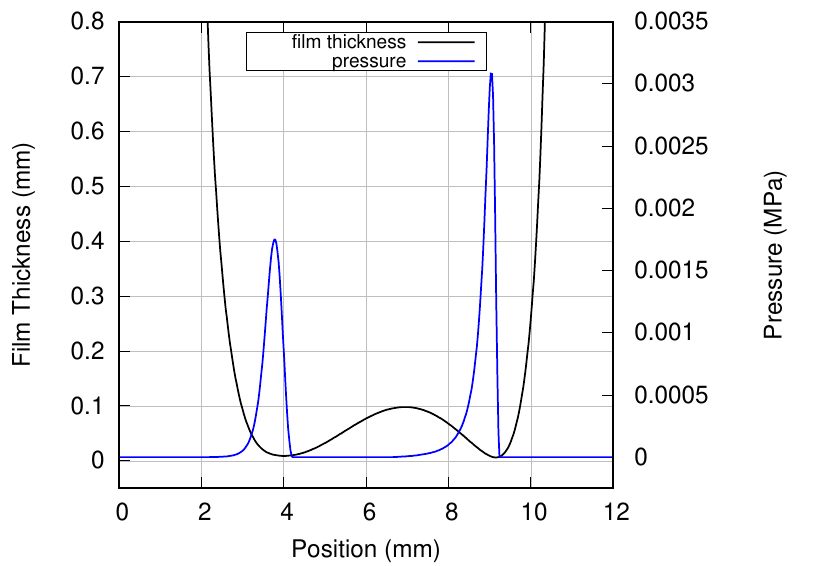}\label{fig:cyl150}}
  \caption{Cylinder on a rigid, flat surface: Pressure distributions and the film thicknesses for elastohydrodynamic lubricated contact problem at two different time steps}
  \label{fig:cylresult}
\end{figure}

In this example, a 2D elastohydrodynamically lubricated line contact problem is analyzed to test the performance of the formulation with regard to large deformations. As illustrated in Fig.~\ref{fig:cylonflatschematic}, we now consider a thin layer of fluid film between the elastic roller (half cylinder) and a rigid flat surface. Two simultaneous motions generate pressure inside the fluid film. The rigid surface is moving with a velocity $u$ in positive $x$-direction, while the cylinder is moving vertically in negative $y$-direction. The pressure in the fluid film will deform the elastic cylinder, and the deformation causes the change in the fluid film profile.
This problem lies in the full film lubrication regime, meaning asperity contact does not exist in this case, which has been confirmed by several researchers with different numerical methods (see, for instance, the work by \cite{hamrock_pan_1988}). However, in most of these research works, the elastic deformation of the cylinder was evaluated by a semi-analytical integral equation, and the fluid phase was solved by either finite difference methods or finite element methods. These types of semi-analytical approaches can only be applied for small deformation problems in which there is an analytical relation between the applied pressure and the deformation for the solid phase. Another drawback of the semi-analytical approach is that the shear stress in the fluid film is not considered when computing the solid phase deformation. In such approaches, the shear stress is usually obtained from post-processing after solving the pressure field in the fluid film.\\
In the present study, a Neo-Hookean material law with Young’s modulus $\young=10$ MPa and Poisson’s ratio $\poisson=0.3$ is utilized for the cylinder. Low Young’s modulus and especially the low structural rigidity of the cylinder makes this problem highly elastic, which gives low pressure magnitudes and cause the effect of pressure on viscosity negligible. Therefore, in this example, an isoviscous lubricant with $\vis=4.0\times 10^{-8}$ MPa.s is considered. The penalty parameter for the cavitation region is taken to be $10^8$ s/mm.\\
The radius of the cylinder is 4 $mm$ and the wall thickness is 0.1 $mm$. The dimensions of the flat do not matter for the problem, since the flat is assumed to be long enough in negative $x$ direction and rigid. The velocity evolution $u$ of the flat is defined by a constant acceleration of 2 mm/$s^2$ and an initial velocity of zero. The load is applied as prescribed displacement at the two uppermost surfaces of the half cylinder until the total reaction force reaches the desired value. In this regard, the cylinder has a constant velocity of \textendash$0.2$ mm/s prescribed as displacement-controlled move on the upper surface in (negative) y-direction. The problem is simulated as a quasi two-dimensional one, i.e. with one element in wall thickness direction.
The rigid flat surface is defined as the master surface and the bottom surface of the cylinder is defined as the slave surface. The mesh (two dimensional mesh) for the lubrication phase is therefore inherited from the cylinder surface. The last boundary condition applied on the problem is on the lubrication domain. The pressure is set to zero on the in- and outflow side of the lubrication domain. The slave surface has $4800$ elements, which refers to the full outer boundary of the cylinder.\\
In Fig.~\ref{fig:cylresult}, the lubricant film thickness and pressure distribution in the lubricant film at the 55th and 150th timestep with $\Delta \mtime = 0.05 s$ are shown. One of the first points standing out is the graph of the film thickness at 150th timestep, caused by the low stiffness of the cylinder. The cylindrical geometry is largely distorted and the deformations dominate the fluid film profile. Fig.~\ref{fig:cylmesh}(b) shows a considerable difference compared to the initial configuration Fig.~\ref{fig:cylmesh}(a). The increase of the film thickness resulting from the concave cylinder surface shape at the center of the contact area also goes in hand with a change of the pressure distribution, as shown in Fig.~\ref{fig:cylresult}(b). In particular, the film thickness curve has two minima and almost exactly at those points, the peak pressures can be seen.
\subsection{Elastic pin on rigid plane}
\begin{figure}[h!]
  \centering
  \subfloat{\includegraphics[width=0.45\textwidth]{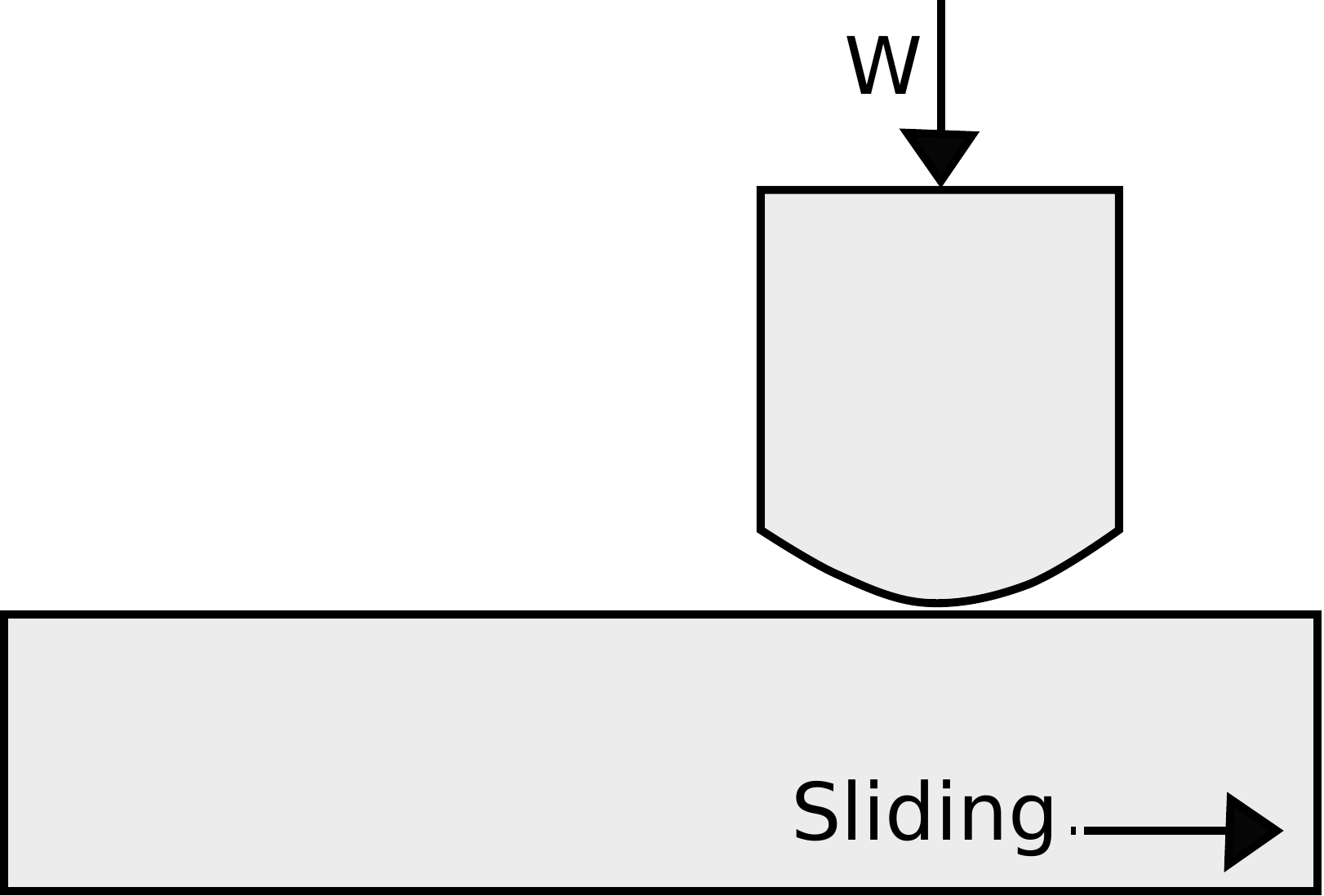}\label{fig:pingeo}}
  \caption{Elastic pin on rigid plane: Schematic of problem}
  \label{fig:pinschematic}
\end{figure}

\begin{figure}[!tbp]
\centering
\renewcommand{\arraystretch}{0}
\addtolength{\floatsep}{-5mm}
\begin{tabular}{|@{}c@{}c@{}|}
  \hline
  \includegraphics[width=0.5\textwidth]{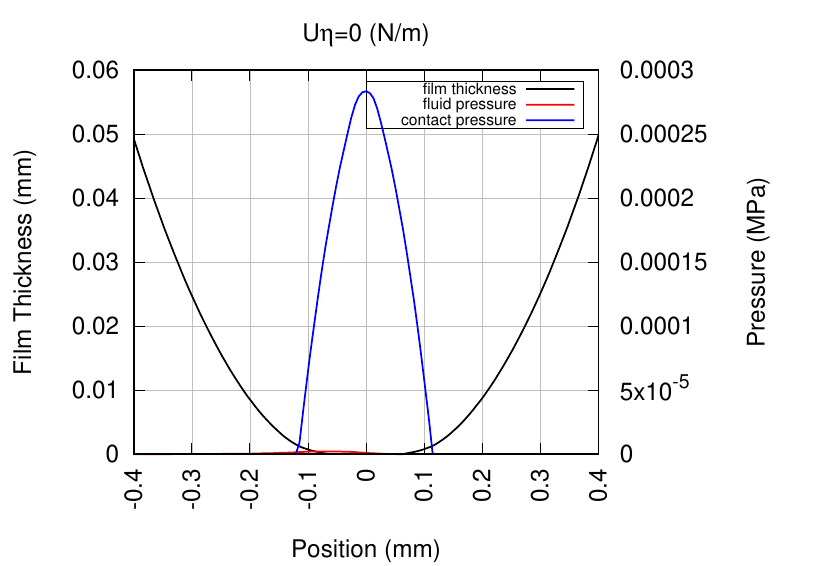}&%
  \includegraphics[width=0.5\textwidth]{{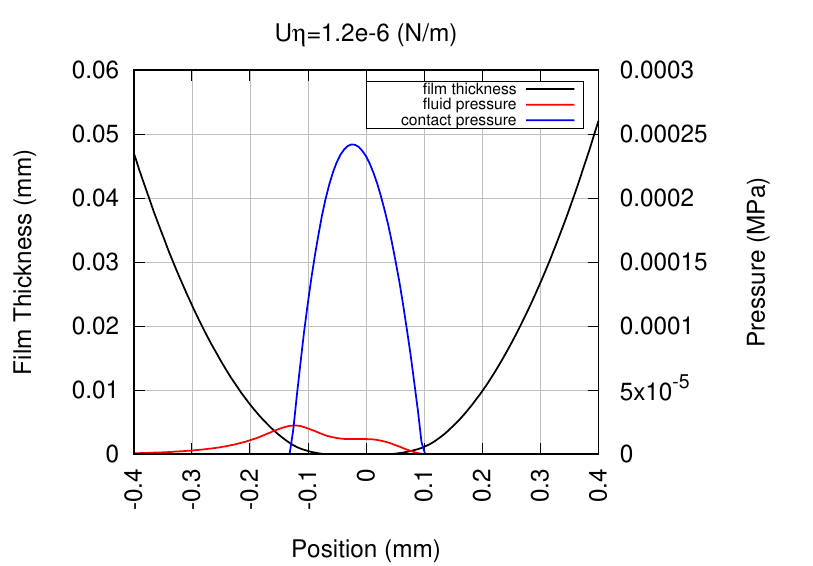}}\\
  \includegraphics[width=0.5\textwidth]{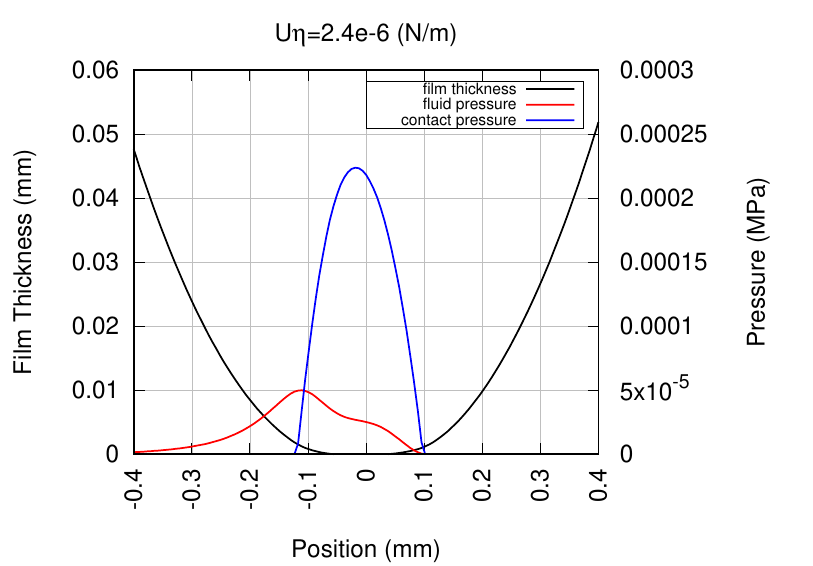}&%
  \includegraphics[width=0.5\textwidth]{{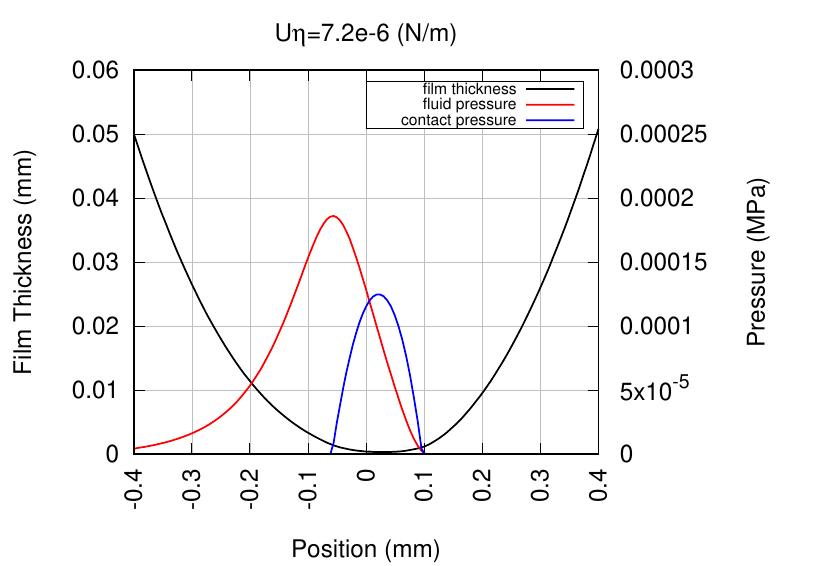}}\\
  \includegraphics[width=0.5\textwidth]{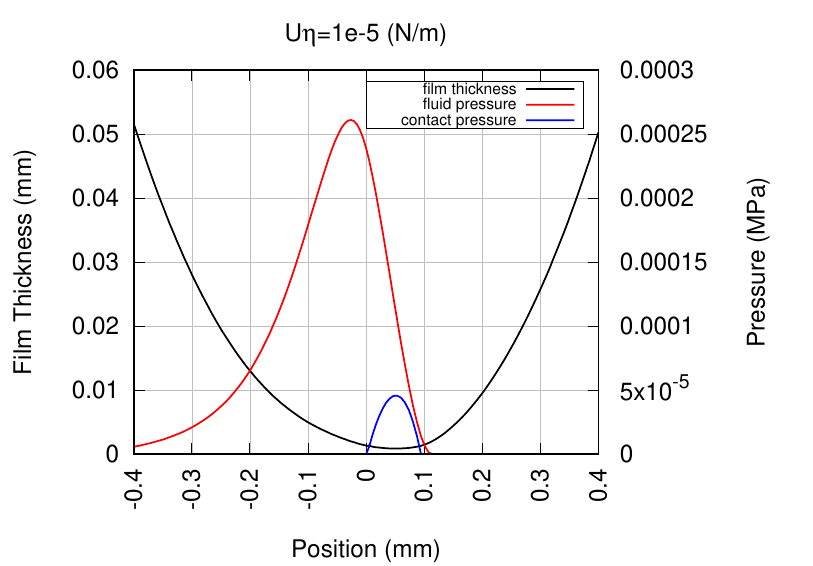}&%
  \includegraphics[width=0.5\textwidth]{{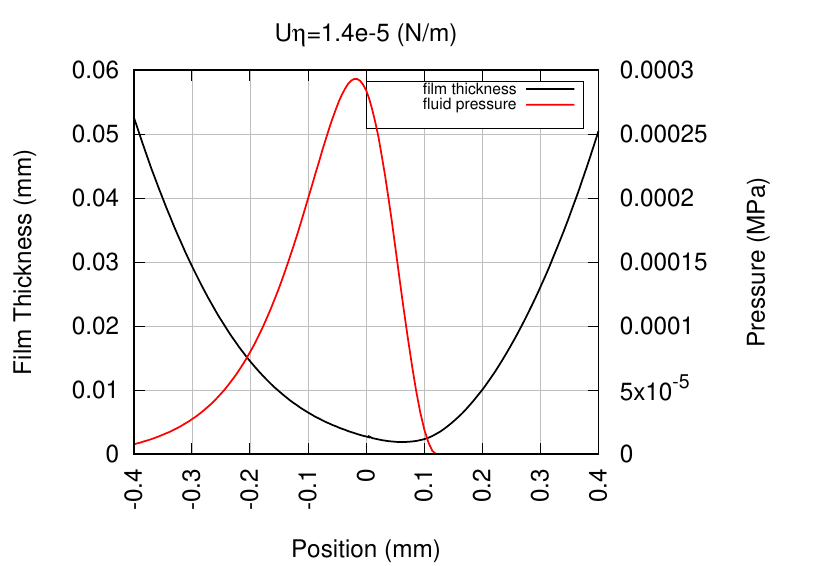}}\\
  \hline
\end{tabular}
\caption{Elastic pin on rigid plane: hydrodynamic pressure, contact pressure and film thickness corresponding to different time steps and the product of rigid plane velocity and lubricant viscosity $U\vis$ (increasing from top left to bottom right) for the lubricated contact problem}
  \label{fig:pinresult}
\end{figure}

\begin{figure}[h!]
  \centering
  \includegraphics[width=0.5\linewidth]{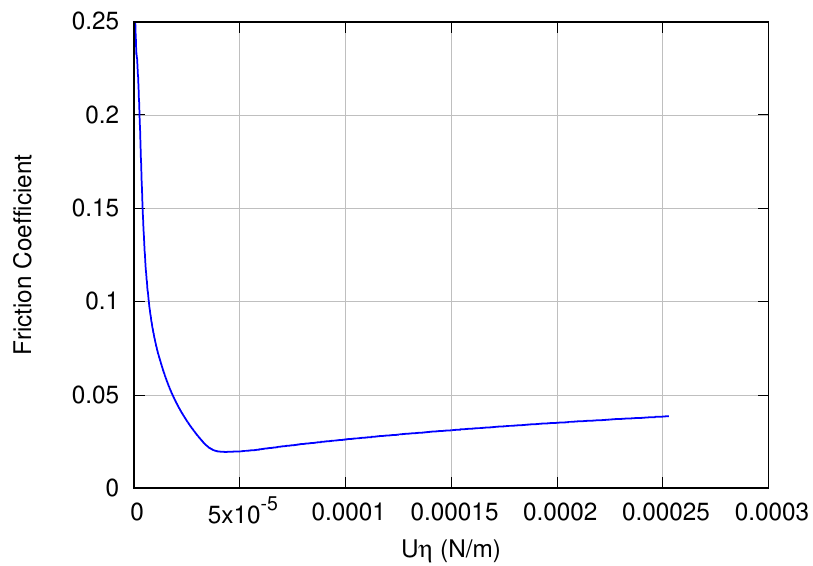}
  \caption{Elastic pin on rigid plane: Friction coefficient as a function of the product of rigid plane velocity $U$ and lubricant viscosity $\vis$}
  \label{fig:pinstribeck}
\end{figure}

As second example, a simplified two dimensional elastic pin on rigid plane problem is studied, which roughly corresponds to the pin-on-disk tribological test. This example aims at illustrating the ability of the framework to correctly model the continuous transition from mixed to full film lubrication.
At the beginning, the hyperelastic pin is pressed onto a rigid plane with a constant normal reaction force $W$, in order to squeeze out the lubricant and let the asperity contact occur (Boundary lubrication). Then, the rigid plane begins moving (sliding) to the right with a constant acceleration $a$, whilst the top of the pin is pressed and kept in the same vertical postion. This movement generates pressure inside the lubricant film which lifts the bottom of the pin up, while the top position is still fixed by the DBCs, and this elastic deformation of the pin increases the film thickness (full film lubrication).
It should be noted that the force $W$ is taken as the normal reaction force resulting from the prescribed DBC at the top surface of the pin until the total reaction force reaches the desired value. In this regard, the top of the pin is completely DBC-controlled (in vertical and horizontal direction) and it has a prescribed vertical displacement with constant velocity of $v$ in the beginning while it is horizontally fixed. See Fig.~\ref{fig:pinschematic} for the schematic of the problem and the geometry used in the computations.
Geometrical, material and process parameters are summarized in Table~\ref{table:1}.
The hyperelastic neo-Hookean material model is adopted for the pin. The dimensions of the plane do not matter for the problem, since the plane is assumed to be long enough in negative $x$ direction and rigid. The rigid plane surface is defined as the master surface and the bottom surface of the pin is defined as the slave surface. The mesh (two dimensional mesh) for the lubrication domain is therefore inherited from the pin surface. As boundary conditions for the lubrication problem the hydrodynamic pressure is prescribed to zero on the left and right end of the lubrication domain. The slave surface has $40$ elements. The range of rigid plane velocities is specified in Table~\ref{table:1} and corresponds to the product of rigid plane velocity and lubricant viscosity $U\vis$ varying between 0 and 1.4e-5 N/m.
\begin{table}[h!]
\centering
\caption{Elastic pin on rigid plane: geometrical, material and process parameters}
\setlength{\tabcolsep}{30pt} 
\begin{tabular}{ l c }
\hline
 Pin radius & 1.5 mm \\
 Pin height and length & 1 mm \\ 
 Plane thickness & 1 mm \\
 Cavitation penalty parameter & $10^8$ s/mm \\
 Young's modulus & 1e-2 MPa \\ 
 Poisson's ratio & 0 \\
 Lubricant viscosity, $\vis$ & 4e-8 MPa s \\ 
 Rigid plane velocity, $U$ & 0-0.35 mm/s \\
 Friction coefficient of dry solid, $\dryfriction$ & 0.25 \\  
 Pin constant velocity, $v$ & 0.05 mm/s \\
 Plane constant acceleration, $a$ & 0.01 mm/$s^2$ \\
 $\regthick$ & 3.0 $\mu m$ \\
 regularization stiffness, $\regstiff$ & 1 MPa/mm \\
 Surface roughness standard deviation, $\stddev$ & 1.0 $\mu m$ \\
\hline    
\end{tabular}
\label{table:1}
\end{table}
Film thickness profiles, and hydrodynamic and asperity contact pressure distributions are illustrated in Fig.~\ref{fig:pinresult} for various individual time steps and representative value of the product of rigid plane velocities and lubricant viscosity $U\vis$. The (horizontal) position is measured along the surface, and the zero value corresponds to the horizontal position of the centre of the pin.
It can be seen in Fig.~\ref{fig:pinresult} that in the very beginning, where $U\vis = 0.0 N/m$, the gap is closed and contact pressure is the only active pressure while there is no sliding, i.e. the hydrodynamic pressure is zero. After the plane starts to move towards the right with constant acceleration, while keeping the constant reaction force on the pin, the hydrodynamic pressure initiates which can be seen in the graph corresponding to $U\vis = 1.2e-6 N/m$ and is found to increase furthur, at $U\vis = 2.4e-6 N/m$ while the contact pressure is diminishing. With increasing the rigid plane velocity, the hydrodynamic pressure develops more and becomes the dominant pressure of the system, which can be seen in the graph related to $U\vis = 7.2e-6 N/m$, and this is accompanied with film thickness growth at the same time. Further increase of rigid plane velocity induces the pin to lift up gradually and the contact pressure to disappear, shown in $U\vis = 1e-5 N/m$. Graph related to $U\vis = 1.4e-5 N/m$ corresponds to the highest rigid plane velocity considered, leading to large lubricant film thickness. In this case the gap between the plane and the pin is clearly visible.
To conclude, the elastic pin on rigid plane example, the simulated friction coefficient as a function of the product of rigid plane velocity $U$ and lubricant viscosity $\vis$ is presented in Fig.~\ref{fig:pinstribeck}. The friction coefficient is calculated as the ratio of the measured horizontal reaction forces and the vertical force $W$ at the top of the pin. It is seen that the solution covers the entire range of the Stribeck curve shown in Fig.~\ref{fig:stribek-mixmodel}, and also that the friction coefficient reaches the dry solid friction coefficient for very low rigid plane velocities.
\subsection{Ball-on-Disk tribometer}
\begin{figure}[h!]
\renewcommand{\arraystretch}{0}
  \centering
  \subfloat[]{\includegraphics[width=0.4\textwidth]{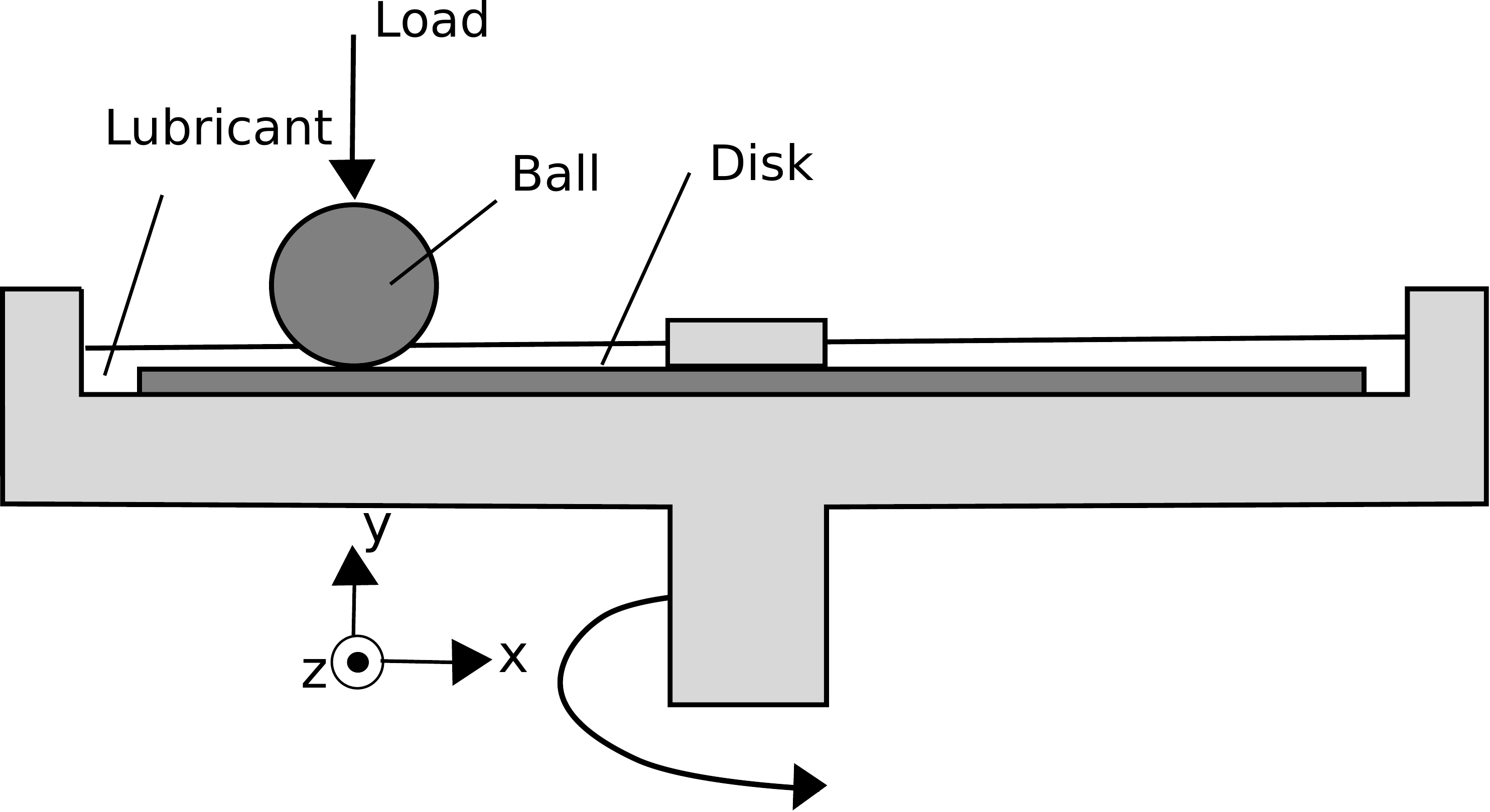}\label{fig:mtmschematic}}
  \hfill
  \subfloat[]{\includegraphics[width=0.35\textwidth]{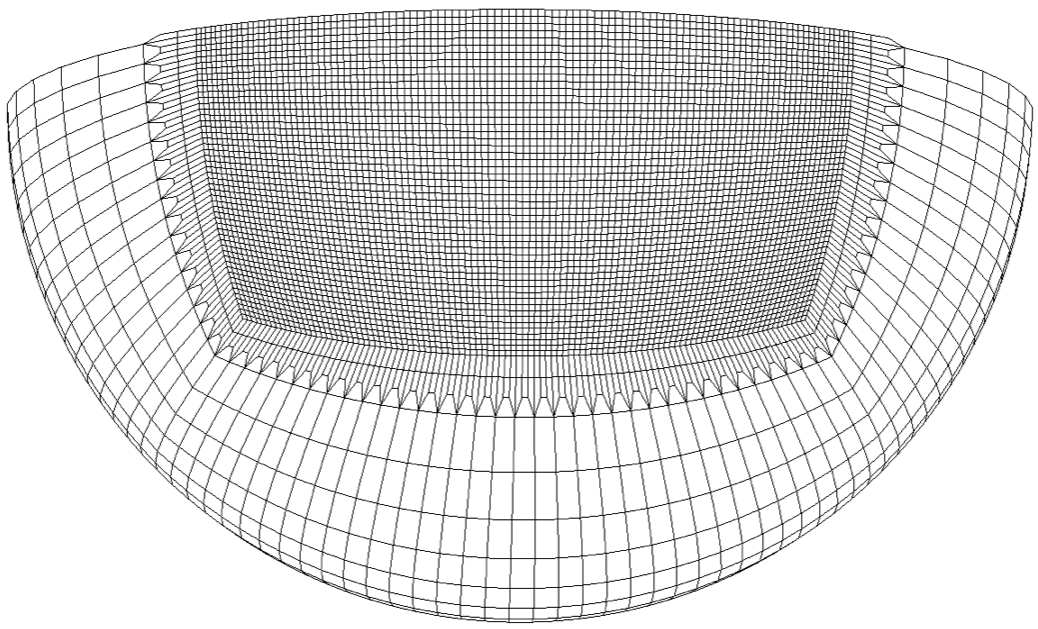}\label{fig:mesh2}}
  \hfill
  \subfloat[]{\includegraphics[width=0.25\textwidth]{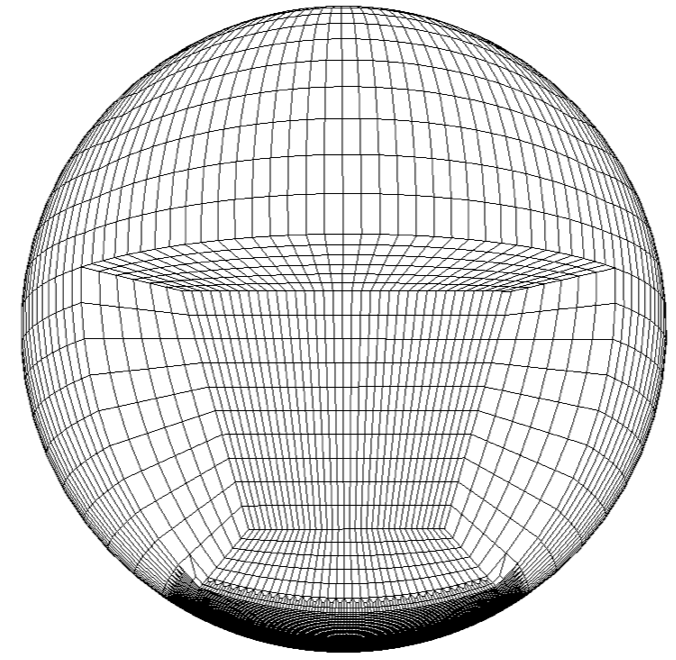}\label{fig:mesh3}}
  \hfill
  \caption{Ball-on disk tribometer:(a) Schematic of experimental setup (Redrawn from \cite{stup_2016}), (b,c) finite element mesh}
  \label{fig:mtmmesh}
\end{figure}

The last example illustrates the application of the present model to a fully coupled large deformation lubricated contact problem. The setup corresponds to a ball-on-disk tribometer studied experimentally in \cite{stup_2016}. Therein, an elastomeric ball is placed in a grip and is loaded by a normal force to be pressed against a rotating rigid flat disk. The setup allows testing of steady-state lubrication in pure sliding only, and the sliding velocity is adjusted by changing the angular velocity of the supporting disk and the radial position of the ball. In a first step, the disk is assumed to be rigid. At the end of this section, the suitability of our model to represent the lubricated contact interaction of two deformable bodies is demonstrated, by modeling the disk as a deformable body as well.\\
In this example, a hyperelastic ball of radius $R$ is loaded by a normal force $W$, thus being pressed against a rotating rigid flat disk. The disk is driven with different values of constant angular velocity which results in sliding velocities $V$ in the range of 62-690 mm/s for a fixed distance of the ball from the axis of rotation. The corresponding radius of the sliding path is 42 mm.
The study is carried out at normal contact force levels $W$ equal to 0.25 N, 0.98 N, 5.13 N and 19.3 N which are applied via prescribed DBC on the top surface of the ball until the total reaction force reaches the desired constant steady-state value. See Fig.~\ref{fig:mtmmesh} for the schematic of experimental setup and the finite-element mesh.\\
The geometrical, material and process parameters are presented in Table~\ref{table:2}. The hyperelastic behaviour of the ball is governed by a neo-Hookean material model. The material properties for the fluid film have been chosen according to five different lubricants, namely distilled water and four silicone oils (Polsil OM 10, OM 50, OM 300 and OM 3000).
For each lubricant and for each normal load, the dry friction coefficient has been taken from experimental results \cite[Fig.~2]{stup_2016}.
The rigid disk surface is defined as the master surface and the bottom surface of the ball is defined as the slave surface. The lubrication domain with coinciding nodes is defined on the lubricated boundary of the slave surface and a steady-state lubrication problem is solved. The mesh (two dimensional mesh) for the lubrication phase is therefore inherited from the ball bottom surface. The cavitation pressure is assumed to be equal to zero, and the Dirichlet boundary condition for the Reynolds equation, $\presf=0$, is prescribed at sufficient distance from the contact zone, which corresponds to the fully flooded condition.
The finite element mesh comprises almost 94,000 elements, and a total of 284,000 unknowns including nodal displacements of the solid body and pressure degrees of freedom on the lubricated contact surface.
\begin{table}[h!]
\centering
\caption{Ball-on-disk tribometer: geometrical, material and process parameters.}
\setlength{\tabcolsep}{30pt} 
\begin{tabular}{ l c }
\hline
 Ball radius & 10.7 mm \\
 Disk radius & 53 mm \\
 Cavitation penalty parameter & $10^8$ s/mm \\
 Ball material-Young's modulus of Rubber (NBR) & 3.5 MPa \\ 
 Poisson's ratio & 0 \\
 Lubricant viscosity at 25 degree Celsius, $\vis$: Distilled water &  0.000891 Pa s \\
 $\vis$: OM 10 & 0.00942 Pa s \\
 $\vis$: OM 50 & 0.0493 Pa s \\
 $\vis$: OM 300 & 0.3395 Pa s \\
 $\vis$: OM 3000 & 2.735 Pa s \\
 Disk sliding velocity, $V$ & 62-690 mm/s \\
 $\regthick$ & 4 $\mu m$ \\
 regularization stiffness, $\regstiff$ & 1e2 MPa/mm \\
 Ball surface roughness standard deviation, $\stddev$ & 1.33 $\mu m$ \\
\hline    
\end{tabular}
\label{table:2}
\end{table}
\subsubsection{Study of Stribeck curve}
\begin{figure}[h!]
\renewcommand{\arraystretch}{0}
  \centering
  {\includegraphics[width=0.32\textwidth]{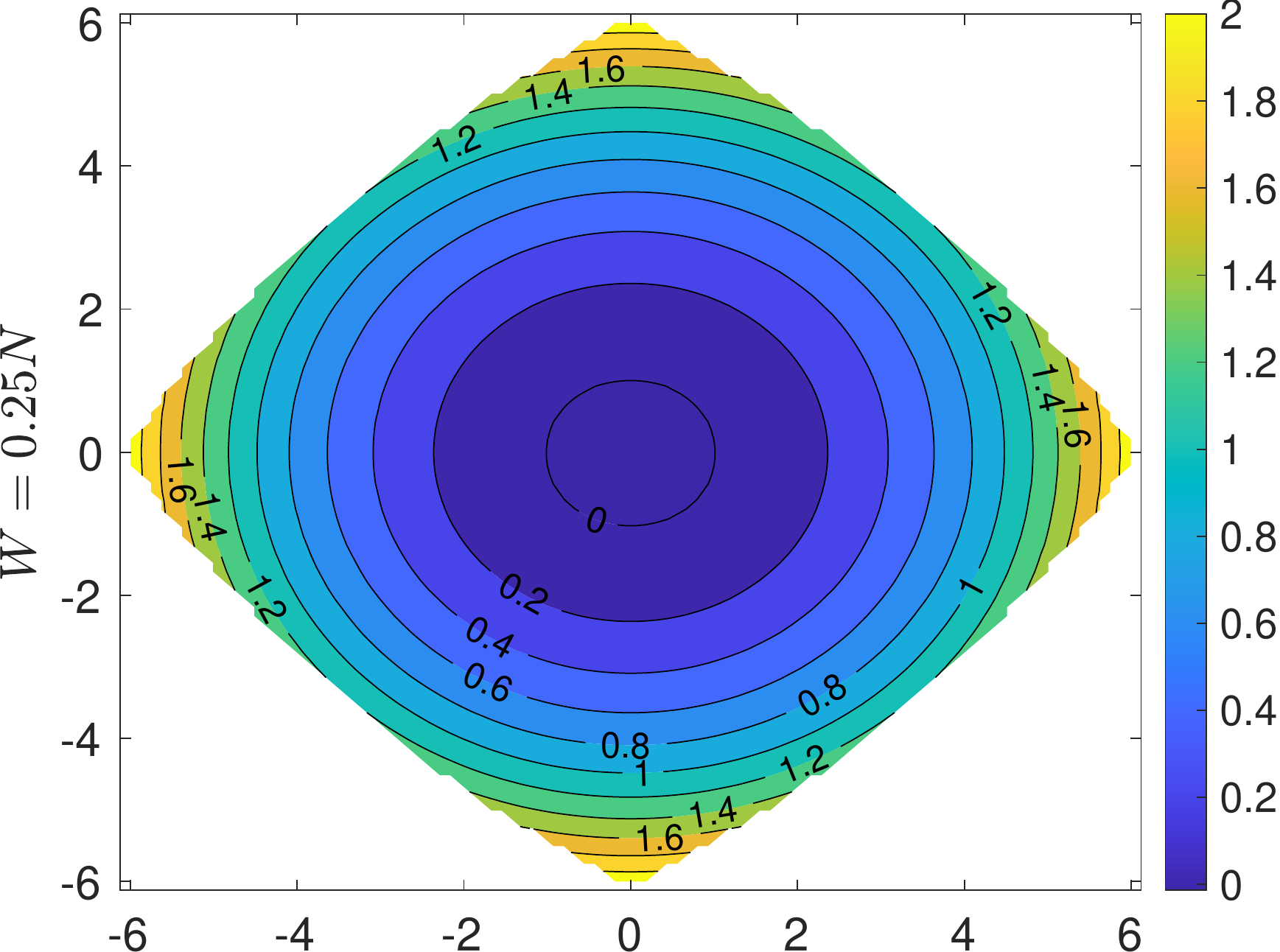}\label{fig:1cf}}
  \hfill
  {\includegraphics[width=0.33\textwidth]{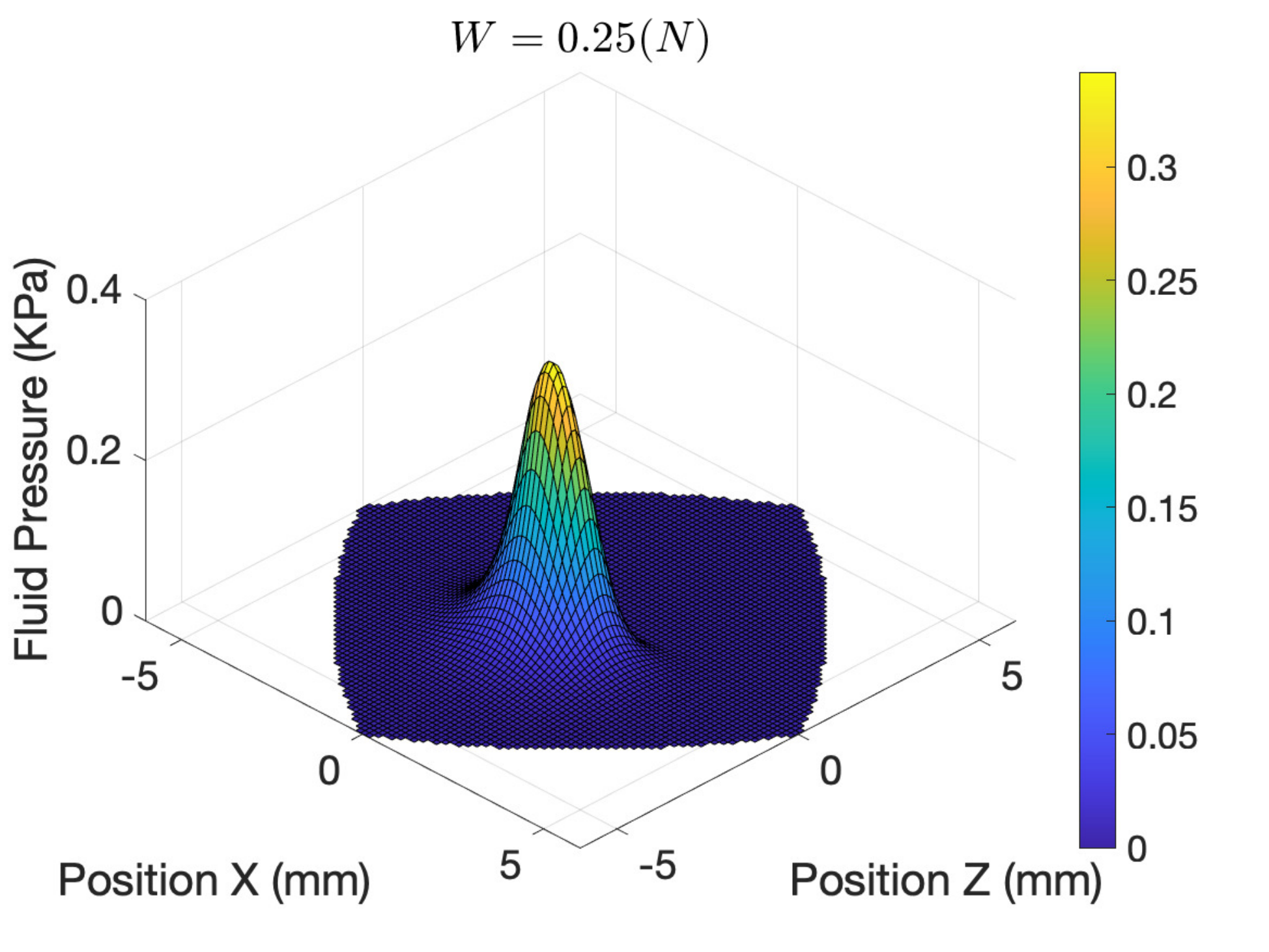}\label{fig:3cf}}
  \hfill
  {\includegraphics[width=0.33\textwidth]{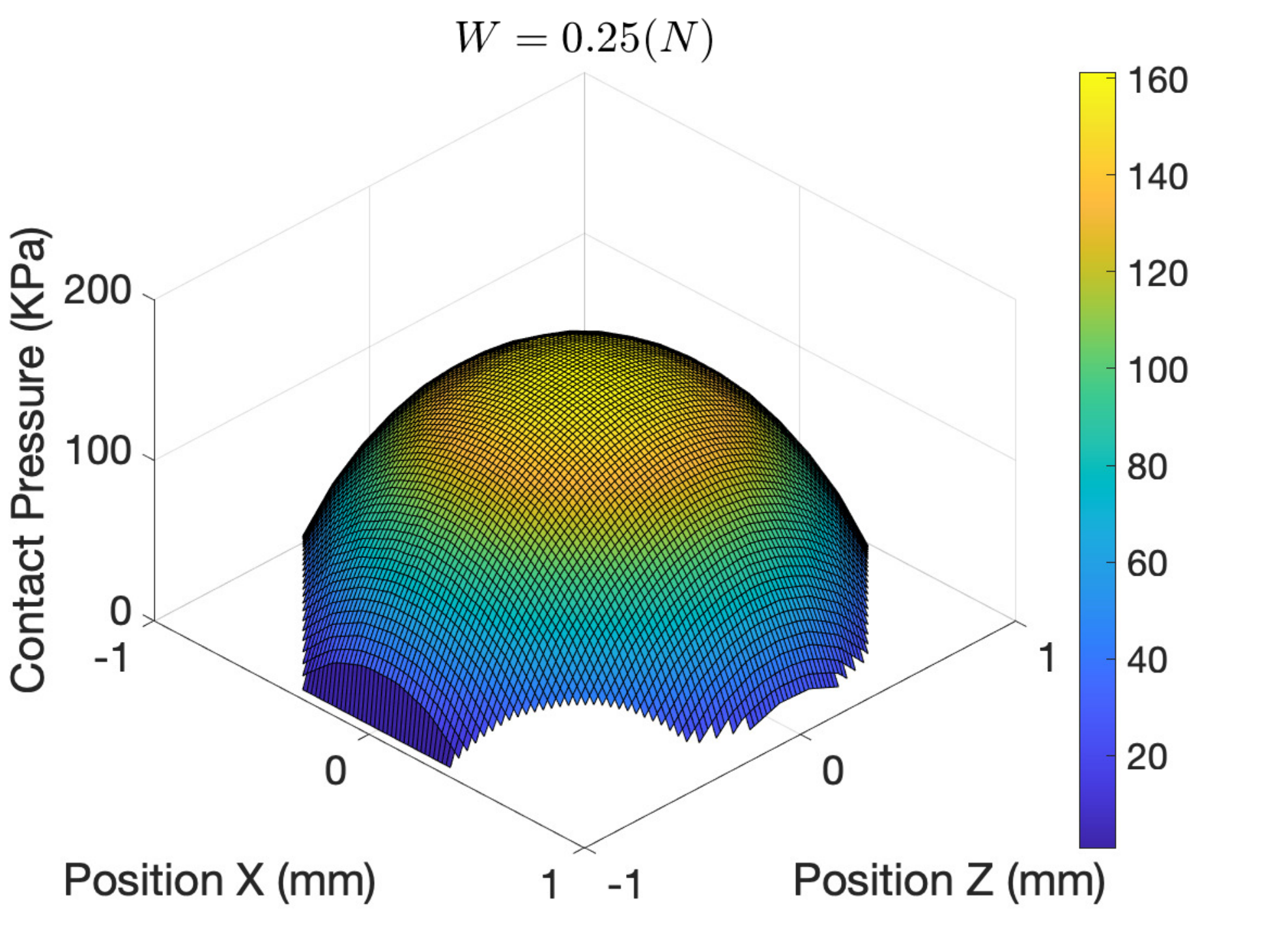}\label{fig:3w9cf}}
  \hfill
  {\includegraphics[width=0.32\textwidth]{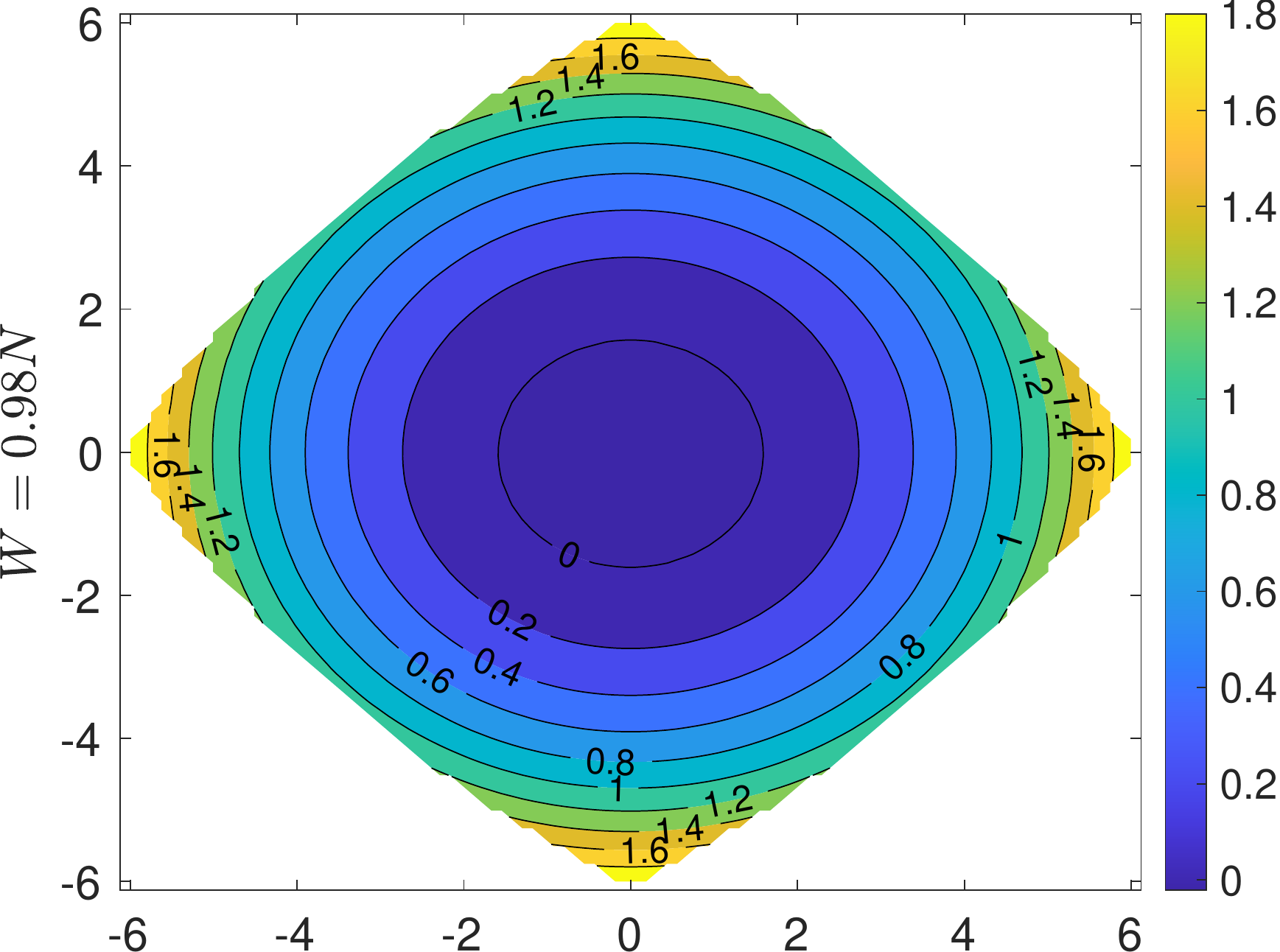}\label{fig:1cf}}
  \hfill
  {\includegraphics[width=0.33\textwidth]{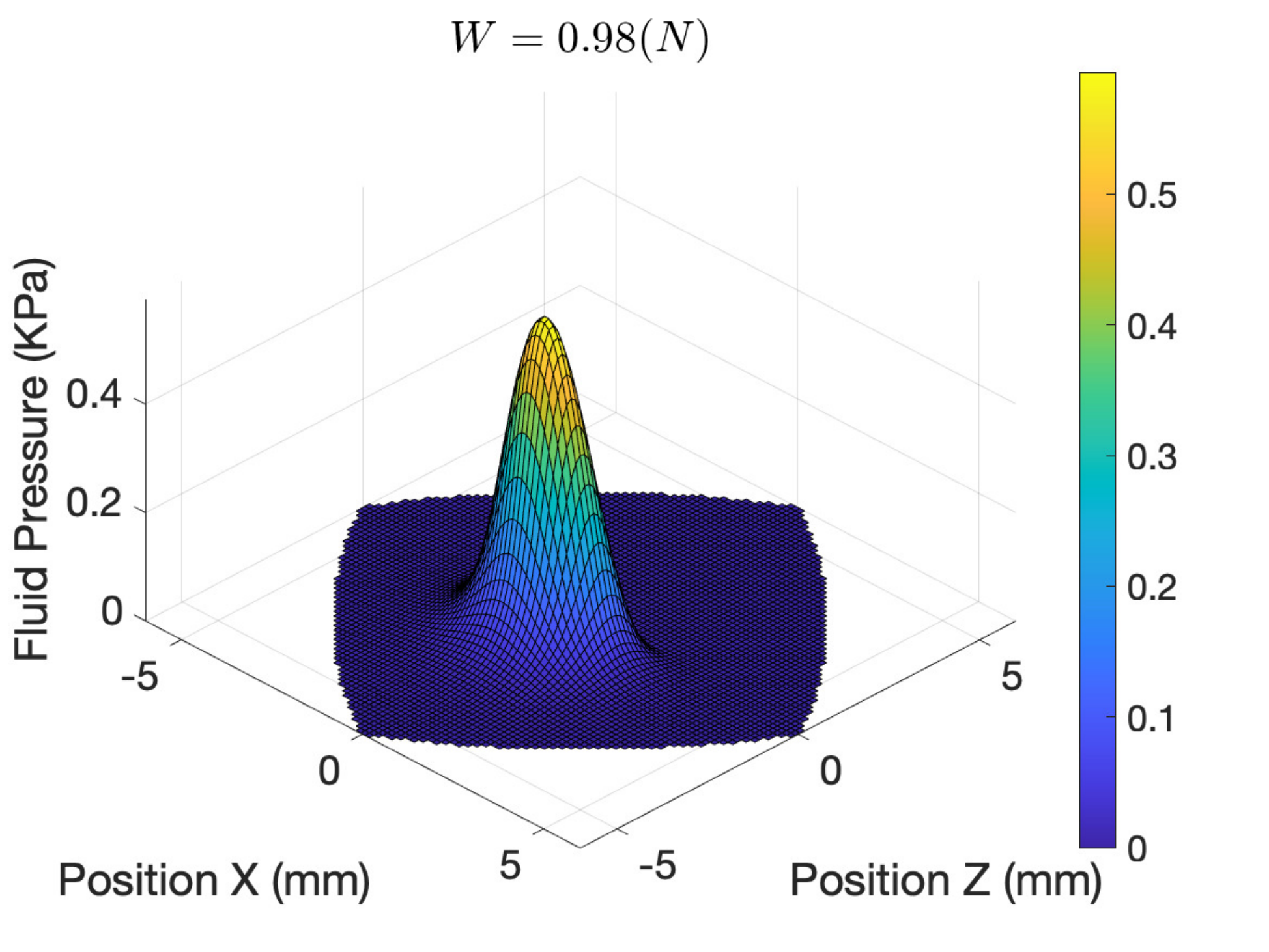}\label{fig:3cf}}
  \hfill
  {\includegraphics[width=0.33\textwidth]{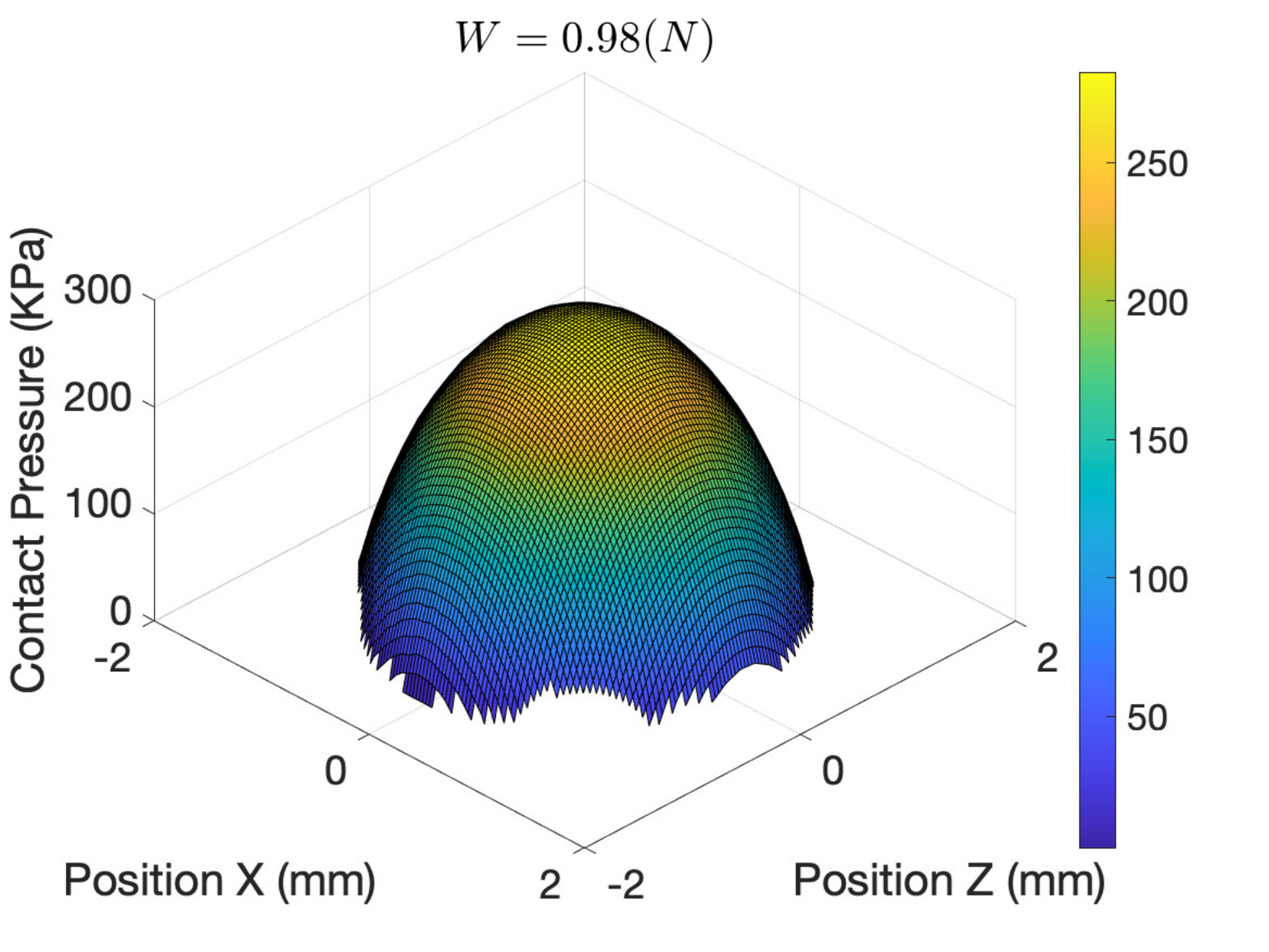}\label{fig:3w9cf}}
  \hfill
  {\includegraphics[width=0.32\textwidth]{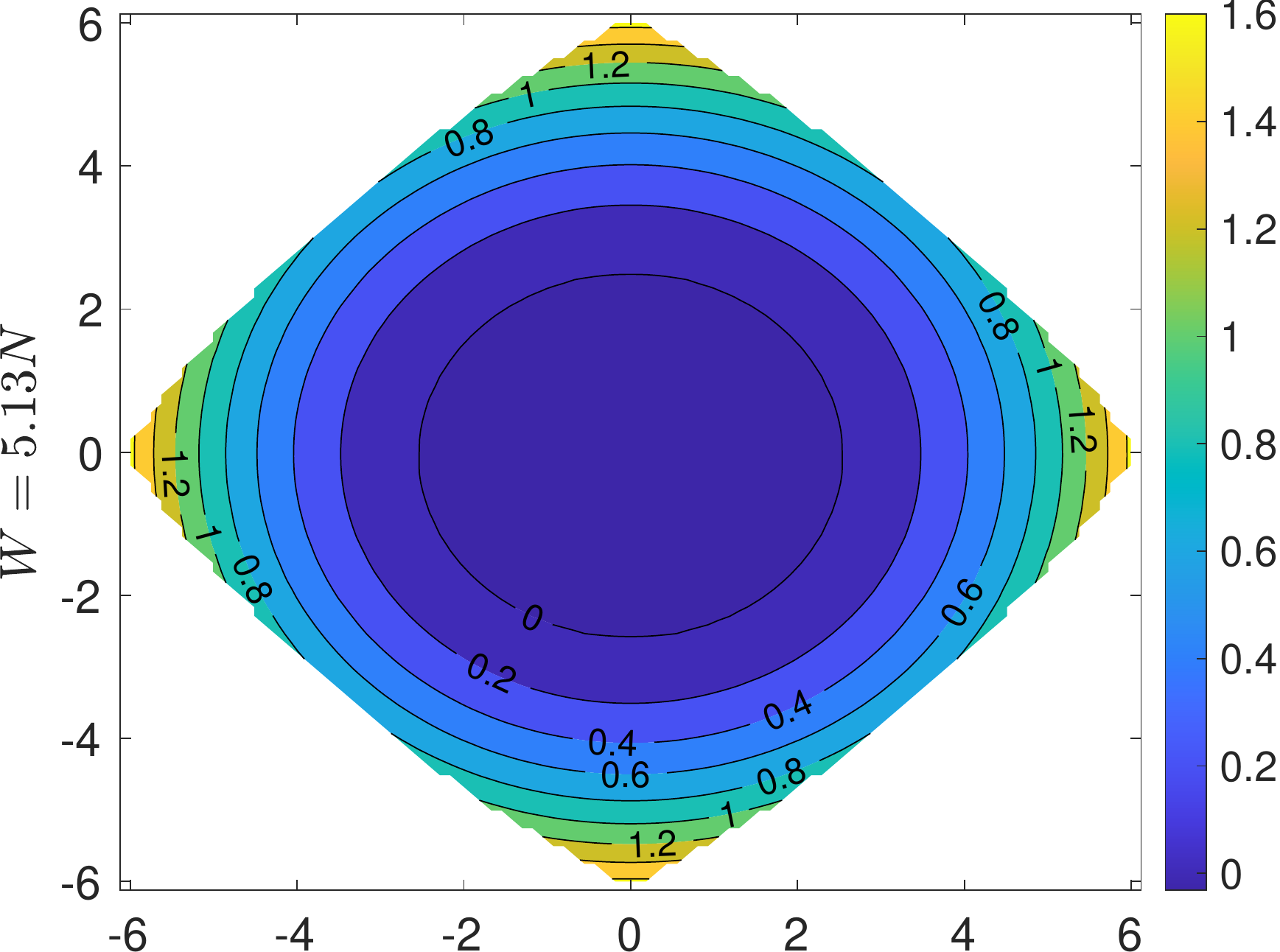}\label{fig:4cf}}
  \hfill
  {\includegraphics[width=0.33\textwidth]{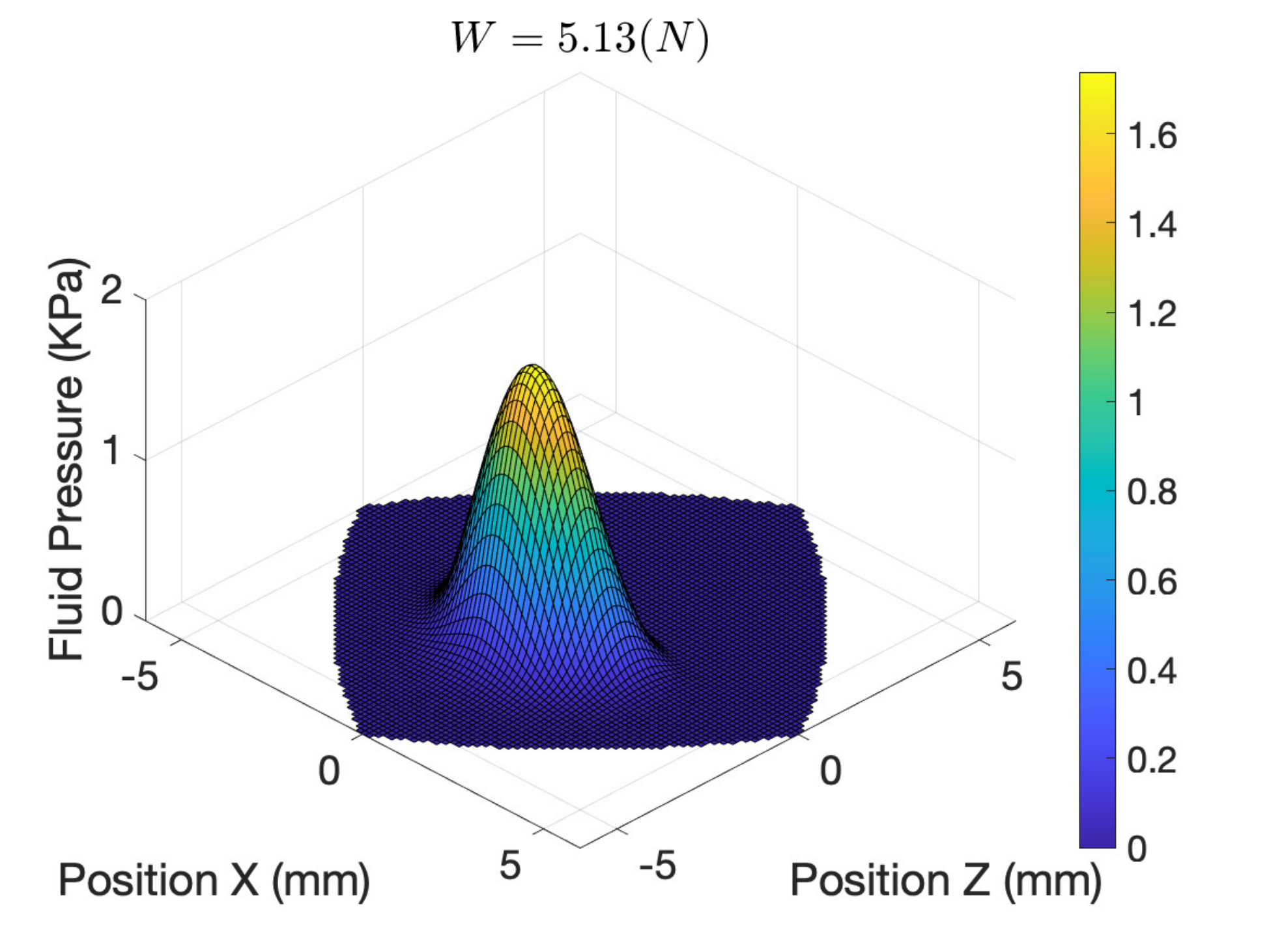}\label{fig:6cf}}
  \hfill
  {\includegraphics[width=0.33\textwidth]{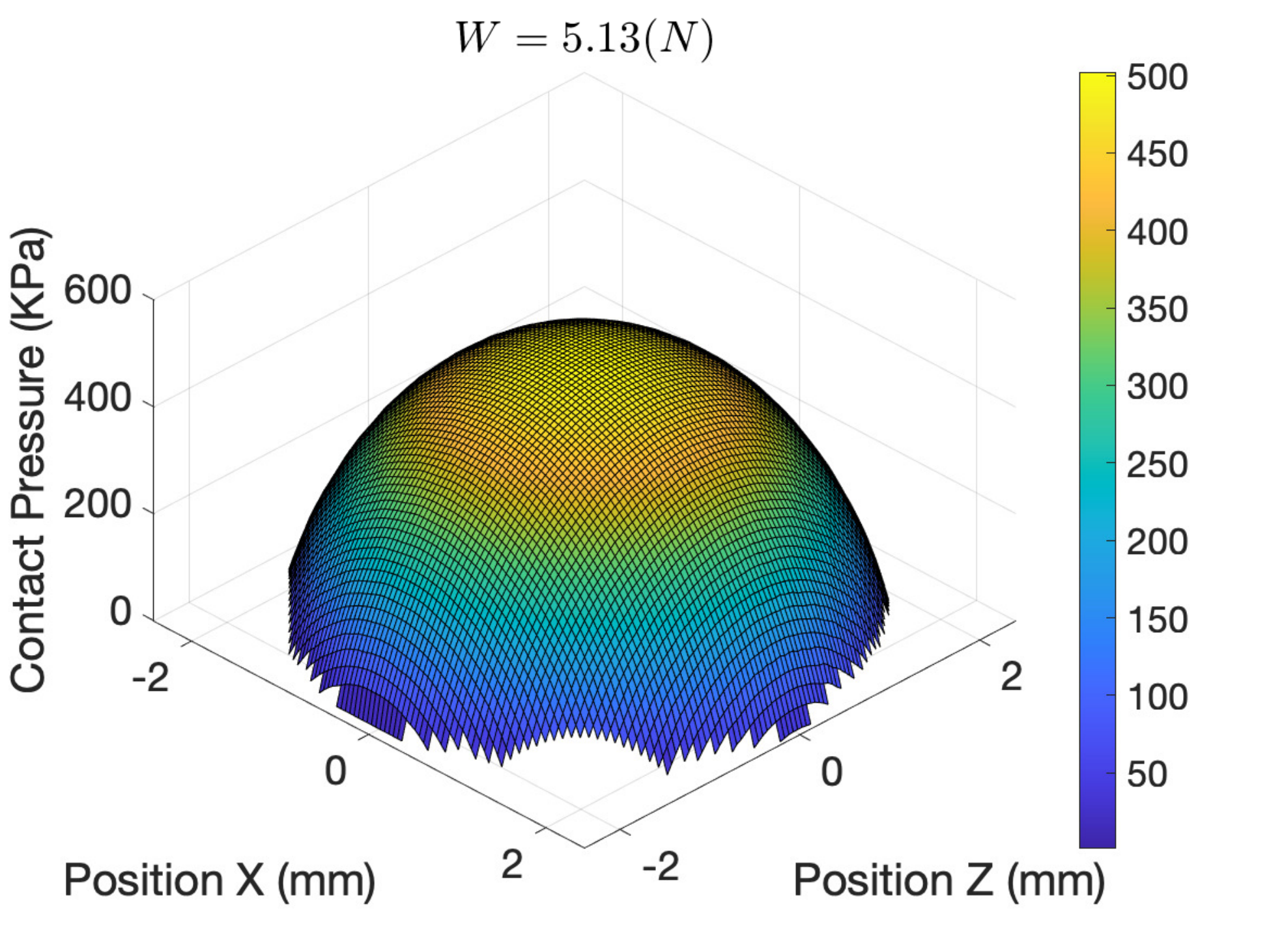}\label{fig:6w9cf}}
  \hfill
  \subfloat[]{\includegraphics[width=0.32\textwidth]{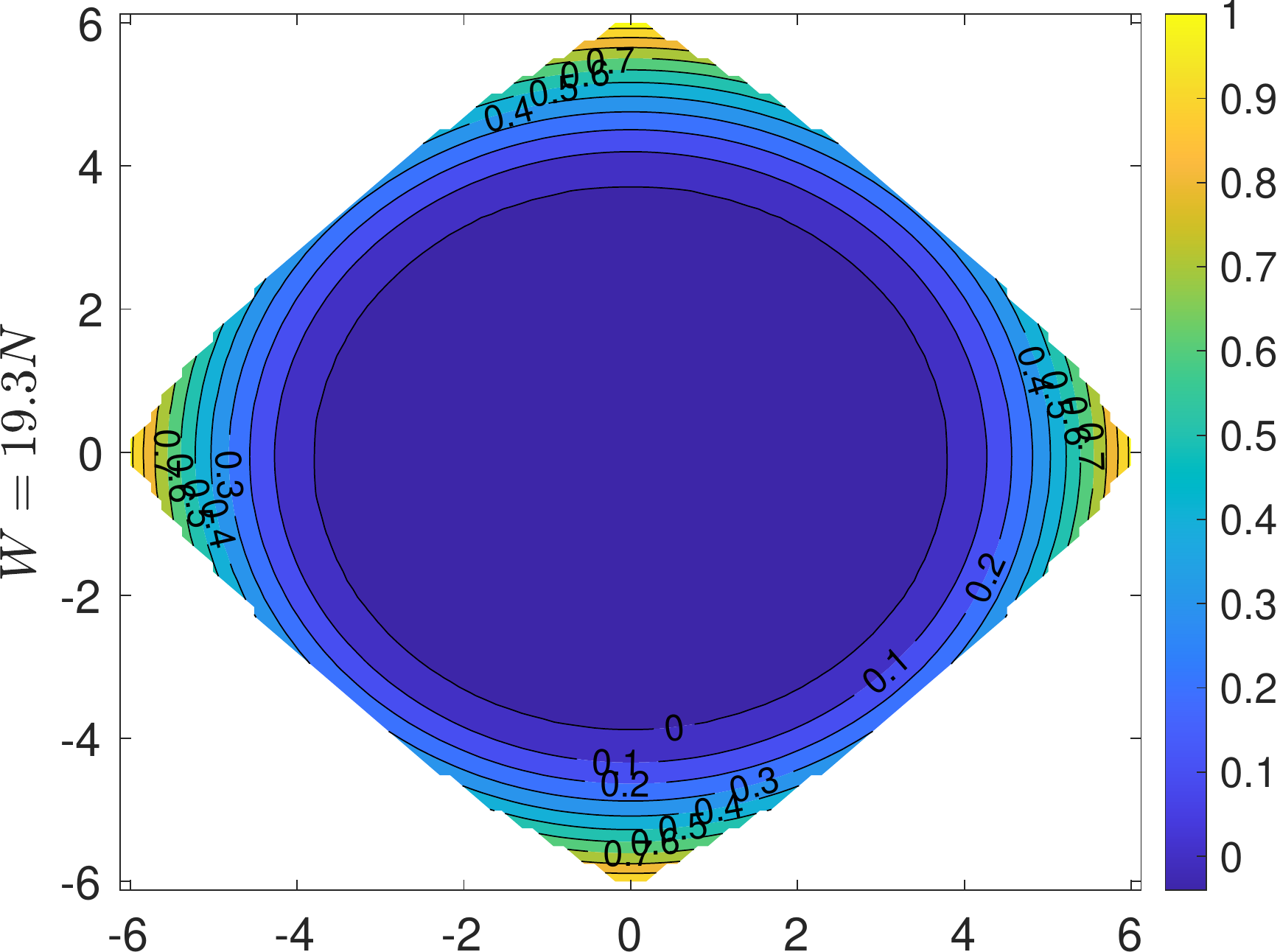}\label{fig:7cf}}
  \hfill
  \subfloat[]{\includegraphics[width=0.33\textwidth]{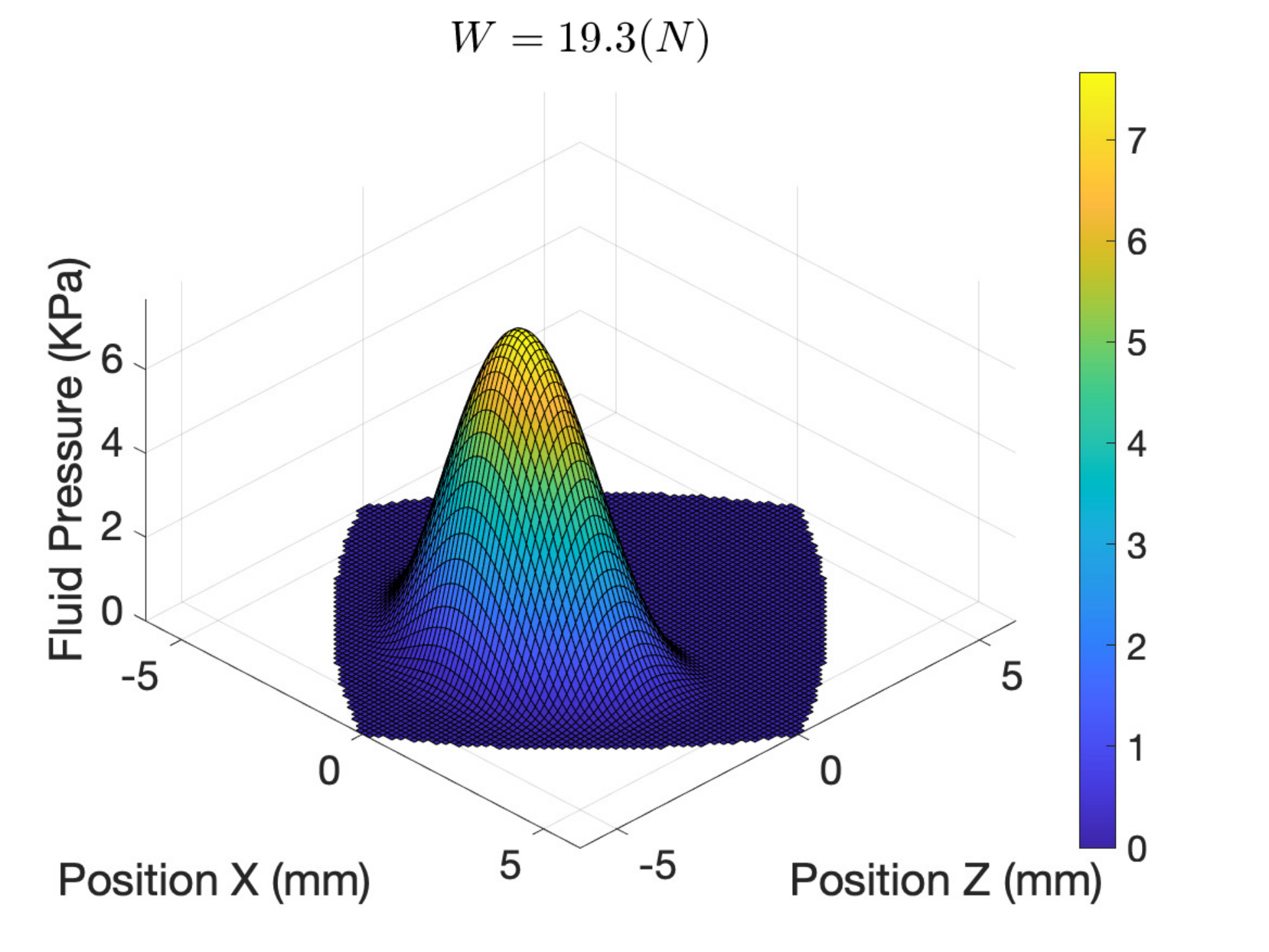}\label{fig:9cf}}
  \hfill
  \subfloat[]{\includegraphics[width=0.33\textwidth]{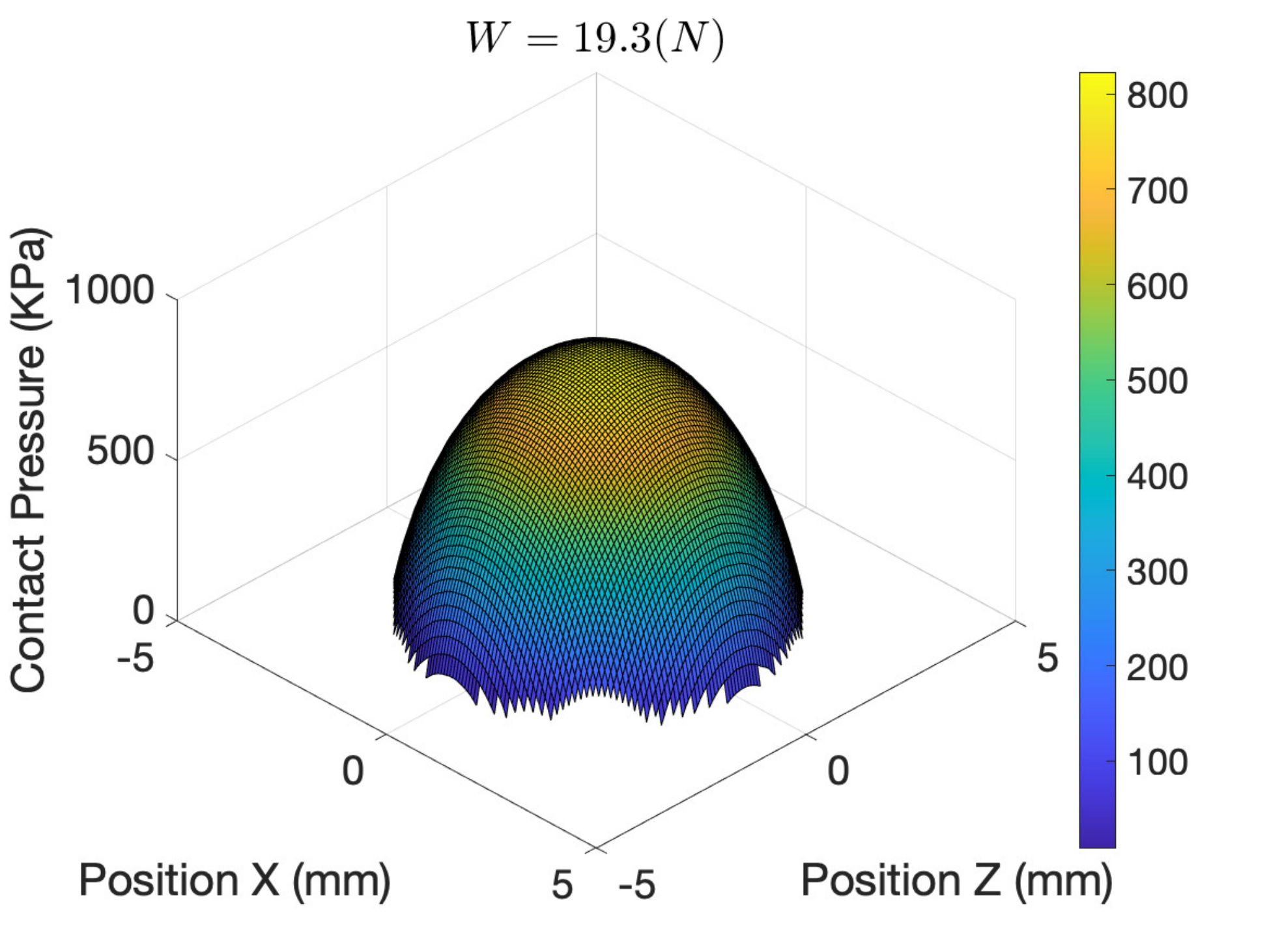}\label{fig:9w9cf}}
  \hfill
  \caption{Ball-on-disk tribometer: (a) Maps of contact gap $\gap$ in $mm$ (position in $mm$). (b) Contour of fluid pressure. (c) Contour of contact pressure. All figures correspond to the lubricant OM 10 in $U\vis = 0.001 N/m$. (Scales are different)}
  \label{fig:mtmcontour}
\end{figure}

\begin{figure}[h!]
\centering
\begin{tabular}{|@{}c@{}c@{}|}
  \hline
  \includegraphics[width=0.48\textwidth]{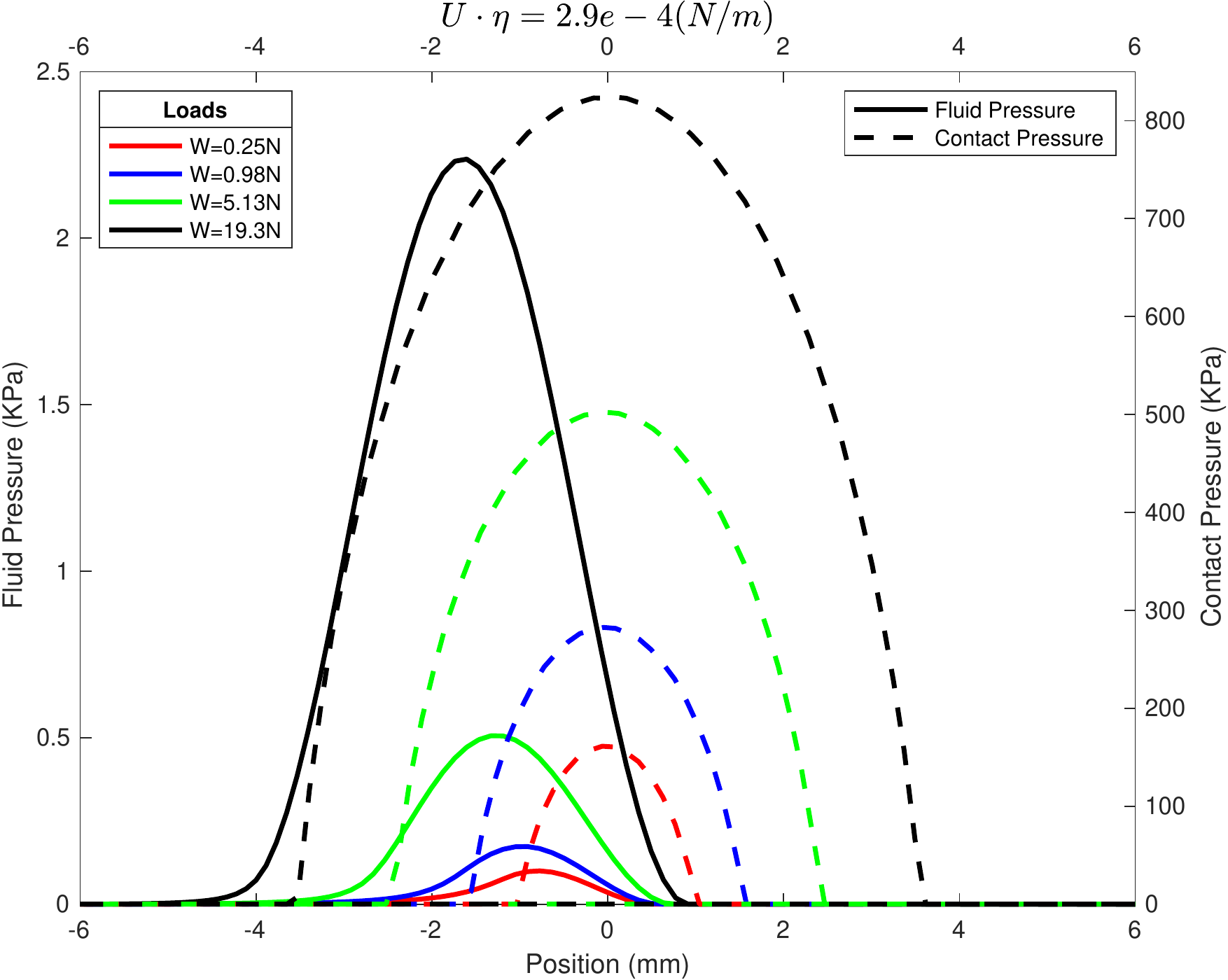}&%
  \includegraphics[width=0.48\textwidth]{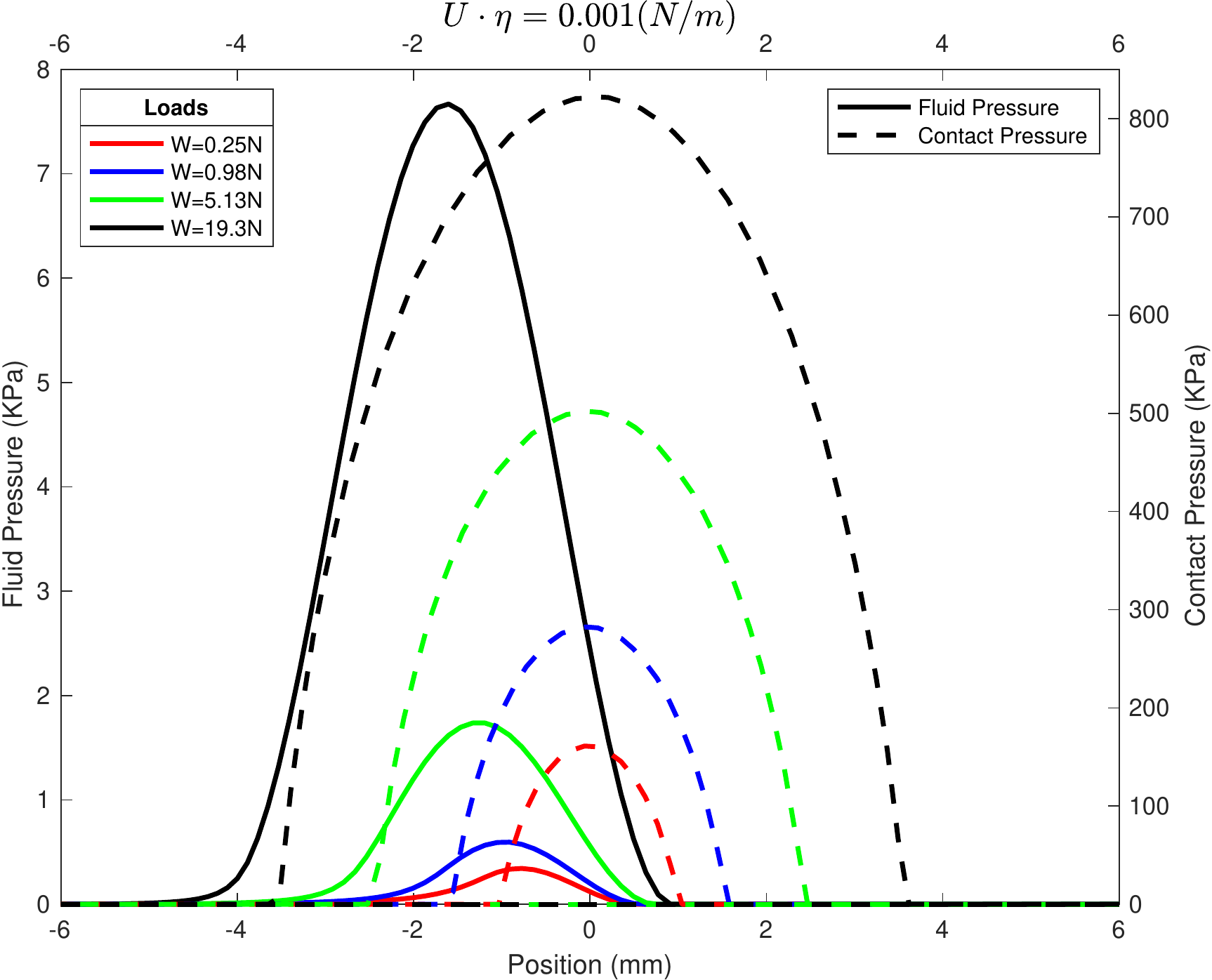}\\
  \hline
\end{tabular}
\caption{Ball-on-disk tribometer: Profile of fluid pressure and contact pressure in the symmetry plane $x=0$ corresponding to different entrainment velocities}
  \label{fig:mtmpp}
\end{figure}

\begin{figure}[h!]
  \centering
  \includegraphics[scale=0.8]{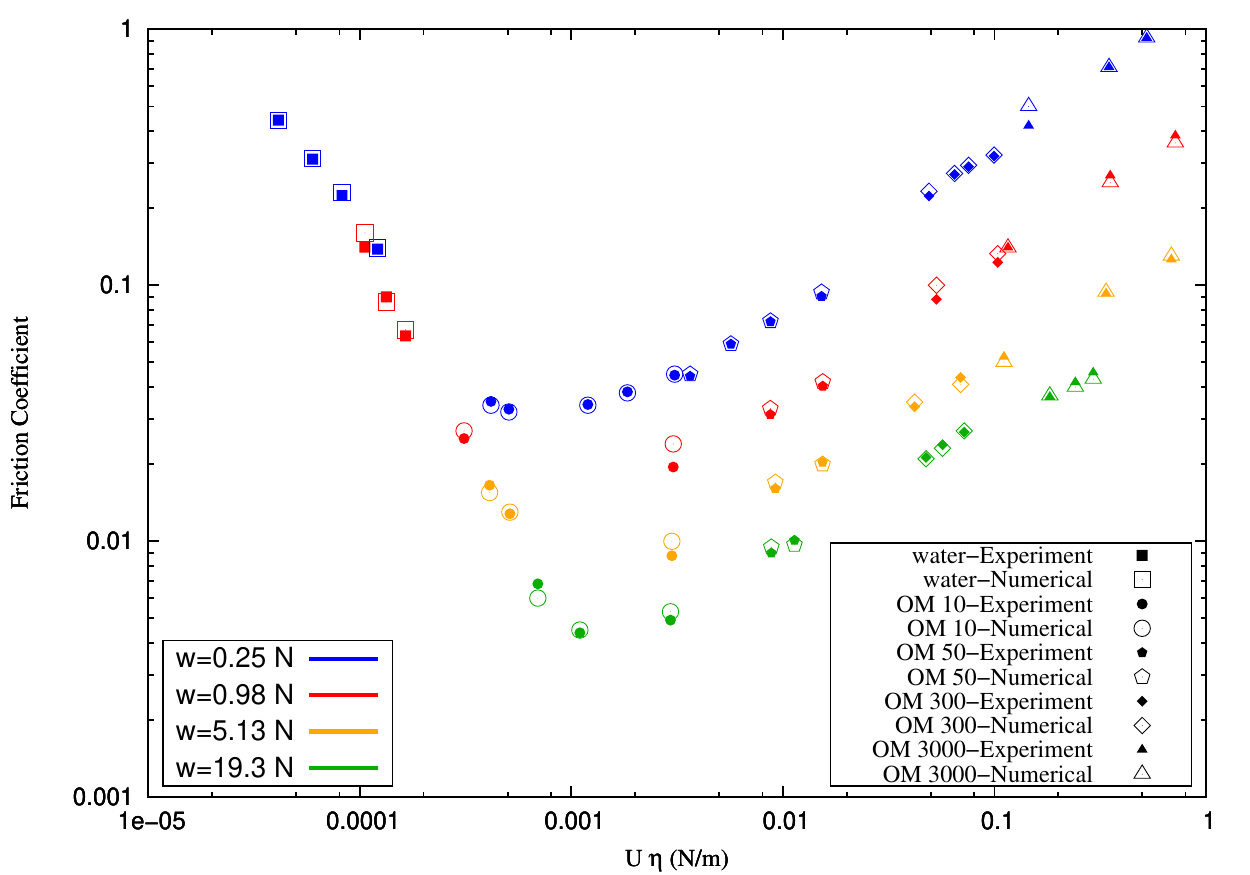}
  \caption{Ball-on-disk tribometer: Friction coefficient as a function of the product of entrainment velocity $U$ and viscosity $\vis$ for five lubricants and a range of loads $W$}
  \label{fig:validation}
\end{figure}

Fig.~\ref{fig:mtmcontour} shows the maps of contact gap $\gap$ for selected values of load $W$ along with the fluid pressure and contact pressure contours, all for a value of $0.001 N/m$ for the product of entrainment velocity and viscosity $U\vis$, where the entrainment velocity is equal to one half of the sliding velocity, $U=V/2$. It can be seen that the contact gap grows with increasing normal load $W$. Increasing normal load $W$ also goes along with raise of the fluid pressure and the contact pressure.
Fluid pressure and contact pressure profiles along the symmetry plane $x=0$ are shown in Fig.~\ref{fig:mtmpp} for different values of load $W$ and also two different values for the product of entrainment velocity and viscosity $U\vis$. It can be seen that the contact pressure and the fluid pressure increase when raising the normal load $W$. It is also clear that the fluid pressure raises with increasing $U\vis$, however, as both values of $U\vis$ correspond to the mixed lubrication regime, the contact pressure is the dominant pressure in both cases.
Fig.~\ref{fig:validation} shows a log-log plot of the calculated friction coefficient as a function of the product of viscosity $\vis$ and entrainment velocity $U$. It should be noted that the friction coefficient is calculated as the ratio of the measured horizontal forces and the vertical force $W$ at the top of the ball.
Here, the prediction of the presented fully nonlinear model is compared to experimental results presented in \cite{stup_2016}.
The results corresponding to a fixed load are denoted by markers of the same colour, and the results corresponding to a fixed lubricant are denoted by markers of the same shape, see the legend in Fig.~\ref{fig:validation}.
It can be seen that the results corresponding to each load form a part of the classical Stribeck curve with a transition between the results corresponding to the lubricants of different viscosity. At high values of $U\vis$, the dependence of the friction coefficient on $U\vis$ appears approximately linear on the log-log plot which indicates that the contact operates in the hydrodynamic lubrication regime. At low values of $U\vis$, the contact operates in the mixed lubrication regime and the friction coefficient increases with decreasing $U\vis$. Considering the dependence on the load, it is apparent that the friction coefficient decreases with increasing load in the whole range of the examined values of $U\vis$. The numerical and experimental results are in very good agreement during the full range of lubrication regimes which validates the ability of the developed framework to model the continuous transition from mixed to full film lubrication.
Indeed, the dependence of the friction coefficient on $U\vis$ (i.e., the slope on the log-log plot) and the dependence on the load are in a good agreement.
\subsubsection{Sensitivity of the formulation for variations of the regularized contact parameters}
\begin{figure}[h!]
  \centering
  \subfloat[]{\includegraphics[width=0.5\textwidth]{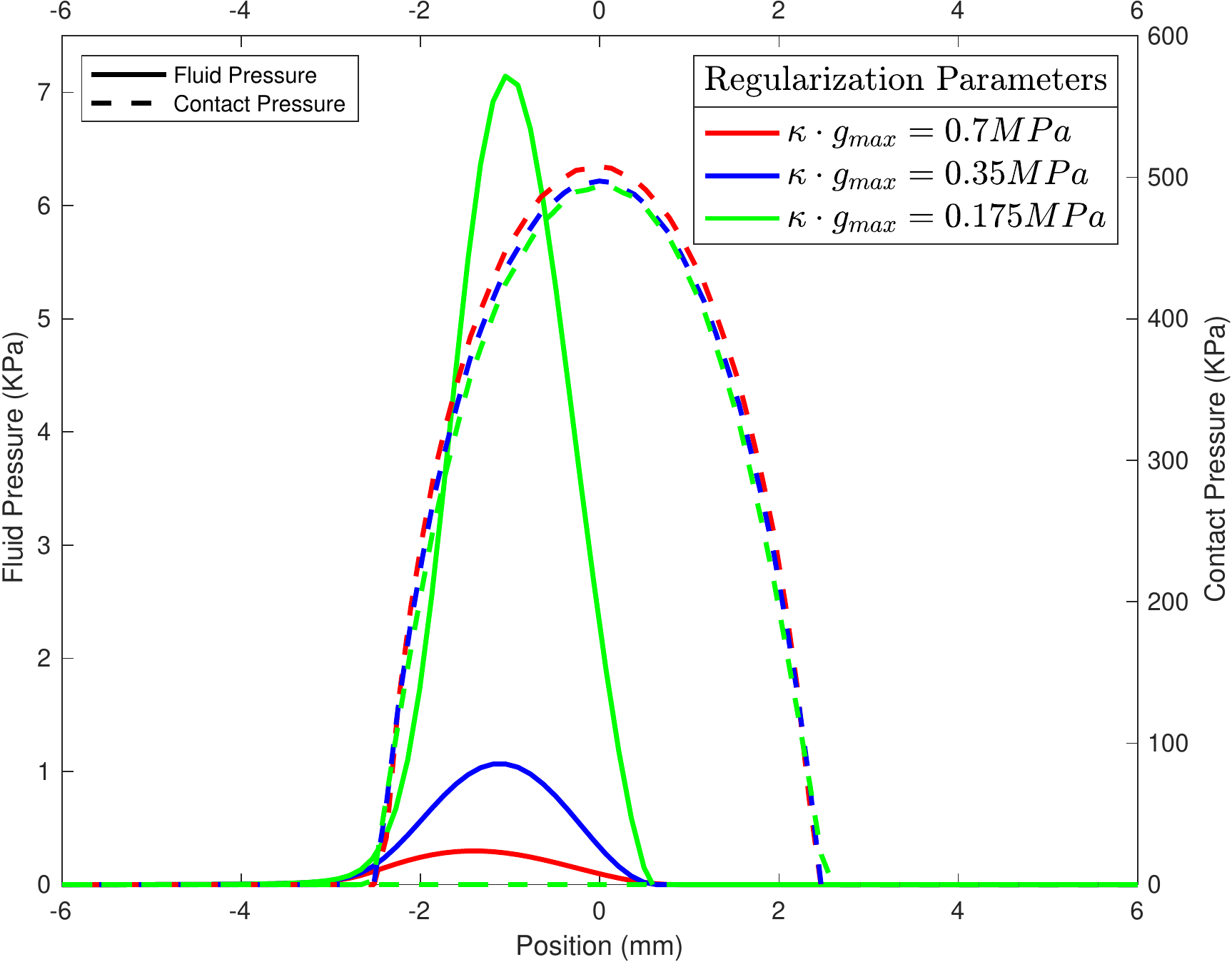}\label{fig:lefthard}}
  \hfill
  \subfloat[]{\includegraphics[width=0.5\textwidth]{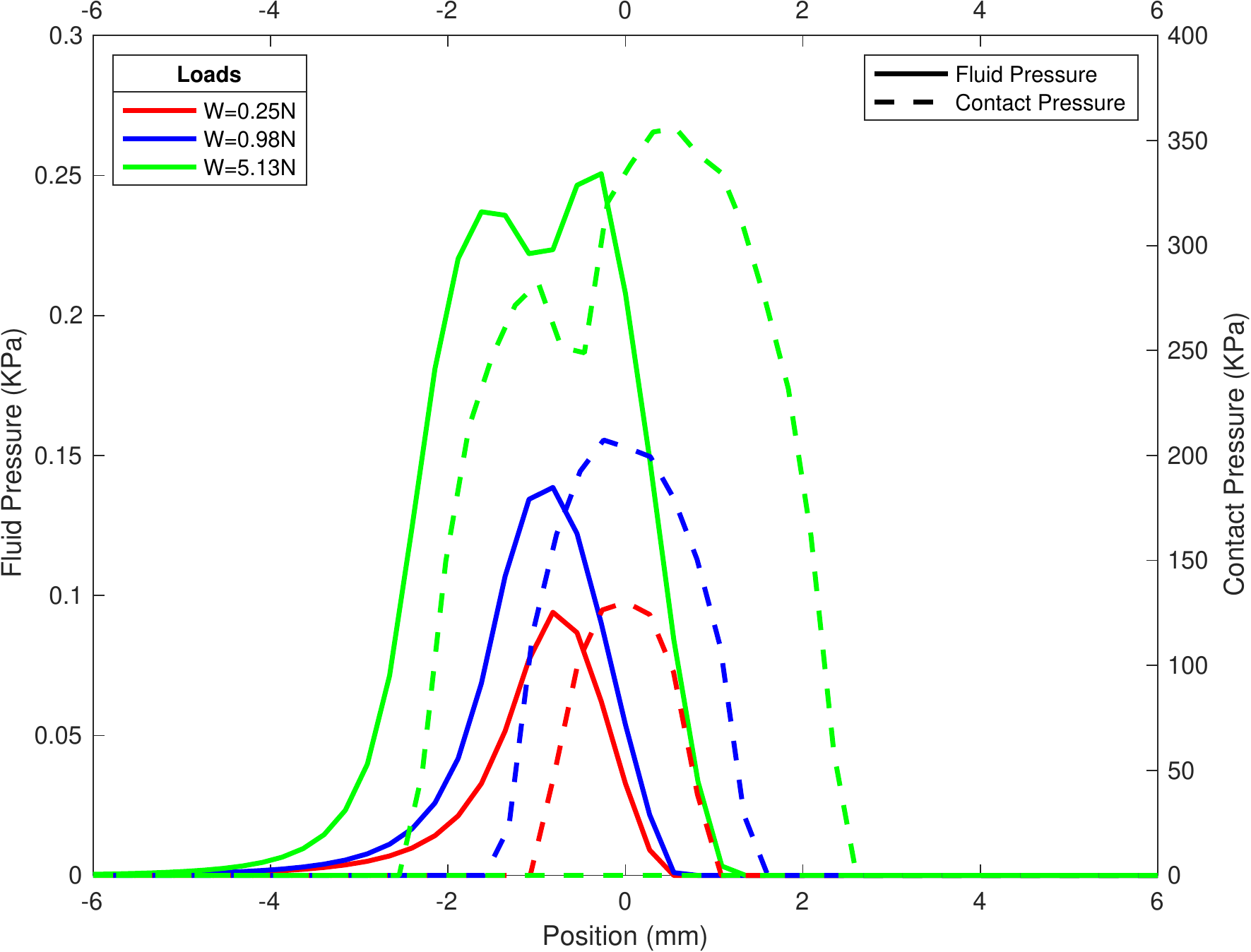}\label{fig:rightsoft}}
  \hfill
  \caption{Ball-on-disk tribometer: Profile of fluid pressure and contact pressure in the symmetry plane $x=0$ corresponding to $U\vis = 2.9e-4 N/m$. (a) Effect of regularization parameters $\regstiff \cdot \regthick$. (b) Two deformable bodies}
  \label{fig:mtmthick}
\end{figure}

In the following, the sensitivity of the formulation for variations of the regularized contact parameters on the lubricated contact model is analyzed. In Section~\ref{model}, regularization thickness $\regthick$ and regularization stiffness $\regstiff$ were introduced. It is seen in Fig.~\ref{fig:regcontact} that the regularization thickness is added to the film thickness in order to keep the lubrication problem resolvable while the asperity contact occurs and the gap closes. A discussion on the choice of these parameters value is presented here. For all computations in this section, the values in Table~\ref{table:2} are used and only specified parameters values are changed according to Table~\ref{table:3}.
\begin{table}[h!]
\centering
\caption{Ball-on-disk tribometer: geometrical, material and process parameters for sensitivity analysis.}
\vspace*{0.2cm}
\setlength{\tabcolsep}{30pt} 
\begin{tabular}{ l c }
\hline
 Lubricant viscosity at 25 degree Celsius, $\vis$: OM 10 & 0.00942 Pa s \\
 Disk sliding velocity, $V$ & 62 mm/s \\
 $\regstiff \cdot \regthick$ & 0.175-0.35-0.5 MPa \\
\hline    
\end{tabular}
\label{table:3}
\end{table}
The viscosity and disk sliding velocity are chosen in a way to represent the mixed lubrication regime, which is relevant to the novelty of this study. The normal load is also selected as 5.13 N, in order to investigate the case with higher asperity contact pressure. Three values for the product of regularization stiffness $\regstiff$ and $\regthick$ are chosen, representing different cases. The lowest value 0.175MPa is equal to 5 percent of ball Young's modulus which corresponds to the case where the surface asperities density is lower, meaning that the asperities are less stiff and consequently are flattened faster. On the other hand, the highest value 0.5MPa is equal to 20 percent of ball Young's modulus which resembles the case where the surface asperities density is higher and surface asperities resist against the deformation. The latter case correlates to the surface with higher roughness.\\
Fig.~\ref{fig:lefthard} illustrates the effect of regularization parameters, $\regstiff \cdot \regthick$ on fluid and contact pressure in the symmetry plane $x=0$.
It can be observed that contact pressure increases smoothly by adding to $\regstiff \cdot \regthick$. From the physical point of view, it completely makes sense that the contact pressure gets higher for the surface with higher roughness. It should also be noted that the increase in the contact pressure is limited, since the study is carried on in low velocity corresponding to mixed lubrication. In the mixed lubrication regime, the asperity contact pressure is already dominant and large. On the other hand, lower values of $\regstiff \cdot \regthick$ represent a smoother surface which can be seen in the higher fluid pressure values obtained. This parameter study reveals that the recommended value of regularization parameters in Section~\ref{model} is reasonable and ensures an accurate and robust solution of the lubrication problem when the contact gap tends to zero.
\subsubsection{Study of two deformable bodies}
Finally, the suitability of our model to represent the lubricated contact interaction of two deformable bodies is demonstrated, by modeling the disk as a deformable body as well. The study is carried out at normal load levels $W$ equal to 0.25 N, 0.98 N and 5.13 N which are applied via prescribed DBC similar to the previous rigid disk study. The geometrical, material and process parameters for computations in this Section are presented in Table~\ref{table:4}. The hyperelastic behaviour of the disk is also governed by a neo-Hookean material model in this Section. The lubricant viscosity and disk sliding velocity are chosen in a way to correspond to the mixed lubrication regime. 
\begin{table}[h!]
\centering
\caption{Ball-on-disk tribometer: geometrical, material and process parameters for study of two deformable bodies.}
\setlength{\tabcolsep}{30pt} 
\begin{tabular}{ l c }
\hline
 Ball radius & 10.7 mm \\
 Disk radius & 53 mm \\
 Cavitation penalty parameter & $10^8$ s/mm \\
 Ball material-Young's modulus of Rubber (NBR) & 3.5 MPa \\
 Disk material-Young's modulus & 4.5 MPa \\ 
 Poisson's ratio & 0 \\
 Lubricant viscosity at 25 degree Celsius, $\vis$: OM 10 & 0.00942 Pa s \\
 Disk sliding velocity, $V$ & 62 mm/s \\
 $\regthick$ & 4 $\mu m$ \\
 regularization stiffness, $\regstiff$ & 1e2 MPa/mm \\
 Ball surface roughness standard deviation, $\stddev$ & 1.33 $\mu m$ \\
\hline    
\end{tabular}
\label{table:4}
\end{table}
The elastic disk surface is defined as the master surface and the bottom surface of the ball is defined as the slave surface. The definition of the lubrication domain and its finite element mesh and boundary condition are carried out identical to the setup in study of the rigid disk before. The finite element mesh for this computation comprises 100,000 elements, and a total of 300,000 unknowns including nodal displacements of the solid body and pressure degrees of freedom on the lubricated contact surface.
\begin{figure}[h!]
\renewcommand{\arraystretch}{0}
  \centering
  \subfloat{\includegraphics[width=0.6\textwidth]{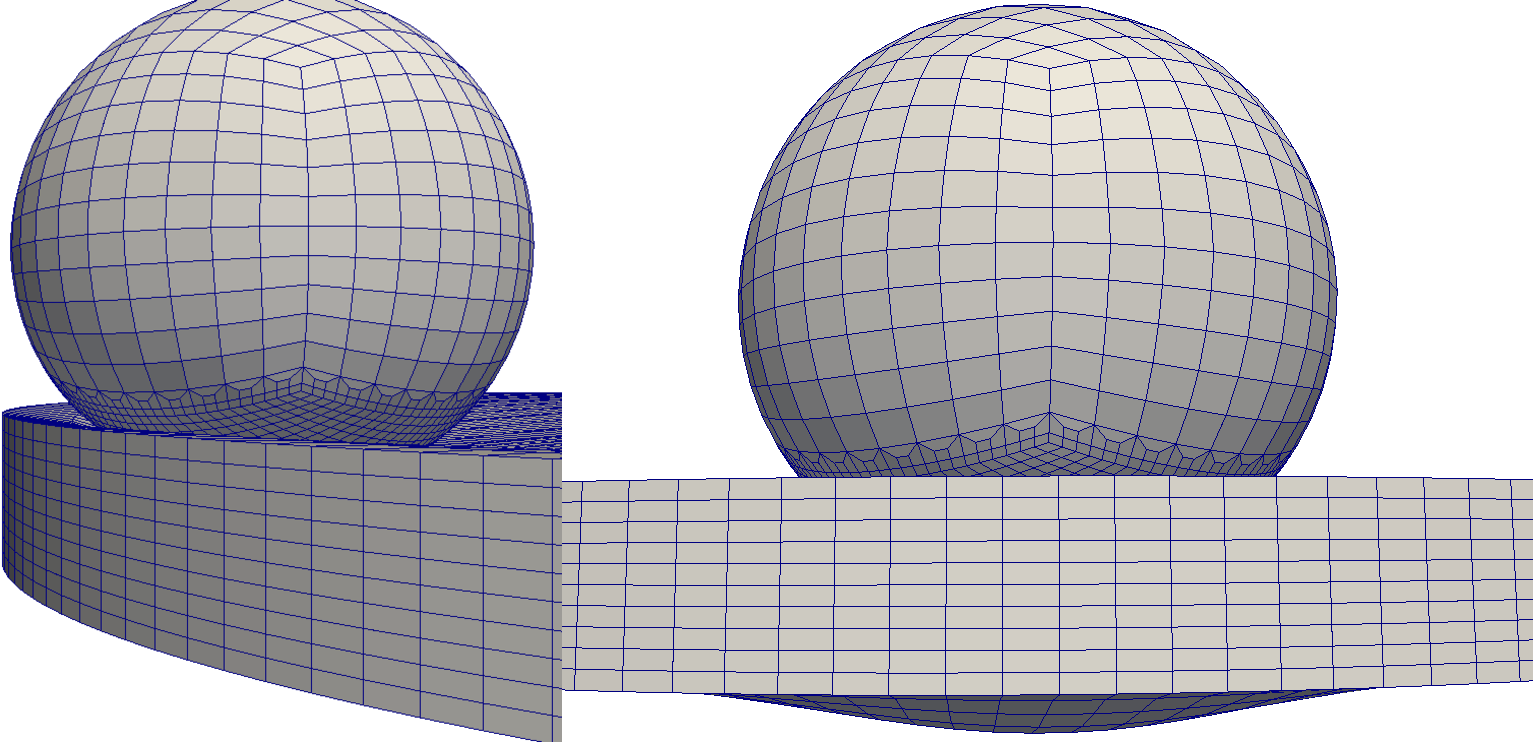}}
  \hfill
  \caption{Ball-on disk tribometer: two deformable bodies after deformation from two point of views}
  \label{fig:mtmmeshsoft}
\end{figure}
\begin{figure}[h!]
\renewcommand{\arraystretch}{0}
  \centering
  \subfloat[]{\includegraphics[width=0.33\textwidth]{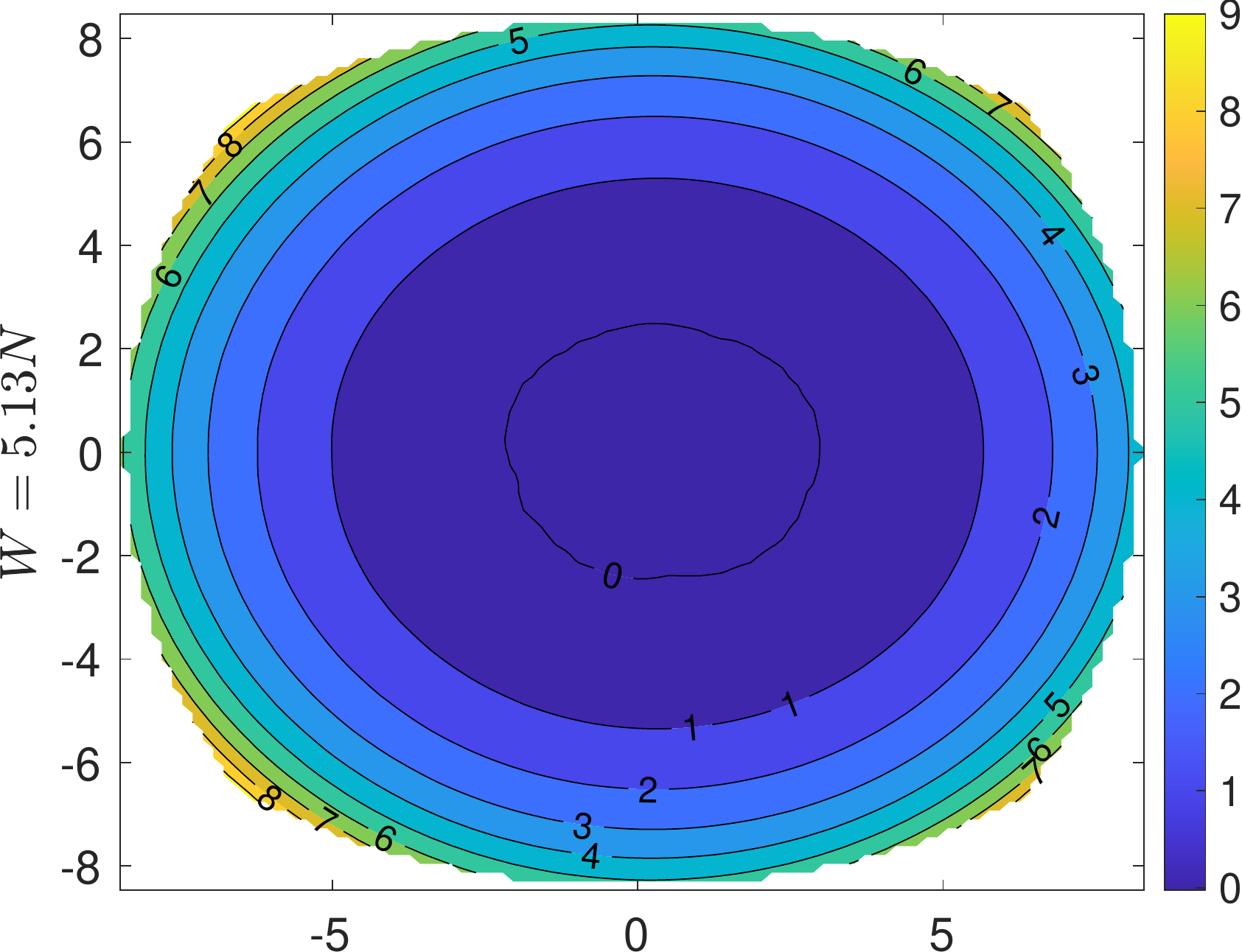}\label{fig:7cf}}
  \hfill
  \subfloat[]{\includegraphics[width=0.33\textwidth]{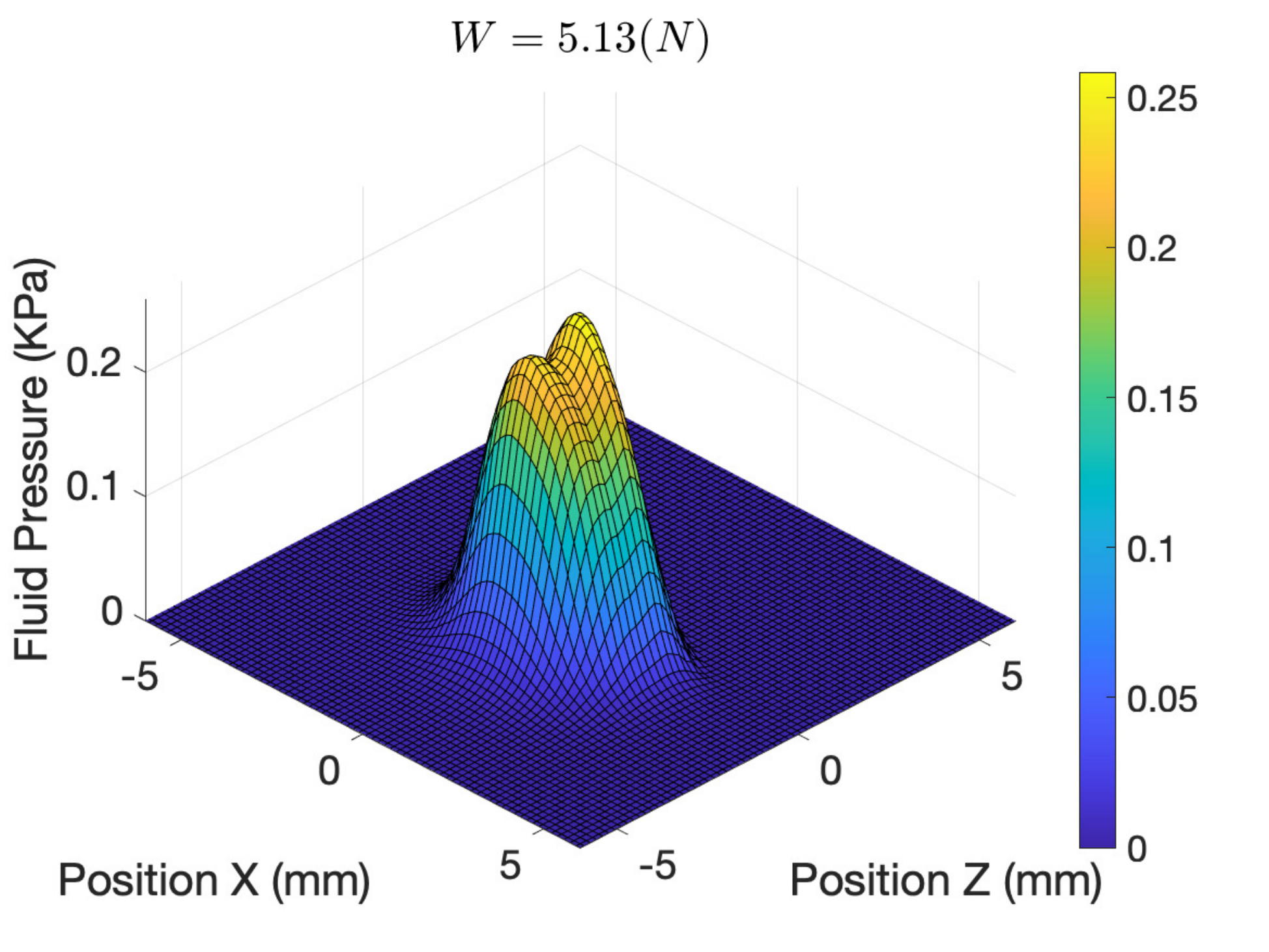}\label{fig:9cf}}
  \hfill
  \subfloat[]{\includegraphics[width=0.33\textwidth]{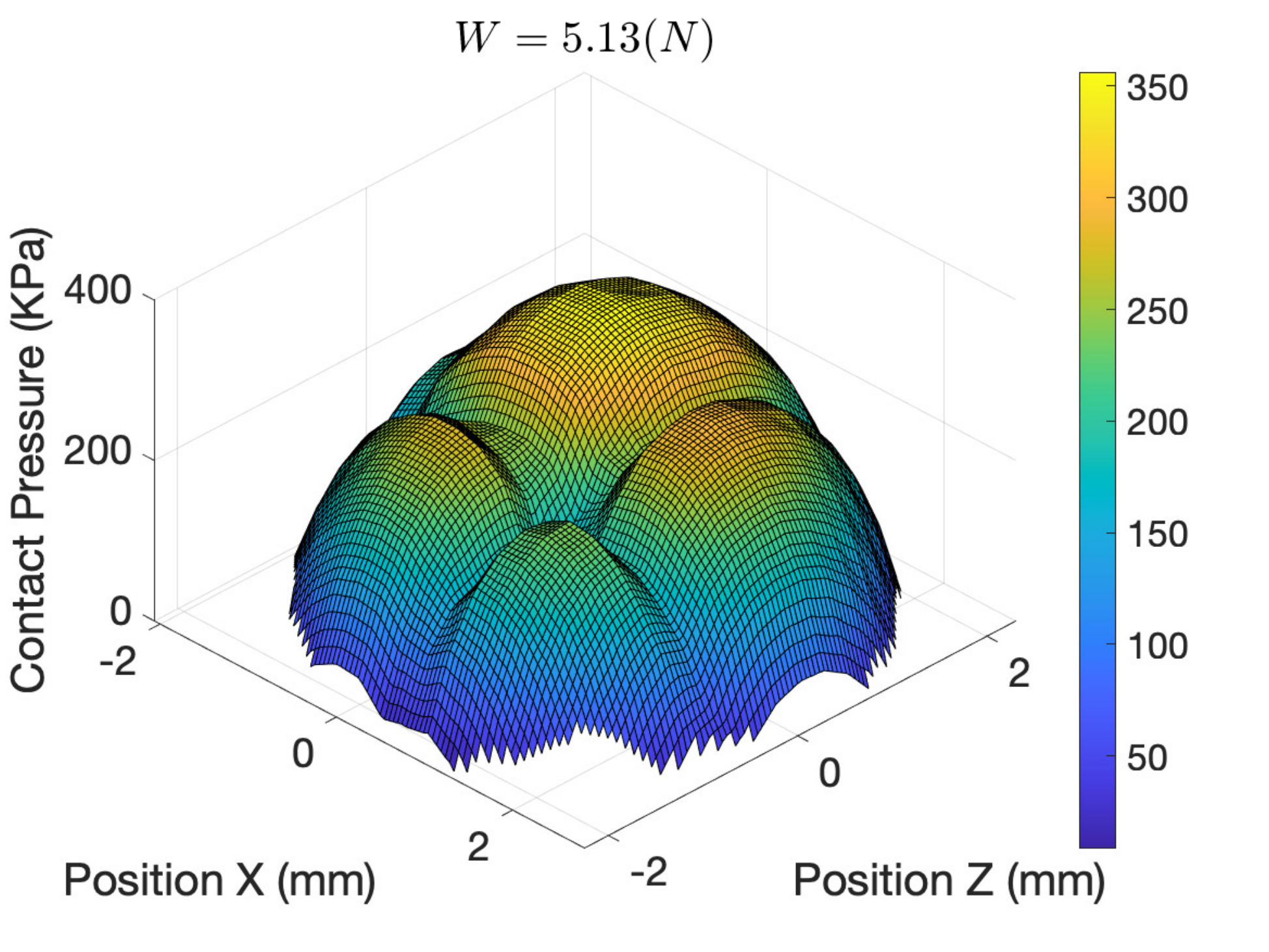}\label{fig:9w9cf}}
  \hfill
  \caption{Ball-on-disk tribometer:(a) Maps of contact gap $\gap$ in $mm$ (position in $mm$). (b) Contour of fluid pressure. (c) Contour of contact pressure. All figures corespond to $U\vis = 2.9e-4 N/m$ (Scales are different)}
  \label{fig:mtmcontoursoft}
\end{figure}
The finite element mesh of two deformable bodies after deformation is shown in Fig.~\ref{fig:mtmmeshsoft}. Fig.~\ref{fig:mtmcontoursoft} shows the maps of contact gap $\gap$ for selected value of load $W$ along with the fluid pressure and the contact pressure contours for the case of $2.9e-4 N/m$ for product of entrainment velocity and viscosity $U\vis$. It can be seen that the pressure contours have more maximum points, which is as a result of large deformation in this case, similar to the first example (Cylinder on a rigid, flat surface).
Fluid pressure and contact pressure profiles along the symmetry plane $x=0$ are shown in Fig.~\ref{fig:rightsoft} for different values of load $W$ and the product of entrainment velocity and viscosity $U\vis=2.9e-4$. It can be seen that the contact pressure and the fluid pressure increase when raising the normal load $W$. The presence of more maximum points as a result of large deformation also is clear in this graph for the contact pressure profile.
\section{Summary / Conclusion}\label{conclusion}
In this paper we propose a novel modeling approach to solve lubricated contact problem across the full range of lubrication regimes. Critically, the model relies on a recently proposed regularization scheme for the mechanical contact constraint combining the advantages of classical penalty and Lagrange multiplier approaches by expressing the mechanical contact pressure as a function of the effective gap between the solid bodies while at the same time limiting the minimal gap value occurring at the (theoretical) limit of infinitely high contact pressures. From a methodological point of view, this is the key ingredient to regularize the pressure field in the averaged Reynolds equation, i.e., to avoid the pressure field's singularity in the limit of vanishing fluid film thickness, and thus to enable a smooth transition between all relevant lubrication regimes. From a physical point of view, this approach can be considered as a model for the elastic deformation of surface asperities, with a bounded magnitude depending on the interacting solids' surface roughness. To apply the model, this framework combines the Reynolds approximation of the thin film fluid equations to describe the fluid behavior on the interface and the equations of finite deformation elastodynamics to govern the solid body behavior without imposing restrictions on the constitutive model. Mortar finite element methods are employed in order to couple the non-matching discretizations at the interface and to allow for an imposition of the fluid film tractions on both surfaces. This coupling is also utilized in order to relate the fluid film thickness and sliding velocity to the displacement field of the solid bodies. Those quantities enter to lubrication equation and lead to a fully-coupled, displacement- and pressure-dependent system of equations. Concerning frictional contact, dual Lagrange Multipliers with regularized contact condition is implemented as the contact constraints which utilize a nonlinear complementary function leading to a semi-smooth Newton method. The resulting non-linear, fully-coupled, 3D equations are solved monolithically with a non-smooth variant of Newton’s method owing to consistent linearization.\\
This model enables the numerical method to allow for a continuous transition throughout changes in the lubrication domain for contacting and non-contacting solid bodies, which grant to investigate entire lubrication regimes. 
Finally, three different numerical examples were presented to show the behavior and capabilities of the presented model. The first was a typical configuration to analyze the behavior of the lubricated contact model in large deformations. The second configuration focused on the contacting and lift-off of elastic pin on rigid plane configuration in order to show the applicability of the model to full range of lubrication regimes. In the final numerical example, to validate the proposed monolithic formulation of the model, we studied ball-on-disk Tribometer and compared the results with experimental work. A sensitivity study on the included regularized contact constraints parameters allowed for a specification of a proper range for these parameters.
\paragraph{Acknowledgment}
This work was supported by the European Education, Audiovisual and Culture Executive Agency (EACEA) under the Erasmus Mundus Joint Doctorate Simulation in Engineering and Entrepreneurship Development (SEED), FPA 2013-0043.
\bibliographystyle{unsrtnat}
\bibliography{mix}

\begin{thebibliography}{55}
\providecommand{\natexlab}[1]{#1}
\providecommand{\url}[1]{\texttt{#1}}
\expandafter\ifx\csname urlstyle\endcsname\relax
  \providecommand{\doi}[1]{doi: #1}\else
  \providecommand{\doi}{doi: \begingroup \urlstyle{rm}\Url}\fi

\bibitem[Ager et~al.(2020)Ager, Seitz, and Wall]{ager_seitz_2019}
C.~Ager, A.~Seitz, and W.A. Wall.
\newblock A consistent and comprehensive computational approach for general
  fluid-structure-contact interaction problems.
\newblock \emph{International Journal for Numerical Methods in Engineering},
  122\penalty0 (19):\penalty0 5279--5312, 2020.

\bibitem[Ager et~al.(2019)Ager, Schott, Vuong, Popp, and Wall]{ager_porous}
C.~Ager, B.~Schott, A.~Vuong, A.~Popp, and W.A. Wall.
\newblock A consistent approach for fluid-structure-contact interaction based
  on a porous flow model for rough surface contact.
\newblock \emph{International Journal for Numerical Methods in Engineering},
  119\penalty0 (13):\penalty0 1345--1378, 2019.

\bibitem[Dowson and Higginson(1977)]{dowson_higginson_1977}
D.~Dowson and G.R. Higginson.
\newblock \emph{Elasto-hydrodynamic lubrication}.
\newblock Pergamon, 1977.

\bibitem[Hamrock(1994)]{hamrock_1994}
B.J. Hamrock.
\newblock \emph{Fundamentals of fluid film lubrication}.
\newblock McGraw-Hill, 1994.

\bibitem[Dowson(1995)]{dowson_1995}
D.~Dowson.
\newblock Elastohydrodynamic and micro-elastohydrodynamic lubrication.
\newblock \emph{Wear}, 190\penalty0 (2):\penalty0 125--138, 1995.

\bibitem[de~Vicente et~al.(2005)de~Vicente, Stokes, and Spikes]{vicente}
J.~de~Vicente, J.R. Stokes, and H.A. Spikes.
\newblock The frictional properties of newtonian fluids in rolling-sliding
  soft-ehl contact.
\newblock \emph{Tribology Letters}, 20\penalty0 (3-4):\penalty0 273--286, 2005.

\bibitem[Adams et~al.(2007)Adams, Briscoe, and Johnson]{adams_2007}
M.J. Adams, B.J. Briscoe, and S.A. Johnson.
\newblock Friction and lubrication of human skin.
\newblock \emph{Tribology Letters}, 26\penalty0 (3):\penalty0 239--253, 2007.

\bibitem[Jones et~al.(2007)Jones, Fulford, Please, McElwain, and
  Collins]{jones_2007}
M.B. Jones, G.R. Fulford, C.P. Please, D.L.S. McElwain, and M.J. Collins.
\newblock Elastohydrodynamics of the eyelid wiper.
\newblock \emph{Bulletin of Mathematical Biology}, 70\penalty0 (2):\penalty0
  323--343, 2007.

\bibitem[Crook(1963)]{crook_1963}
A.W. Crook.
\newblock The lubrication of rollers iv. measurements of friction and effective
  viscosity.
\newblock \emph{Philosophical Transactions of the Royal Society A:
  Mathematical, Physical and Engineering Sciences}, 255\penalty0
  (1056):\penalty0 281--312, 1963.

\bibitem[Crouch and Cameron(1961)]{crouch}
R.F. Crouch and A.~Cameron.
\newblock Viscosity-temperature equation for lubricants.
\newblock \emph{Journal of the Institue of Petroluem}, 47:\penalty0 307--313,
  1961.

\bibitem[Patir and Cheng(1978)]{patir1}
N.~Patir and H.S. Cheng.
\newblock An average flow model for determining effects of three-dimensional
  roughness on partial hydrodynamic lubrication.
\newblock \emph{Journal of Lubrication Technology}, 100\penalty0 (1):\penalty0
  12--17, 1978.

\bibitem[Patir and Cheng(1979)]{patir2}
N.~Patir and H.S. Cheng.
\newblock Application of average flow model to lubrication between rough
  sliding surfaces.
\newblock \emph{Journal of Lubrication Technology}, 101\penalty0 (2):\penalty0
  220--229, 1979.

\bibitem[Houpert and Hamrock(1986)]{houpert1986}
L.G. Houpert and B.J. Hamrock.
\newblock Fast approach for calculating film thicknesses and pressures in
  elastohydrodynamically lubricated contacts at high loads.
\newblock \emph{Journal of Tribology}, 108\penalty0 (3):\penalty0 411--419,
  1986.

\bibitem[Ai and Cheng(1994)]{ai}
A.X. Ai and H.S. Cheng.
\newblock Numerical simulation of elastohydrodynamically lubricated contacts
  with rough surfaces.
\newblock \emph{Applied Mechanics Reviews}, 47:\penalty0 221--227, 1994.

\bibitem[Jiang et~al.(1999)Jiang, Hua, Cheng, Ai, and Lee]{jiang1999}
X.~Jiang, D.Y. Hua, H.S. Cheng, X.~Ai, and S.C. Lee.
\newblock A mixed elastohydrodynamic lubrication model with asperity contact.
\newblock \emph{Journal of Tribology}, 121\penalty0 (3):\penalty0 481--491,
  1999.

\bibitem[Hu and Zhu(1999)]{huzhu}
Y.Z. Hu and D.~Zhu.
\newblock A full numerical solution to the mixed lubrication in point contacts.
\newblock \emph{Journal of Tribology}, 122\penalty0 (1):\penalty0 1--9, 1999.

\bibitem[Evans and Hughes(2000)]{evans2000}
H.P. Evans and T.G. Hughes.
\newblock Evaluation of deflection in semi-infinite bodies by a differential
  method.
\newblock \emph{Proceedings of the Institution of Mechanical Engineers, Part C:
  Journal of Mechanical Engineering Science}, 214\penalty0 (4):\penalty0
  563--584, 2000.

\bibitem[Azam et~al.(2019)Azam, Ghanbarzadeh, Neville, Morina, and
  Wilson]{ghanbar2019}
A.~Azam, A.~Ghanbarzadeh, A.~Neville, A.~Morina, and M.C.T. Wilson.
\newblock Modelling tribochemistry in the mixed lubrication regime.
\newblock \emph{Tribology International}, 132:\penalty0 265--274, 2019.

\bibitem[Nikas(2002)]{nikas_2002}
G.K. Nikas.
\newblock Elastohydrodynamics and mechanics of rectangular elastomeric seals
  for reciprocating piston rods.
\newblock \emph{Journal of Tribology}, 125\penalty0 (1):\penalty0 60--69, 2002.

\bibitem[Stupkiewicz(2009)]{stup2009}
S.~Stupkiewicz.
\newblock Finite element treatment of soft elastohydrodynamic lubrication
  problems in the finite deformation regime.
\newblock \emph{Computational Mechanics}, 44\penalty0 (5):\penalty0 605--619,
  2009.

\bibitem[O\"ngu\"n et~al.(2008)O\"ngu\"n, Andre, Bartel, and Deters]{ongun}
Y.~O\"ngu\"n, M.~Andre, D.~Bartel, and L.~Deters.
\newblock An axisymmetric hydrodynamic interface element for finite-element
  computations of mixed lubrication in rubber seals.
\newblock \emph{J. Engineering Tribology}, 222\penalty0 (3):\penalty0 471--481,
  2008.

\bibitem[Schmidt et~al.(2010)Schmidt, Andre, and Poll]{schmidt_2010}
T.~Schmidt, M.~Andre, and G.~Poll.
\newblock A transient 2d-finite-element approach for the simulation of mixed
  lubrication effects of reciprocating hydraulic rod seals.
\newblock \emph{Tribology International}, 43\penalty0 (10):\penalty0
  1775--1785, 2010.

\bibitem[Yang and Laursen(2009)]{yang2009}
B.~Yang and T.A. Laursen.
\newblock A mortar-finite element approach to lubricated contact problems.
\newblock \emph{Computer Methods in Applied Mechanics and Engineering},
  198\penalty0 (47-48):\penalty0 3656--3669, 2009.

\bibitem[Shvarts et~al.(2019)Shvarts, Vignollet, and Yastrebov]{andrei_2019}
A.G. Shvarts, J.~Vignollet, and V.A. Yastrebov.
\newblock Computational framework for monolithic coupling for thin fluid flow
  in contact interfaces.
\newblock \emph{Computer Methods in Applied Mechanics and Engineering}, 379,
  2019.

\bibitem[Sitzmann et~al.(2014)Sitzmann, Willner, and Wohlmuth]{sitzmann2014}
S.~Sitzmann, K.~Willner, and B.I. Wohlmuth.
\newblock A dual lagrange method for contact problems with regularized contact
  conditions.
\newblock \emph{International Journal for Numerical Methods in Engineering},
  99\penalty0 (3):\penalty0 221--238, 2014.

\bibitem[Sitzmann et~al.(2015)Sitzmann, Willner, and Wohlmuth]{sitzmann_2015}
S.~Sitzmann, K.~Willner, and B.I. Wohlmuth.
\newblock A dual lagrange method for contact problems with regularized
  frictional contact conditions: Modelling micro slip.
\newblock \emph{Computer Methods in Applied Mechanics and Engineering},
  285:\penalty0 468--487, 2015.

\bibitem[Cryer(1971)]{cryer1971}
C.W. Cryer.
\newblock The method of christopherson for solving free boundary problems for
  infinite journal bearings by means of finite differences.
\newblock \emph{Mathematics of Computation}, 25\penalty0 (115):\penalty0
  435--435, 1971.

\bibitem[Rohde and McAllister(1975)]{rohde_mcallister_1975}
S.M. Rohde and G.T. McAllister.
\newblock A variational formulation for a class of free boundary problems
  arising in hydrodynamic lubrication.
\newblock \emph{International Journal of Engineering Science}, 13\penalty0
  (9-10):\penalty0 841--850, 1975.

\bibitem[Wu(1986)]{wu_1986}
S.R. Wu.
\newblock A penalty formulation and numerical approximation of the
  reynolds-hertz problem of elastohydrodynamic lubrication.
\newblock \emph{International Journal of Engineering Science}, 24\penalty0
  (6):\penalty0 1001--1013, 1986.

\bibitem[Laursen(2002)]{laursen_2002}
T.A. Laursen.
\newblock \emph{Computational contact and impact mechanics}.
\newblock Springer, 2002.

\bibitem[Wriggers and Laursen(2006)]{wriggers_2006}
P.~Wriggers and T.A. Laursen.
\newblock \emph{Computational Contact Mechanics}.
\newblock Springer, 2006.

\bibitem[Puso et~al.(2008)Puso, Laursen, and Solberg]{puso_2008}
M.A. Puso, T.A. Laursen, and J.~Solberg.
\newblock A segment-to-segment mortar contact method for quadratic elements and
  large deformations.
\newblock \emph{Computer Methods in Applied Mechanics and Engineering},
  197\penalty0 (6-8):\penalty0 555--566, 2008.

\bibitem[H\"ueber et~al.(2008)H\"ueber, Stadler, and Wohlmuth]{hueber_2008}
S.~H\"ueber, G.~Stadler, and B.I. Wohlmuth.
\newblock A primal-dual active set algorithm for three-dimensional contact
  problems with coulomb friction.
\newblock \emph{SIAM Journal on Scientific Computing}, 30\penalty0
  (2):\penalty0 572--596, 2008.

\bibitem[Hesch and Betsch(2009)]{hesch_2009}
C.~Hesch and P.~Betsch.
\newblock A mortar method for energy-momentum conserving schemes in
  frictionless dynamic contact problems.
\newblock \emph{International Journal for Numerical Methods in Engineering},
  77\penalty0 (10):\penalty0 1468--1500, 2009.

\bibitem[Popp et~al.(2009)Popp, Gee, and Wall]{popp2009}
A.~Popp, M.W. Gee, and W.A. Wall.
\newblock A finite deformation mortar contact formulation using a primal-dual
  active set strategy.
\newblock \emph{International Journal for Numerical Methods in Engineering},
  79\penalty0 (11):\penalty0 1354--1391, 2009.

\bibitem[Popp et~al.(2010)Popp, Gitterle, Gee, and Wall]{popp2010}
A.~Popp, M.~Gitterle, M.W. Gee, and W.A. Wall.
\newblock A dual mortar approach for 3d finite deformation contact with
  consistent linearization.
\newblock \emph{International Journal for Numerical Methods in Engineering},
  83\penalty0 (11):\penalty0 1428--1465, 2010.

\bibitem[Gitterle et~al.(2010)Gitterle, Popp, Gee, and Wall]{gitterle2010}
M.~Gitterle, A.~Popp, M.W. Gee, and W.A. Wall.
\newblock Finite deformation frictional mortar contact using a semi-smooth
  newton method with consistent linearization.
\newblock \emph{International Journal for Numerical Methods in Engineering},
  84\penalty0 (5):\penalty0 543--571, 2010.

\bibitem[Fischer and Wriggers(2005)]{fischer_2005}
K.A. Fischer and P.~Wriggers.
\newblock Frictionless 2d contact formulations for finite deformations based on
  the mortar method.
\newblock \emph{Computational Mechanics}, 36\penalty0 (3):\penalty0 226--244,
  2005.

\bibitem[Popova and Popov(2015)]{popov}
E.~Popova and V.L. Popov.
\newblock The research works of coulomb and amontons and generalized laws of
  friction.
\newblock \emph{Friction}, 3:\penalty0 183--190, 2015.

\bibitem[Ozaki et~al.(2020)Ozaki, Matsuura, and Maegawa]{ozaki}
S.~Ozaki, T.~Matsuura, and S.~Maegawa.
\newblock Rate-, state-, and pressure-dependent friction model based on the
  elastoplastic theory.
\newblock \emph{Friction}, 8:\penalty0 768--783, 2020.

\bibitem[Kraus et~al.(2021)Kraus, Lenzen, and Merklein]{KRAUS2021}
M.~Kraus, M~Lenzen, and M.~Merklein.
\newblock Contact pressure-dependent friction characterization by using a
  single sheet metal compression test.
\newblock \emph{Wear}, 476\penalty0 (203679), 2021.

\bibitem[Stupkiewicz and Marciniszyn(2009)]{stupseal2009}
S.~Stupkiewicz and A.~Marciniszyn.
\newblock Elastohydrodynamic lubrication and finite configuration changes in
  reciprocating elastomeric seals.
\newblock \emph{Tribology International}, 42\penalty0 (5):\penalty0 615--627,
  2009.

\bibitem[Stupkiewicz et~al.(2016)Stupkiewicz, Lengiewicz, Sadowski, and
  Kucharski]{stup_2016}
S.~Stupkiewicz, J.~Lengiewicz, P.~Sadowski, and S.~Kucharski.
\newblock Finite deformation effects in soft elastohydrodynamic lubrication
  problems.
\newblock \emph{Tribology International}, 93:\penalty0 511--522, 2016.

\bibitem[Chung and Hulbert(1993)]{chung_1993}
J.~Chung and G.M. Hulbert.
\newblock A time integration algorithm for structural dynamics with improved
  numerical dissipation: The generalized-a method.
\newblock \emph{Journal of Applied Mechanics}, 60\penalty0 (2):\penalty0
  371--375, 1993.

\bibitem[Hughes(2000)]{hughes_2000}
T.J.R. Hughes.
\newblock \emph{The finite element method: linear static and dynamic finite
  element analysis}.
\newblock Dover Publications, 2000.

\bibitem[Zienkiewicz et~al.(2013{\natexlab{a}})Zienkiewicz, Taylor, and
  Fox]{zienkiewicz_fox_2013}
O.C. Zienkiewicz, R.L. Taylor, and D.D. Fox.
\newblock \emph{Finite element method for solid and structural mechanics}.
\newblock Butterworth Heinemann, 2013{\natexlab{a}}.

\bibitem[Zienkiewicz et~al.(2013{\natexlab{b}})Zienkiewicz, Taylor, and
  Zhu]{zienkiewicz_zhu_2013}
O.C. Zienkiewicz, R.L. Taylor, and J.Z. Zhu.
\newblock \emph{The finite element method: Its basis and fundamentals}.
\newblock Butterworth Heinemann, 2013{\natexlab{b}}.

\bibitem[Curnier et~al.(1995)Curnier, He, and Klarbring]{Curnier1995}
A.~Curnier, Q.C. He, and A.~Klarbring.
\newblock Continuum mechanics modelling of large deformation contact with
  friction.
\newblock \emph{Contact Mechanics}, pages 145--158, 1995.

\bibitem[Wohlmuth(2000)]{wohlmuth2000}
B.I. Wohlmuth.
\newblock A mortar finite element method using dual spaces for the lagrange
  multiplier.
\newblock \emph{SIAM Journal on Numerical Analysis}, 38:\penalty0 989--1012,
  2000.

\bibitem[H\"ueber and Wohlmuth(2005)]{hueber_2005}
S.~H\"ueber and B.I. Wohlmuth.
\newblock A primal-dual active set strategy for non-linear multibody contact
  problems.
\newblock \emph{Computer Methods in Applied Mechanics and Engineering},
  194\penalty0 (27-29):\penalty0 3147--3166, 2005.

\bibitem[Flemisch and Wohlmuth(2007)]{wohlmuth2007}
B.~Flemisch and B.I. Wohlmuth.
\newblock Stable lagrange multipliers for quadrilateral meshes of curved
  interfaces in 3d.
\newblock \emph{Computer Methods in Applied Mechanics and Engineering},
  196:\penalty0 1589--1602, 2007.

\bibitem[Farah et~al.(2014)Farah, Popp, and Wall]{farah2014}
P.~Farah, A.~Popp, and W.A. Wall.
\newblock Segment-based vs. element-based integration for mortar methods in
  computational contact mechanics.
\newblock \emph{Computational Mechanics}, 55:\penalty0 209–228, 2014.

\bibitem[Qi and Sun(1993)]{qi_sun_1993}
L.~Qi and J.~Sun.
\newblock A nonsmooth version of newton's method.
\newblock \emph{Mathematical Programming}, 58\penalty0 (1-3):\penalty0
  353--367, 1993.

\bibitem[bac(2021)]{baciweb}
Baci: a comprehensive multi-physics simulation framework, 2021.
\newblock URL \url{https://baci.pages.gitlab.lrz.de/website/}.

\bibitem[Hamrock et~al.(1988)Hamrock, Pan, and Lee]{hamrock_pan_1988}
B.J. Hamrock, P.~Pan, and R.T. Lee.
\newblock Pressure spikes in elastohydrodynamically lubricated conjunctions.
\newblock \emph{Journal of Tribology}, 110\penalty0 (2):\penalty0 279--284,
  1988.

\end{thebibliography}
\end{document}